\documentclass[aps,prd,10pt,notitlepage,nofootinbib,superscriptaddress,showkeys,showpacs]
{revtex4-1}

\usepackage{amsfonts, amsmath, amssymb, amsthm,latexsym}

\usepackage{color}
\usepackage{stmaryrd,wasysym,upgreek,mathrsfs,dsfont}
\usepackage{graphicx}
\usepackage[babel=true]{csquotes}
\usepackage{amsmath}
\usepackage{bbm}
\usepackage{hhline}

\DeclareMathOperator{\Tr}{Tr}

\newcommand{\be}{\begin{equation}}
\newcommand{\ee}{\end{equation}}
\newcommand{\bea}{\begin{eqnarray}}
\newcommand{\eea}{\end{eqnarray}}

\newcommand{\bbbone}{{\mathds{1}}}

\newcommand{\bee}{\begin{equation}}

\newcommand{\cB}{\mathcal{B}}
\newcommand{\cF}{\mathcal{F}}

\newcommand{\cE}{\mathcal{E}}
\newcommand{\cV}{\mathcal{V}}

\newcommand{\cI}{\mathcal{I}}
\newcommand{\cJ}{\mathcal{J}}

\newcommand{\cS}{\mathcal{S}}
\newcommand{\cH}{{\cal{H}}}

\newcommand{\bM}{\mathbb{M}}

\newcommand{\bN}{\mathbb{N}}

\newcommand{\un}{\mathds{1}}
\newcommand{\unk}{\bbbone^{\otimes (k+1)}}

\newcommand{\prf}{{\noindent \bf Proof\; \; }}

\newtheorem{lemma}{Lemma}
\newtheorem{definition}{Definition}
\newtheorem{theorem}{Theorem}

\begin{document}

\title{Intermediate Field Representation\\
for Positive Matrix and Tensor Interactions}

\author{{\bf Luca Lionni}}\email{luca.lionni@th.u-psud.fr}
\affiliation{Laboratoire de Physique Th\'eorique, CNRS UMR 8627, Universit\'e Paris XI, 91405 Orsay Cedex, France, EU}
\affiliation{LIPN, UMR CNRS 7030, Institut Galil\'ee, Universit\'e Paris 13, Sorbonne Paris Cit\'e, 99, avenue Jean-Baptiste Cl\'ement, 93430 Villetaneuse, France, EU}
\author{{\bf Vincent Rivasseau}}\email{vincent.rivasseau@th.u-psud.fr}
\affiliation{Laboratoire de Physique Th\'eorique, CNRS UMR 8627, Universit\'e Paris XI, 91405 Orsay Cedex, France, EU}

\begin{abstract}
In this paper we introduce an intermediate field representation
for random matrices and random tensors with positive (stable)
interactions of degree higher than 4. This representation 
respects the symmetry axis responsible for positivity. It is non-perturbative and allows to prove that such models 
are Borel-Le Roy summable of the appropriate order in their coupling constant. However we have not been able yet
to associate a convergent Loop Vertex Expansion to this representation, hence our Borel
summability result is not of the optimal expected form when the size $N$ of
the matrix or of the tensor tends to infinity. 
\end{abstract}

\noindent  Pacs numbers: 02.10.Ox, 04.60.Gw, 05.40-a

\maketitle

\section{Introduction}

The functional integrals of quantum field theory are often considered simply as \emph{formal} expressions. 
Certainly they are the generating functions for Feynman graphs and their amplitudes
in the sense of \emph{formal power series}.  However their non-perturbative content is essential to their
physical interpretation, in particular for investigating stability of the vacuum and  the phase structure of the theory. 

It is perhaps the main result of the constructive quantum field theory program \cite{Erice,GJ,VR} that the functional integrals
of many (Euclidean) quantum field theories with quartic interactions are the Borel sum of their renormalized perturbative series 
\cite{EMS,MS,Feldman:1986ax}. This is a crucial fact because Borel summability means 
that there is a \emph{unique} non-perturbative definition of the theory, independent of the particular cutoffs used as intermediate tools.
It is less often recognized that such a statement also means that all information about the theory
is in fact \emph{contained} in the list of coefficients of the renormalized perturbative series. It includes in particular 
all the so-called ``non-perturbative" issues. Of course to extract such information 
often requires an analytic continuation beyond the domains which constructive theory 
currently controls.

Since Borel summability is such an essential aspect of local quantum field theory with quartic interactions, one should try to generalize
it both to higher order interactions and to non-local ones. This paper is a small step in these two important research directions.

Generalized quantum field theories with non-local interactions
might indeed hold the key to a future \emph{ab initio} theory of quantum gravity. 
To get rid of the huge symmetry of general relativity under diffeomorphisms (change of coordinates), discretized 
versions of quantum gravity based on random tensor models have received recently  increased attention
\cite{Rivasseau:2013uca}. 
Random matrix and tensor models can indeed be considered as a kind of simplification of Regge calculus \cite{Regge}, which one could call 
simplicial gravity or \emph{equilateral} Regge calculus \cite{ambjorn}. Other important discretized approaches to quantum gravity
are the causal dynamical triangulations \cite{scratch,Ambjorn:2013apa} and
group field theory \cite{boulatov,laurentgft,Krajewski:2012aw,Geloun:2009pe}, in which either causality constraints or holonomy and simplicity constraints are added 
to bring the discretization closer to the usual formulation of  general relativity in the continuum. 

Random matrices are relatively well-developed and have been used 
successfully for discretization of two dimensional quantum gravity\cite{mm,Kazakov,matrix}. They have interesting 
field-theoretic counterparts, such as the renormalizable Grosse-Wulkenhaar model \cite{Grosse:2004yu,Grosse:2004by,Disertori:2006uy,Disertori:2006nq,Grosse:2009pa,Grosse:2012uv,Grosse:2014lxa,Grosse:2015fka}.

Tensor models extend matrix models and 
were therefore introduced as promising candidates for an \emph{ab initio} quantization of gravity 
in rank/dimension higher than two \cite{ADT2,sasa,gross,ambjorn}.
However their study is much less advanced since they lacked for a long time an analog of  't Hooft 
$1/N$ expansion for random matrix models \cite{Hooft} to probe their large $N$ limit.
Their recent modern reformulation
\cite{Gurau:2009tw,Gurau:2011xp,Gurau:2011kk,Bonzom:2012hw} considered
\emph{unsymmetrized} random tensors, a crucial improvement. 
Such tensors in fact have a large and truly tensorial symmetry, typically in the complex case
a $U(N)^{\otimes d}$ symmetry at rank $d$ instead of the single $U(N)$ of symmetric tensors. 
This larger symmetry allows to probe their large $N$ limit through 
$1/N$ expansions of a new type \cite{Gurau:2010ba,Gurau:2011aq,Gurau:2011xq,Bonzom:2012wa,Bonzom:2015axa,Bonzom:2016dwy}.

Random tensor models can be further divided into fully invariant models, in which both propagator and interaction are invariant, 
and field theories in which the interaction is invariant but the propagator is not \cite{BenGeloun:2011rc}. 
This propagator can incorporate or not a gauge invariance of the Boulatov group field theory type. In such field theories the
use of tensor invariant interactions is the critical ingredient allowing in many cases for their successful renormalization \cite{BenGeloun:2011rc,Geloun:2012fq,Samary:2012bw,Geloun:2013saa,Carrozza:2012uv,Carrozza:2013wda}. Surprisingly
the simplest just renormalizable models turn out to be asymptotically free \cite{BenGeloun:2012pu,BenGeloun:2012yk,Geloun:2012qn,Samary:2013xla,Rivasseau:2015ova}.

In all examples of random matrix and tensor models, the key issue is to understand in detail the limit
in which the matrix or the tensor has many entries. Accordingly, 
the main constructive issue is not simply Borel summability but uniform Borel summability
with the right scaling in $N$ as $N \to \infty$. 
In the field theory case the corresponding key issue is to prove Borel summability of the
\emph{renormalized} perturbative expansion without cutoffs. 

Recent progress has been fast on both fronts \cite{Rivasseau:2016rgt}. 
On one hand, \emph{uniform} Borel summability in the coupling constant has been proven for vector, matrix and tensor \emph{quartic} 
models  \cite{Rivasseau:2007fr,MNRS,Gurau:2013pca,Delepouve:2014bma,Gurau:2014lua}, based on the loop vertex expansion (LVE) \cite{Rivasseau:2007fr,MR1,Rivasseau:2013ova}, 
which combines an intermediate field representation with the use of a \emph{forest formula} \cite{BK,AR1}. On the other hand,
Borel summability of the \emph{renormalized} series has been proved
for the simplest super-renormalizable tensor field theories \cite{Delepouve:2014hfa,Lahoche:2015yya,Lahoche:2015zya}, using 
typically a multi-scale loop vertex expansion (MLVE) \cite{Gurau:2013oqa}, which combines
an intermediate field representation with the use of a  \emph{two-level jungle formula} \cite{AR1}.

What are the next steps in this program? One obvious direction is to generalize these results to more difficult super-renormalizable 
and to just renormalizable quartic models. But in matrix models as well as tensor models it can be also important 
to consider higher order interactions. They allow for multi-critical points \cite{Bonzom:2012np} and are also essential in the search for 
interesting new models with enhanced $1/N$ expansions \cite{BLR,Bonzom:2016dwy}.

However it remains a difficult issue to generalize uniform Borel summability theorems 
to higher-than-quartic positive interactions. Proving that such higher 
order models admit an intermediate field integral representation involving non-singular determinants is a first step in that direction. Such a representation has been indeed the cornerstone of all results mentioned above in the quartic case. 

In a previous paper \cite{LR}  the authors have derived such a representation for the (zero dimensional, combinatorial) 
scalar $(\phi\bar\phi)^{k}$ model. In the first part of this paper we provide the generalization of this representation to random matrix models 
with even positive interaction  $\frac{\lambda}{N^{k-1}} \Tr(MM^\dagger)^{k}$ of arbitrary order $k$. Our main result is to
prove in this representation Borel-Le Roy summability of the right order $m=k-1$ for the perturbative expansion in powers of $\lambda$.
However the lower bound on the Borel radius which we prove shrinks as $N \to \infty$, unlike the true radius which 
is expected not to shrink as $N \to \infty$.

We then turn to tensor models. We introduce first 
a definition of positivity for connected tensor interactions of any order. It is based on existence of a Hermitian symmetry axis which allows
to write the interaction as as scalar product in a certain tensor space of arbitrarily high rank. Then we prove that such positive random 
tensor models admit an intermediate field representation which we call Hermitian since it respect in a certain sense this symmetry axis. 
This Hermitian representation is different from the one introduced in \cite{BLR}. In contrast with the latter, it
holds in a non-perturbative sense. Our main result is to establish, in that Hermitian representation,
Borel-Le Roy summability of the right order for the initial perturbative expansion, although again not with the right expected scaling
of the Borel radius as $N \to \infty$.

We can therefore consider this paper both as a first step in the extension of the quartic constructive methods to higher order interactions, and
as a constructive counterpart of the perturbative intermediate field-type representation of general tensor 
models through stuffed Walsh maps introduced in \cite{BLR}. 

Our main results (Theorems \ref{Borelmatrix}, \ref{shrinkmatrix}, \ref{inttensrep} and \ref{Boreltensor}) prove that in the case of positive interactions this 
new intermediate field representation contains exactly the same non-perturbative information than the initial
representation.

However we are not fully satisfied with the current situation. Indeed we cannot prove, in any of the representations, 
that the Borel radius of analyticity scales in the expected optimal way for $N$ large. More precisely we conjecture that 
Borel summability holds uniformly in $N$ after a suitable rescaling of the coupling constant by a certain optimal power of $N$.
However let us stress the difficulty of the problem. First, even in the simpler matrix case we have not been able yet to prove this last result, which is postponed to a future study. Second in the case of  tensor models, even an \emph{explicit formula} for the optimal perturbative rescaling is not yet known for the most general invariants \cite{BLR}.

The plan of this paper essentially extends the one of \cite{LR}, as we follow the same strategy.
In section \ref{prereq} we provide the mathematical prerequisites about the contour integrals and kind of Borel theorems we shall use.
In section \ref{matrep} we consider the matrix case and give its Hermitian intermediate field representation. In section
\ref{tenrep} we provide a similar Hermitian intermediate field representation for a general positive tensor interaction. Several cases are treated 
explicitly in detail: sixth order interactions (which are all planar) and a particular non-planar tenth order interaction. 

\section{Prerequisites}\label{prereq}

This section essentially reproduces for self-contained purpose material already contained in \cite{LR}.

\subsection{Imaginary Gaussian Measures}\label{imag}
The ordinary normalized Gaussian measure of covariance $C>0$ on a real variable $\sigma$
will be noted as $d\mu_C$
\bee  d \mu_C (\sigma) =  \frac{1}{\sqrt{2 \pi  C}}   e^{- \frac{\sigma^2}{2C}} .
\ee

Consider a function $f(z)$ which is analytic in the strip $\Im z \le \delta$ and exponentially bounded in that 
domain by $K  e^{\eta \vert z \vert} $ for some $0 \le  \eta  < \delta$, where $K$ is some constant.

\begin{definition}
We define the \emph{imaginary Gaussian integral} of $f$ with covariance $\pm i  C$, where $C >0$, by
\be  \int d \mu_{\pm i  C} (x) f (x)  := \int_{C_{\pm ,\epsilon}}  \frac{e^{- z^2  /\pm 2 i C} dz}{\sqrt{\pm 2 \pi i C}} f(z)  \label{imgauss}
\ee
where the contour $C_{\pm , \epsilon}$
can be for instance chosen as  the graph in the complex plane (identifying ${\mathbb C}$ to ${\mathbb R}^2$) 
of the real function $\Re z = x \to \Im z =y = \pm \epsilon \tanh (x)$
for any $\epsilon \in ]C\eta, \delta [$.
\end{definition}

\begin{figure}[!htb]
\centering
\includegraphics[width=10cm]{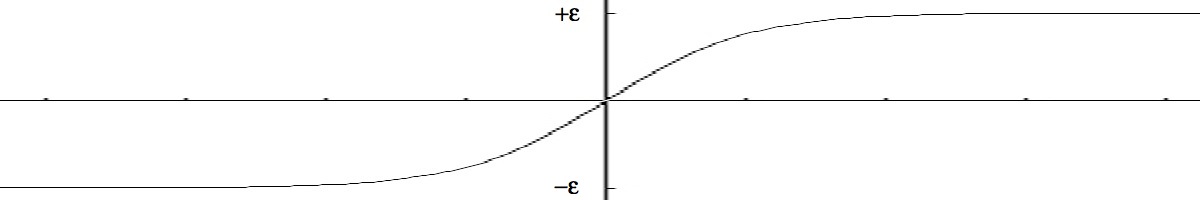}
\caption{The integration contour $C_{+,\epsilon}$.}
\label{contour+}
\end{figure}

Remark indeed that from our hypotheses on $f$, the integral \eqref{imgauss} is well defined 
and absolutely convergent for $C\eta < \epsilon < \delta$, and by Cauchy theorem, independent
of $\epsilon \in ]C\eta, \delta [$. The contour $C_{+, \epsilon}$ is shown in Figure \ref{contour+}. Of course the choice of the $\tanh $
function is somewhat arbitrary, as by Cauchy theorem,
many contours of similar shape would lead to the same integral.

Notice that such imaginary Gaussian integrals oscillate, in contrast with the ordinary ones. They
are not \emph{positive}, hence in particular they are not \emph{probability measures}.
Also remark that although the result of integration does not depend 
on the contour, actual bounds on the result typically depend on choosing particular contours for which $\epsilon$ is not too small.

Remark also that if $f$ is a polynomial, the Gaussian rules of integration (found by Isserlis and called Wick theorem in physics)
apply to such imaginary Gaussian integrals. More precisely, defining $(2n -1)!! := (2n-1) (2n-3) \cdots 5.3.1$, we have
\be   \int d \mu_{\pm i C}(x) x^{2n}=   (\pm i C)^n (2n-1)!! \ .
\ee
This is easy to check since a polynomial
is an entire function and we can deform the contour into $z = x+ ix$, in which case we recover an ordinary Gaussian integration.
Similarly
\be   \int d \mu_{\pm iC}(x) e^{ax}=   e^{\pm i Ca^2 /2}  , \label{contexp}
\ee
the integral being absolutely convergent for any contour such that $C\vert a \vert < \epsilon$ \footnote{We can extend these
formulas to the case $C=0$ by defining $d\mu$ in this case to be the Dirac measure at the origin.}.

We can also define \emph{imaginary complex} normalized Gaussian integrals $ d \mu^c_{\pm i C}(z) $ 
of covariance $\pm i$ for a complex variable $z$.
This is in fact just the same as considering a pair of identical imaginary normalized Gaussian integrals of the previous type, one for the real part
and the other for the imaginary part of $z$.
Let us make this explicit. An ordinary complex Gaussian measure is usually written as
\be d \mu^c_{C}(z)  := \frac{e^{- \vert z \vert^2  /C} dz d \bar z}{\pi  C} \label{intcompcont1}
\ee
so that 
\be \int d \mu^c_{C}(z)   z^m \bar z^{n}  = \delta_{mn} C^n  n!  \label{intcompcont2}
\ee
What we call imaginary complex normalized Gaussian integrals $ d \mu^c_{\pm i C}(z) $  are defined by 
writing $z = x+iy$ and separately writing two complex line integrals, one for $x$ and one for $y$.
Let us rewrite a function $f(z)$ as a function of the real and imaginary parts of $z$ as $f(x+iy)= \phi (x,y)$.
Assuming that $\phi$ admits an analytic continuation both in $x$ and $y$ 
in the product strip $\Im x \le \delta, \Im y \le \delta$ and is exponentially bounded in that 
domain by $K  e^{\eta ( \vert x \vert +  \vert y \vert )} $ for some $0 \le  \eta  < \delta$, where $K$ is some constant, we define
\be \int d \mu^c_{\pm i C}(z) f(z) = \frac{1}{\pm i \pi C} \int_{-\infty}^{+\infty} dx    \int_{-\infty}^{+\infty} dy e^{-  [(x \pm  i \epsilon \tanh x)^2 + (y \pm  i \epsilon \tanh y)^2]/\pm iC } \phi (x \pm  i \epsilon \tanh x , y \pm  i \epsilon \tanh y)
\ee
for which we have indeed the integration rules generalizing \eqref{intcompcont1}-\eqref{intcompcont2}
\be   \int d \mu^c_{\pm iC}(z) z^m \bar z^{n}=  \delta_{mn} (\pm i C)^n  n!   \label{compper1}
\ee
and
\be   \int d \mu^c_{\pm i C}(z) e^{a z + b \bar z }=   e^{\pm i a b C}  ,
\ee
again the last integral being absolutely convergent if $C\sup\{\vert a \vert, \vert b \vert \} < \epsilon$.

Following \cite{LR}, we would also like to introduce some notations for
pairs of imaginary Gaussian integrals with opposite imaginary covariances.
More precisely suppose we have two variables $a$ and $b$, one with covariance $-i$ and the other $+i$
and corresponding integration contours in the complex plane of the type above. This integration being denoted
as $d\mu_{\pm i}(a, b) := d\mu_{-i} (a) d\mu_{i} (b) $, we can perform the simple change of variables

\begin{eqnarray}
\label{eqref:ChdeVr}
\alpha=\frac{a+b}{\sqrt2} , \qquad\qquad \beta=\frac{a-b}{\sqrt2},
\end{eqnarray}
and define the  imaginary conjugate pair Gaussian integral $d\mu_X (\alpha, \beta)$ by its moments
\begin{eqnarray}
\label{eqref:CrossMeas1}
 <\alpha \beta>_X=-i,\  \ \quad<\alpha^2>_X=0,\  \ \quad<\beta^2>_X=0 .
\end{eqnarray}
Hence this pair integral has two by two covariance  
$X= \begin{pmatrix} 0  & - i \\ - i & 0  \end{pmatrix}$. Remark that the two dimensional surface of
integration in  ${\mathbb C}^2$ remains defined by the one of the $a,b$ variables.
Hence it is parametrized by $(\frac{\alpha+\beta}{\sqrt2}, \frac{\alpha-\beta}{\sqrt2}) \in  C_{- , \epsilon} \times C_{+, \epsilon}$.

We should now define a more general class of Gaussian
integrals combining a finite number of ordinary Gaussian 
measures and imaginary conjugate pair integrals.

\begin{definition} \label{defmixed}
We call mixed Gaussian integral the product of finitely many ordinary real Gaussian measures and finitely 
many Gaussian imaginary conjugate pair integrals. 
\bee  d\nu(\xi) = \prod_{i=1}^p  d\mu_1  (\sigma_i)  \prod_{j=1}^q  d\mu_X (\alpha_j, \beta_j).
\ee
\end{definition}
We have a similar notion also for the complex integrals, which have twice as many real variables than the real ones.

Finally we can define such imaginary Gaussian measures for vectors, matrix or tensor variables 
$\vec x = (x_1, \cdots x_n)$ (real or complex) and a positive covariance matrix $C$ by deforming 
separately each integration contour for all real components, 
staying in the analyticity domain of the function $f$ where it remains exponentially bounded. 
For instance the imaginary Gaussian unitary ensemble with covariance $\pm i$ is defined as 
a product of normalized imaginary Gaussian integrals of covariance $\pm i$ 
on each of the $N^2$ real components $\{ H_{ii}, 1 \le i \le n, 
H_{ij}, 1 \le  i<j \le N \}$ of a Hermitian matrix $H$.

\subsection{Borel-LeRoy-Nevanlinna-Sokal Theorem}
\hspace{0.3cm}

We note $R^q f$ the $q$-th order Taylor remainder of a smooth function $f(\lambda)$. It writes 
\bea  R^q  f = \lambda^{q}
\int_0^1 \frac{(1-t)^{q-1}}{(q-1)!} f^{(q)} ( t \lambda) dt.
\eea

\begin{theorem}(Borel-LeRoy-Nevanlinna-Sokal)\label{blrsok} \\

A power series $\sum_{n=0}^\infty\frac{a_n}{n!}\lambda^n$ is Borel-Le Roy summable of order $m$ 
to the function $f(\lambda)$  if the following conditions are met:

\begin{itemize}
 \item For some real number $\rho>0$, $f(\lambda)$ is analytic in the domain $D^m_\rho=\{\lambda\in \mathbb C: 
\Re \lambda^{-1/m}> \rho^{-1}\}$.
\item The function $f(\lambda)$ admits $\sum_{n=0}^\infty a_n \lambda^n$ as a strong 
asymptotic expansion to all orders as $|\lambda|$ $\rightarrow 0$ with uniform estimate in $D^m_\rho$:
\begin{equation}\label{borelrem}
\left|  R^q f  \right|\leqslant A B^q \Gamma(mq)|\lambda|^{q}.
\end{equation}
where $A$ and $B$ are some constants and $\Gamma$ the usual Euler function.
\end{itemize}
Then the Borel-Le Roy transform of order $m$, which is
\begin{equation}
B^{(m)}_f(u)=\sum_{n=0}^\infty \frac{a_n}{\Gamma(mn+1)}u^n,
\end{equation}
is holomorphic for $|u|<B^{-1}$, it admits an analytic continuation to the strip 
$\{u\in \mathbb C: |\Im u|< R, \Re u>0\}$ for some $R >0$, and for $\lambda \in D^m_\rho$
\begin{equation}
 f(\lambda)=\frac{1}{m\lambda}\int_{0}^{\infty}B^{(m)}_f (u) e^{-(\frac{u}{\lambda})^{\frac{1}{m}}}\bigl(\frac{u}{\lambda}\bigr)^{\frac{1}{m-1}} du .
\end{equation}
\end{theorem}
For $m=1$ remark that $D^1_\rho$ is simply the usual disk of diameter $\rho$ 
tangent at the origin to the imaginary axis on the positive real part side of the complex plane.
The proof of this theorem is a simple rewriting exercise on the usual Nevanlinna theorem  \cite{Sok} but in the variable $\lambda^{1/m}$.\\

\section{Matrix models}\label{matrep}
\hspace{0.3cm}

\begin{definition} \label{hifrepre}
We say that a function $Z$ has an Hermitian Intermediate Field (HIF) representation
if it can be written as a convergent integral of the type 
\be
\label{PrtF1}
Z=\int  d\nu(\xi)e^{-\Tr\ln\bigl[  \un- \bM(\xi) \bigr] },
\ee
in which $d \nu ( \xi)$ is a mixed Gaussian measure in the sense of the previous section, 
$\bM(\xi)=iC\mathbb{H}(\xi)$, 
the matrix $\mathbb{H}$ is linear in term of all the $\xi$ variables, is Hermitian in term of their real parts (i.e. if we set $\epsilon=0$
in all contours),
and $C$ is a mixed covariance, i.e. a sum of blocks made of diagonal 1 and finitely many 
$X= \begin{pmatrix} 0  & - i \\ - i & 0  \end{pmatrix}$ factors.
\end{definition}

We consider now a complex, size-$N$, one-matrix model with $U(N)^{\otimes 2}$ invariant interaction of order $2k$, where 
$N \in \mathbb {N}^\star $. We always write $k-1=m$ in what follows since it will be the order of Borel summability to consider,
and use the notation $O(1)$ for some constants independent of $N$ and $k$ whose exact values are not essential. The partition function of
the model is
\be \label{directrep}
Z_k(\lambda,N) :=\int d\mu^c(M)\exp\bigl[-\frac{\lambda}{N^{m}}\Tr(MM^\dagger)^k\bigr],
\ee
where $d\mu^c$ is the normalized Gaussian measure $d\mu^c(M)=dMdM^\dagger e^{-\Tr MM^\dagger}$, $dMdM^\dagger=\pi^{-N^2}\prod_{i,j=1}^Nd  M_{ij}d\bar M_{ij}$. The associated free energy is
\be
F_k(\lambda,N) := N^{-2} \log Z_k(\lambda,N).
\ee

It is a famous consequence of 't Hooft $1/N$ expansion \cite{Hooft} that the $N^m=N^{k-1}$ scaling
in \eqref{directrep} is the correct one to ensure a non-trivial perturbative limit of $F_k(\lambda,N)$ when $N \to \infty$.
More precisely with this particular scaling each (connected, vacuum) Feynman amplitude with $q$ vertices scales as $\lambda^q N^{-2g}$, where $g$
is the genus of the surface dual to the Feynman (ribbon) graph. Therefore in the sense of \emph{formal power series},
$\lim_{N \to \infty} F_k(\lambda,N) = F_k^{planar}(\lambda) $ exists and is the generating function of \emph{planar} ribbon graphs 
with (bipartite) vertices of degree $2k$ and coupling constant $\lambda$.

However it is much more difficult to perform the planar limit  $\lim_{N \to \infty} F_k(\lambda,N) = F_k^{planar}(\lambda) $
not in the sense of formal power series but in a constructive sense, 
that is as a uniform limit of functions analytic in a well-defined domain.
In the case of a quartic matrix interaction, the Loop Vertex Expansion 
which combines an intermediate field representation with a forest formula 
was introduced precisely to settle this issue \cite{Rivasseau:2007fr,Gurau:2014lua}. 
For instance it allows to prove the following:
\begin{theorem}\label{Borelquarticmatrix}
There exist some $\rho>0$ such that all functions $F_2(\lambda,N)$ 
are analytic and Borel-summable (of ordinary order $m=1$) in $\lambda$ in an $N$-independent cardioid domain $Card_\rho$
defined in polar coordinates by $\lambda = \rho e^{i \phi}$, $\phi \in ]-\pi , \pi[ $ and
$\rho < [ \cos (\phi/2 )]^{2} $. This domain is shown in Figure \ref{cardioidfig}.

\begin{figure}[ht] 
\centerline{\includegraphics[width=5cm,angle=0]{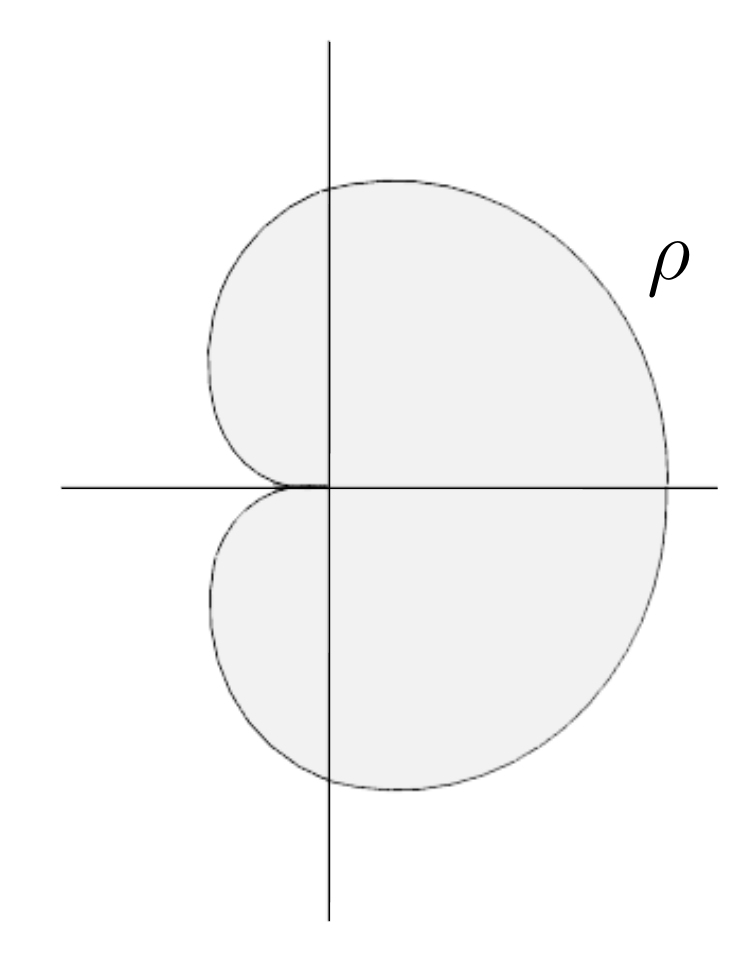}}\label{cardioidfig}
\caption{A cardioid domain}
\end{figure}

Furthermore in that domain the sequence $F_2(\lambda,N)$ uniformly 
converges as $N \to \infty$ to the function $F_2^{planar}(\lambda) $. This limit is not only Borel summable
 in $Card_\rho$
but also analytic in a disk centered at the origin containing $Card_\rho$.
\end{theorem}
The proof of this Theorem can be found in \cite{Gurau:2014lua}. It states in particular 
that for $k=2$ the analyticity domain for $F_2(\lambda,N)$ does not shrink as $N \to\infty$.

However a constructive version of `t Hooft expansion remains to our knowledge a completely 
open issue in the case $k \ge 3$, hence for higher than quartic matrix trace interactions. 
We shall prove in this section the following result in this direction.

\begin{theorem}\label{Borelmatrix}
The partition function $Z_k(\lambda,N)$ and free energy $F_k(\lambda,N)$ are 
Borel-LeRoy summable of order $m=k-1$, in the sense of Theorem \ref{blrsok}.
More precisely they are analytic in $\lambda$ in the \emph{shrinking} domain 
$D^m_{\rho_m(N)} =\{\lambda\in \mathbb C:  \Re \lambda^{-1/m}> [\rho_m(N)]^{-1}\}$ with
$ \rho_m (N) = N^{-1-2/m} \rho_m$, $\rho_m >0$ independent of $N$. In that domain 
$Z_k(\lambda,N)$ has an HIF representation in the sense of Definition \ref{hifrepre}
\be
\label{HIFIGrep}
Z_k(\lambda,N)=\int  d\nu(\xi)e^{-N \Tr\ln\bigl[  1-g_k \mathbb{M}_k(\xi) \bigr] },
\ee
in which  

\begin{itemize}

\item the Gaussian measure $d \nu ( \xi)$ is of the mixed Gaussian type in the sense of Definition \ref{defmixed}

\item
$g_k=(\frac{\lambda}{N^{m}})^{\frac{1}{2k}}$, 

\item
the matrix  $\mathbb{M}_k(\xi)=iC_k.\mathbb{H}_k(\xi)$ is a $(k+1)N\times (k+1)N$ matrix,

\item
$C_k$ is a mixed covariance , i.e. a sum of blocks of type $\begin{pmatrix} 1  & 0 \\ 0 & 1  \end{pmatrix}$ 
and $\begin{pmatrix} 0  & - i \\ - i & 0  \end{pmatrix}$ factors, 

\item
$\mathbb{H}_k$ is linear in the $\{\xi \}$ variables and is 
Hermitian when these variables are taken on undeformed contours, i.e. at $\epsilon = 0$.

\end{itemize}
\end{theorem}

This theorem is proven in the following subsections.  We conjecture
that for $k\ge 3$, just as for $k=2$, there should exist an analyticity domain common to all functions $Z_k(\lambda,N)$ 
and $F_k(\lambda,N)$ of the type  $\{\lambda\in \mathbb C: \Re \lambda^{-1/m}>  \rho^{-1}_m\}$ for some $\rho_m >0$ independent of $N$,
hence which does not shrink as $N \to \infty$.
However for the moment we do not know how to prove this stronger
result, either using \eqref{directrep} or \eqref{HIFIGrep}. We feel \eqref{HIFIGrep} may be 
a better starting point to attack this conjecture since in the case $k=2$, the only one in which the conjecture is proved, the proof 
uses such an intermediate field representation rather than the direct representation \cite{Rivasseau:2007fr,Gurau:2014lua}.\\

\subsection{Formal intermediate field decomposition}
\hspace{0.cm}

To prove Theorem \ref{Borelmatrix} let us first introduce the Hermitian
intermediate field representation for $Z_k$,  
following the strategy of \cite{LR}. 
This calls for a repetitive use of the Hubbard-Stratonovich decomposition applied to matrices
\be
\label{eqref=HS1}
e^{-g\Tr(AB)}=\int d\mu^c(\tau)e^{i\sqrt g\,\Tr(A\tau + B\tau^\dagger)},
\ee
and variations for intermediate fields of imaginary covariances $\pm i$
\be
\label{eqref=HS2}
e^{ig\Tr(AB)}=\int d\mu^c_{\pm i}(\tau)e^{i\sqrt g\,\Tr(A\tau \mp B\tau^\dagger)}.
\ee

Throughout this subsection we shall consider all Gaussian imaginary integrals in the $\epsilon \to 0$ \emph{formal limit}.
Therefore these integrals are not absolutely convergent, and the mathematically inclined reader may consider the 
computations of this subsection just as heuristic. 
However in the next subsection, the $\epsilon$ regulator will be reintroduced for
all imaginary covariances, resulting in well defined convergent integrals confirming rigorously all these 
heuristic computations.\\

\begin{figure}[h!]
\includegraphics[scale=0.7]{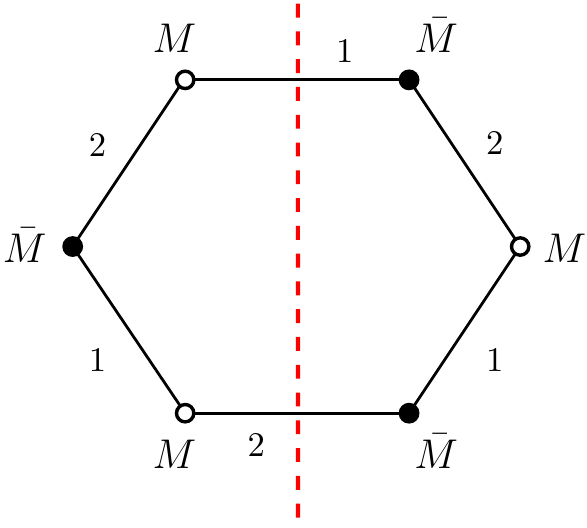}\hspace{2cm}\includegraphics[scale=0.9]{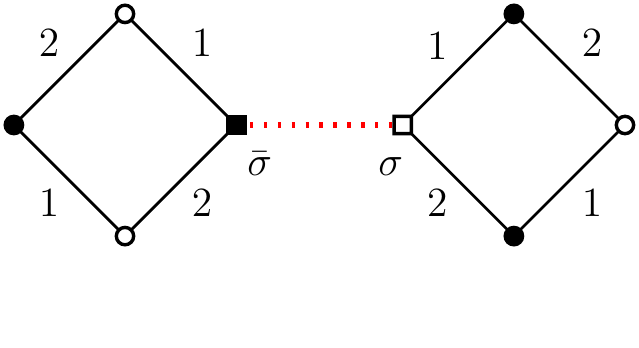}
\caption{\label{fig:mtx3n} $k=3$ Matrix invariant and first intermediate field step.}
\end{figure}

In all following sections, we shall use a graphical representation of the trace invariants in the action and of their iterated splittings. While it could seem a bit trivial in this section, it will prove very useful in the section treating tensorial invariants. We shall represent an integrated matrix $M$ by a white vertex and its complex conjugate by a black vertex. Edges between vertices carry a color, 1 or 2, and represent the contraction of the first or the second index between the two vertices. A matrix trace invariant of the form $\Tr((MM^\dagger)^k)$ is therefore pictured as a bipartite cycle, 
with alternating colors 1 and 2. This is shown in the $k=3$ case on the left of Figure \ref{fig:mtx3n},
\begin{figure}[h!]
\includegraphics[]{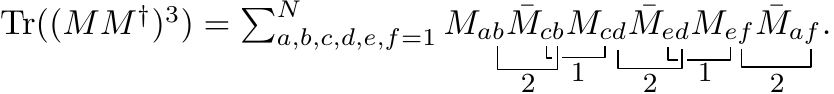}
\label{eqref:trace} 
\end{figure}

Intermediate fields will be represented by black and white squares or triangles. As we shall see later, the 
associated graphs are not generally bipartite cycles of alternating colors.\\

Using relation (\ref{eqref=HS1}), we first split the interaction term in two (Fig. \ref{fig:mtx3n}), depending on the parity of $k$, 
\bea
&e^{-\frac{\lambda}{N^{k-1}}\Tr(MM^\dagger)^{2p+1}}&=\int d\mu^c(\sigma)e^{i\sqrt \frac{\lambda}{N^{m}} \Tr[(MM^\dagger)^p(\sigma M^\dagger + M\sigma^\dagger)]}, \\
&e^{-\frac{\lambda}{N^{k-1}}\Tr(MM^\dagger)^{2p}}&=\int d\mu^c(\sigma)e^{i\sqrt \frac{\lambda}{N^{m}} \Tr[(MM^\dagger)^p(\sigma +\sigma^\dagger)]}.
\eea

Defining $g_k=(\frac{\lambda}{N^{k-1}})^{\frac{1}{2k}}$, we now rewrite the interactions as 
\bea
\sqrt\frac{\lambda}{N^{k-1}}\Tr[(MM^\dagger)^p(\sigma M^\dagger  + M\sigma^\dagger)]
&=&\frac{1}{2}\Tr\bigl[((g_k^2MM^\dagger)^p+g_kM\sigma^\dagger)(g_k\sigma M^\dagger +(g_k^2 MM^\dagger)^p)\nonumber\\
&&\qquad+((g_k^2MM^\dagger)^p-g_kM\sigma^\dagger)(g_k\sigma M^\dagger -(g_k^2MM^\dagger)^p)\bigr],
\\
\nonumber\\
\sqrt\frac{\lambda}{N^{k-1}}\Tr[(MM^\dagger)^p(\sigma + \sigma^\dagger)]
&=&\frac{1}{2}\Tr\bigl[((g_k^2MM^\dagger)^{p-1}M+g_k\sigma^\dagger M)(g_kM^\dagger\sigma+M^\dagger(g_k^2MM^\dagger)^{p-1})\nonumber\\
&&\qquad+((g_k^2MM^\dagger)^{p-1}M-g_k\sigma^\dagger M)(g_kM^\dagger\sigma-M^\dagger(g_k^2MM^\dagger)^{p-1})\bigr].\nonumber
\eea

Now applying (\ref{eqref=HS2}) with complex intermediate fields $a_1$ and $b_1$ of covariances $-i$ and $+i$ respectively, 
\bea
&e^{-\frac{\lambda}{N^{k-1}}\Tr(MM^\dagger)^{2p+1}}&=\int d\mu^c(\sigma)d\mu^c_{\pm i}(a_1,b_1)e^{\frac{i}{\sqrt 2} \Tr\bigl[(g_k^2MM^\dagger)^p(a_1+b_1 + a_1^\dagger +b_1^\dagger)+g_kM\sigma^\dagger(a_1^\dagger-b_1^\dagger)+g_k\sigma M^\dagger  (a_1-b_1)\bigr]}\\
&e^{-\frac{\lambda}{N^{k-1}}\Tr(MM^\dagger)^{2p}}&=\int d\mu^c(\sigma)d\mu^c_{\pm i}(a_1,b_1)e^{\frac{i}{\sqrt 2}  \Tr\bigl[  (g_k^2MM^\dagger)^{p-1}(M(a_1^\dagger +b_1^\dagger)  
+ (a_1+b_1)M^\dagger)+g_k\sigma^\dagger M(a_1^\dagger-b_1^\dagger)+g_kM^\dagger\sigma (a_1-b_1)\bigr]},\nonumber
\eea
where the measure is $d\mu^c_{\pm i}(a_1,b_1)=d\mu^c_{-i}(a_1)d\mu^c_{+i}(b_1)$.\\

Changing variables for 
\be
\alpha_1=\frac{a_1+b_1}{\sqrt 2}\quad ,\qquad\beta_1=\frac{a_1-b_1}{\sqrt 2},
\ee
and complex conjugates, one finds that 
\bea
&e^{-\frac{\lambda}{N^{k-1}}\Tr(MM^\dagger)^{2p+1}}&=\int d\mu^c(\sigma)d\mu^c_{X}(\alpha_1,\beta_1) 
e^{i \Tr\bigl[(g_k^2MM^\dagger)^p(\alpha_1+\alpha_1^\dagger)+g_kM\sigma^\dagger\beta_1^\dagger+g_k\sigma M^\dagger  \beta_1\bigr]}\nonumber\\
&e^{-\frac{\lambda}{N^{k-1}}\Tr(MM^\dagger)^{2p}}&=\int d\mu^c(\sigma)d\mu^c_{X}(\alpha_1,\beta_1)
e^{i  \Tr\bigl[ (g_k^2 MM^\dagger)^{p-1}  (\alpha_1M^\dagger+M\alpha_1^\dagger) +g_k\sigma^\dagger M\beta_1^\dagger+g_kM^\dagger\sigma \beta_1\bigr]},
\eea
the Gaussian measure $d\mu^c_X$ being defined by its moments, that all vanish apart from 
\be
\label{eqref:cov}
\forall j,k \in \{1,...,N\}, \quad<\alpha_{1\mid jk}\bar\beta_{1\mid jk}>_X=<\bar\alpha_{1\mid jk}\beta_{1\mid jk}>_X=-i.
\ee

\begin{figure}[h!]
\includegraphics[scale=.7]{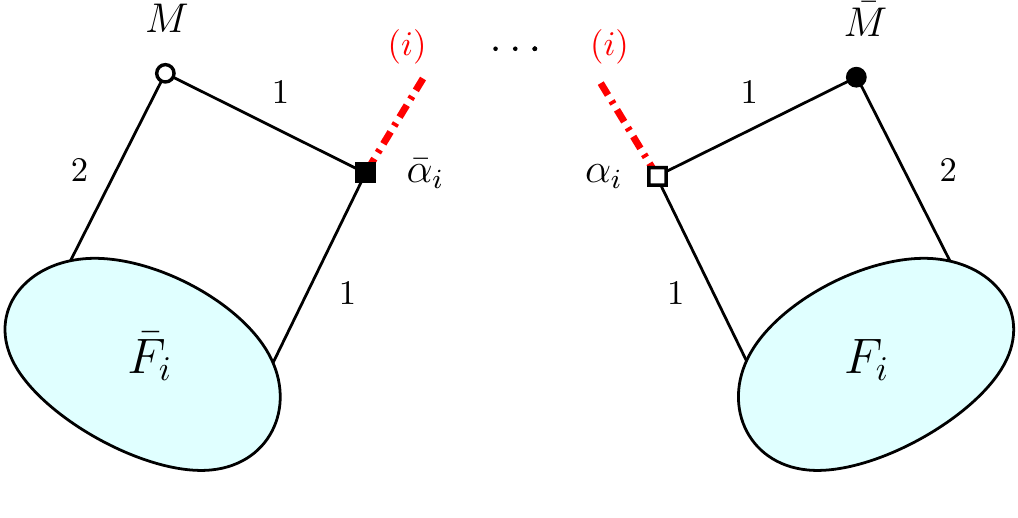}\hspace{2cm}\includegraphics[scale=.7]{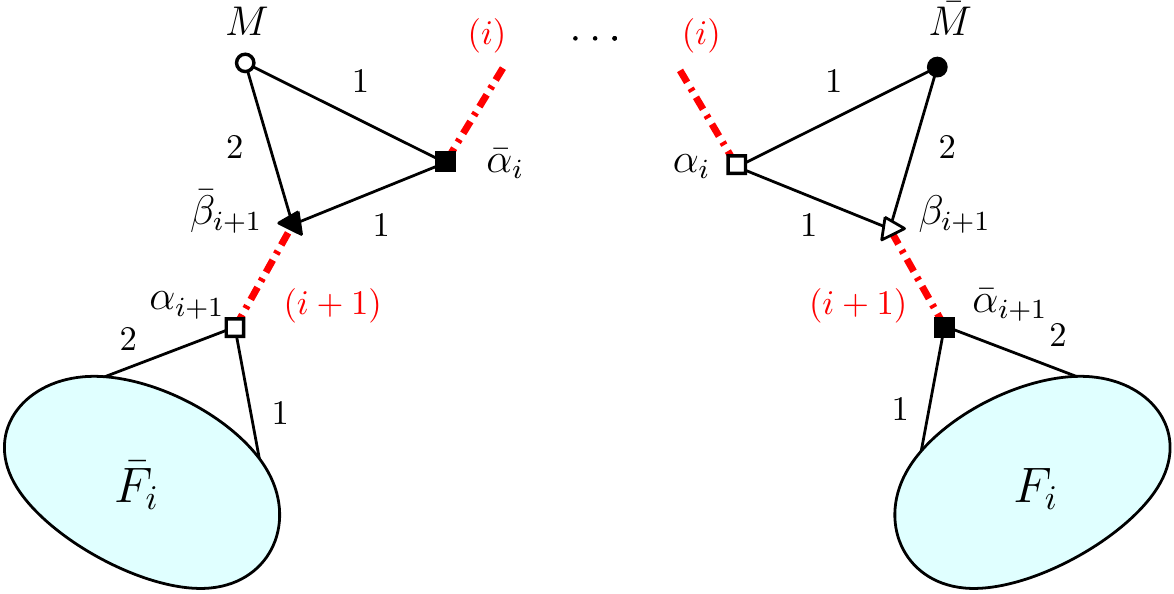}
\caption{\label{fig:odd_split} Step $i$ of the intermediate field decomposition, when $F_i$ has an odd number of vertices. In the even case, the edge between $M$ and $\bar \alpha_i$ and symmetric are of color 2, while the edges toward $F_i$ are both of color 1.}
\end{figure}

Inductively applying the same reasoning leads to the following expressions for the partition function,
\bea
Z_{k, odd}(\lambda,N) &=& \int   d\mu^c(\phi)d\mu^c(\sigma)\prod_{j=1}^{k-2}d\mu^c_{X}(\alpha_j,\beta_j)
e^{  i g_k\Tr\bigl[   M^\dagger\bigl(\beta_1\sigma 
+ \alpha_{1}\beta_{2}  +  \beta_{3}\alpha_{2}   +  \cdots + \beta_{k-2}\alpha_{k-3}  +g_k \alpha_{k-2} M \bigr) 
+ \, c.t.  \bigr]},\nonumber\\
Z_{k, even}(\lambda,N) &=& \int   d\mu^c(\phi)d\mu^c(\sigma)\prod_{j=1}^{k-2}d\mu^c_{X}(\alpha_j,\beta_j)
e^{  ig_k  \Tr\bigl[  M^\dagger \bigl(\sigma\beta_1 
+\beta_{2}\alpha_{1} + \alpha_{2}\beta_{3}  +   \cdots  + \beta_{k-2} \alpha_{k-3}   + g_k \alpha_{k-2} M \bigr) 
+\, c.t.  \bigr]},
\eea
where $c.t.$ stands for conjugate transpose. The $i$'th splitting introduces the matrix intermediate fields $\alpha_{i+1}$ and $\beta_{i+1}$, with complex covariances as in \eqref{eqref:cov} (respectively represented by a square vertex and a triangle vertex) and is represented in Figure \ref{fig:odd_split} (this may also apply for $\alpha_0=\sigma$). Note that the graphs representing the interactions have plain lines.
\\

The partition function rewrites in both cases as
\be
Z_k(\lambda,N)=\int   d\mu^c(M)d\mu^c(\sigma)\prod_{j=1}^{k-2}d\mu^c_{X}(\alpha_j,\beta_j)e^{  ig_k  \Tr\bigl[ M^\dagger R_k+R_k^\dagger M\bigr]},
\ee
where, defining $\alpha_0:=\sigma$, $\beta_{k-1}:=M$ and $\beta_0 = \alpha_{-1} =0$,
\be
R_k(\sigma,\{\alpha_j,\beta_j\})=\sum_{j=1}^{\lfloor\frac{k}{2}\rfloor}(\alpha_{k-2j}\beta_{k-2j+1} + \beta_{k-2j}  \alpha_{k-2j-1}),
\ee 
 

We now integrate over a subset of the intermediate fields using relation
\be
\int d\mu^c_X(\alpha,\beta)e^{i\Tr[A\alpha+B\beta+C\alpha^\dagger+D\beta^\dagger]}=e^{i\Tr[AD+BC]}.
\ee
We choose to integrate over $M$ and all $\alpha_{k-1-2j}$, $\beta_{k-1-2j}$, for 
$j\in\{0,...,\lfloor\frac{k}{2}\rfloor\}$, i.e. 
\begin{itemize}
\item for $k$ odd, over the $k-1$  matrix fields $M$, $\sigma$ and all even $\alpha_{2j}$, $\beta_{2j}$, for $j\in\{1,..,\frac{k-3}{2}\}$ and complex conjugates. 

\item for $k$ even, over the $k-1$ matrix fields $M$ and all odd $\alpha_{2j-1}$, $\beta_{2j-1}$, for $j\in\{1,..,\frac{k-2}{2}\}$ and complex conjugates. 
\end{itemize}

Each integration step is done independently of the others. It gives
\bea
\int d\mu^c_X(\alpha_{k-1-2j},\beta_{k-1-2j})e^{ig_k\Tr\bigl[M^\dagger(\beta_{k-2j}\alpha_{k-1-2j} + \alpha_{k-2(j+1)}\beta_{k-1-2j})+c.t.\bigr]}
=\ e^{ig_k^2\Tr\bigl[M^\dagger\beta_{k-2j} \alpha_{k-2(j+1)}^\dagger M + c.t.\bigr]}, 
\eea
except for the $\sigma$ integration for $k$ odd, which gives
\bea
\int d\mu^c(\sigma)e^{ig_k\Tr\bigl[M^\dagger\beta_{1}  \sigma +  c.t. \bigr]} =\ e^{-g_k^2\Tr\bigl[M^\dagger\beta_{1}  \beta_{1}^\dagger M \bigr]}. \ 
\eea

Let $\xi=(...\alpha_{k-2j}, \beta_{k-2j},...)$ be the vector containing the $k-1$ remaining variables, $\xi_{odd}=(\alpha_1,\beta_1,..., \alpha_{k-2}, \beta_{k-2})$ and $\xi_{even}=(\sigma, \alpha_2,\beta_2,..., \alpha_{k-2} , \beta_{k-2})$. Note that the indices of the remaining intermediate fields have the parity of $k$. The partition function therefore rewrites 

\be
\label{PrtF1a}
Z_k(\lambda,N)=\int  d\nu(\xi)e^{-N \Tr\ln\bigl[  \un -g_k^2 \bigl(i H_k(\xi) -\eta(k) \beta_1\beta_1^\dagger\bigr)\ \bigr] },
\ee
where $\bbbone$ is the $N$ by $N$ identity matrix, $d \nu$ factorizes over the measures $d\mu_X^c$ of each $\alpha,  \beta $ pair plus the measure $d\mu^c (\sigma)$ 
for $k$ even, and $\eta(k)$ is 0 for $k$ even and 1 for $k$ odd, and where we denoted $H_k$ the $N\times N$ Hermitian matrix 
\be
H_k(\xi)=\sum_{j=0}^{\lfloor\frac{k}{2}\rfloor}\beta_{k-2j} \alpha_{k-2(j+1)}^\dagger + c.t. \ .
\ee

\begin{figure}[h!]
\hspace{1.5cm}\includegraphics[scale=0.7]{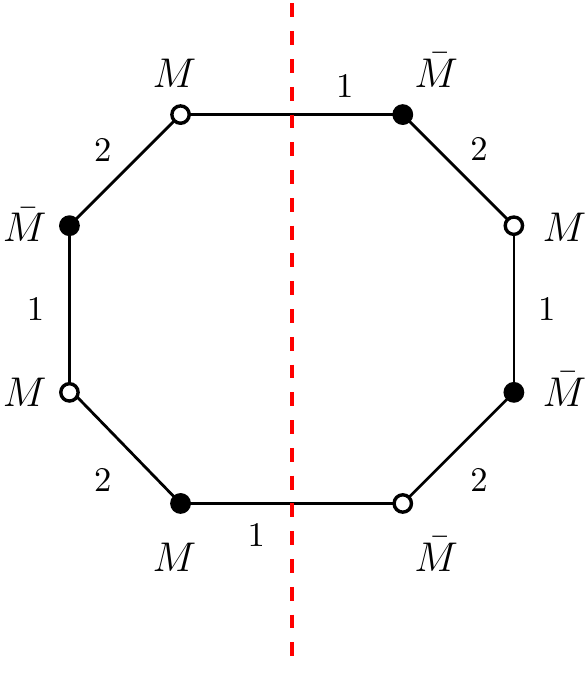}\hspace{2cm}\includegraphics[scale=0.7]{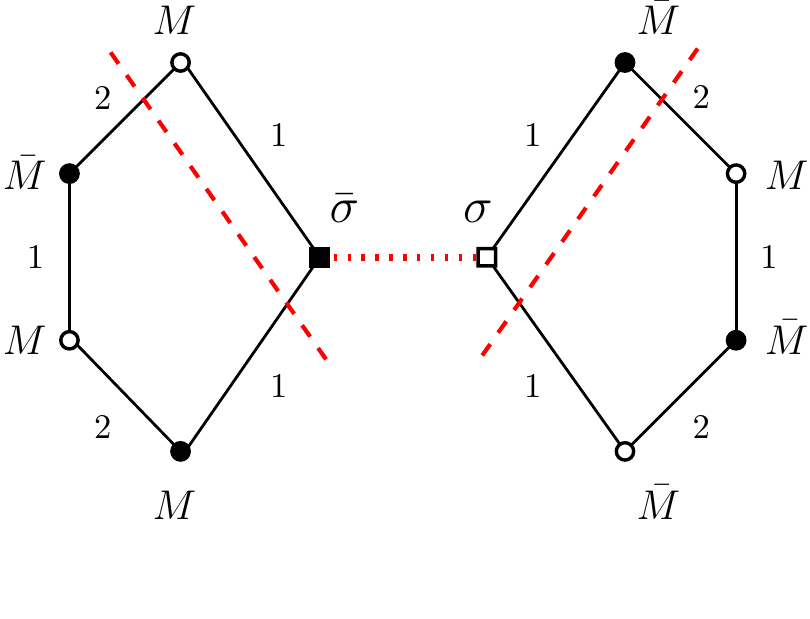}
\caption{\label{fig:k4}  $k=4$ matrix trace invariant and initial intermediate field step.}
\end{figure}

\begin{figure}[h!]
\includegraphics[scale=0.8]{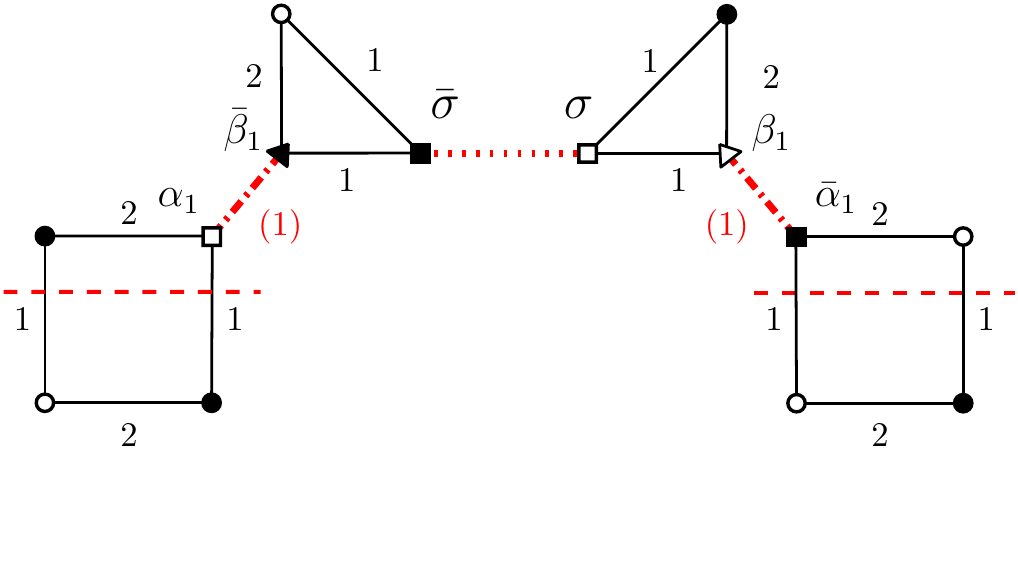}\hspace{1.2cm}\includegraphics[scale=0.8]{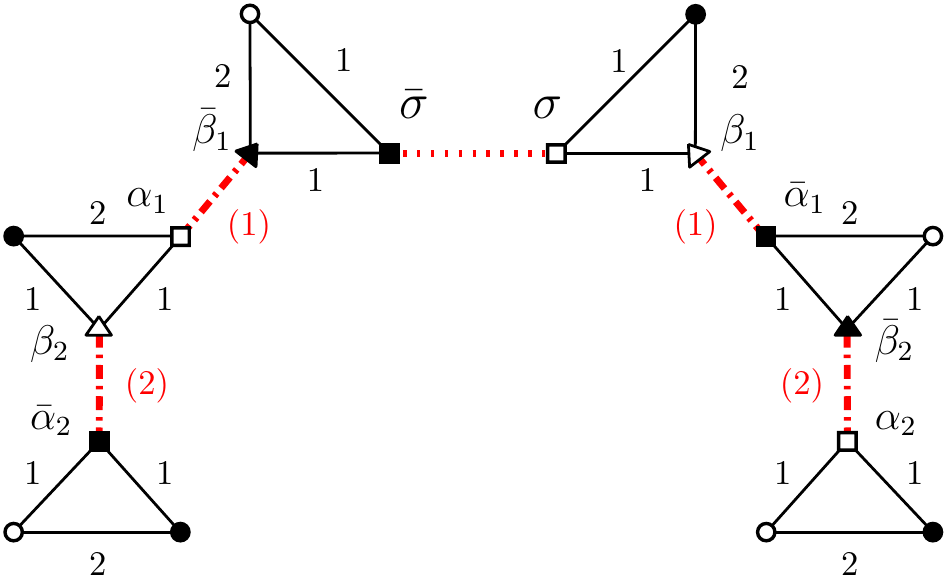}
\caption{\label{fig:k4_cut1}  Successive intermediate field splittings for matrices}
\end{figure}

The detailed graphical representation of the successive intermediate field splittings is summarized in Figures \ref{fig:k4} and \ref{fig:k4_cut1} for the $k=4$ case.\\\\

\noindent{\bf Hermitian Intermediate Field Representation}\\

{\it Expression \eqref{PrtF1a} can be reformulated using a $(k+1)N\times (k+1)N$ determinant
\be
\label{PrtF2}
Z_k(\lambda,N)=\int  d\nu(\xi)e^{-N \Tr\ln\bigl[  \unk -g_k \mathbb{M}_k(\xi) \bigr] },
\ee
where $\mathbb{M}_k(\xi)=iC_k.\mathbb{H}_k(\xi)$, $C_k$ is the complex symmetric covariance of the integrated fields
\begin{eqnarray}
\label{eqref=Codd}
\newcommand*{\temp}{\multicolumn{1}{c|}{}}
\newcommand*{\tempi}{\multicolumn{1}{c|}{-i\bbbone}}
\newcommand*{\tempo}{\multicolumn{1}{c|}{0}}
C_{odd}=\left(\begin{array}{ccccccc}
\bbbone&\multicolumn{1}{c|}{0} &&&&& \\ 
0&\multicolumn{1}{c|}{\bbbone} &&&&& \\ 
\cline{1-4}
& \temp & 0 &  \tempi &  &\Large\mbox{{$0$}} & \\ 
& \temp & - i\bbbone & \tempo &  & & \\ 
\cline{3-6} 
&&& \temp & 0 & \tempi & \\
& &\Large\mbox{{$0$}}& \temp & -i \bbbone& \tempo & \\
\cline{5-6} 
&&&&&&\ddots
\end{array}\right) , \quad
C_{even}=\left(\begin{array}{cccccc}
\multicolumn{1}{c|}{\bbbone} &&&&& \\ 
\cline{1-3}
 \temp & 0 &  \tempi &  &\Large\mbox{{$0$}} & \\ 
 \temp & - i \bbbone& \tempo &  & & \\ 
\cline{2-5} 
&& \temp & 0 & \tempi & \\
 &\Large\mbox{{$0$}}& \temp & -i \bbbone& \tempo & \\
\cline{4-5} 
&&&&&\ddots
\end{array}\right).
\end{eqnarray}
The matrix $\mathbb{H}_k$ is Hermitian, and has two different forms depending whether $k$ is odd or even 
\begin{equation}
\mathbb{H}^{odd}_k= \left(
\begin{array}{c|ccccc}
  0 &   \beta_1 &  \alpha_1 &\cdots & \alpha_{k-2}  & \bbbone \\ \hline
  \beta_1^\dagger & &&\raisebox{-30pt}{{\huge\mbox{{$0$}}}} \\[-6ex]
  \alpha_1^\dagger& \\[-0.5ex]
  \vdots & \\[-0.5ex]
  \alpha_{k-2}^\dagger& \\
\bbbone& \\
[-0.5ex]  
\end{array}
\right), \quad
\mathbb{H}^{even}_k= \left(
\begin{array}{c|ccccc}
  0 &  \sigma & \beta_2  &  \cdots & \alpha_{k-2} & \bbbone  \\ \hline
   \sigma^\dagger & &&\raisebox{-30pt}{{\huge\mbox{{$0$}}}} \\[-6ex]
  \beta_2^\dagger& \\[-0.5ex]
  \vdots & \\[-0.5ex]
  \alpha_{k-2}^\dagger& \\
\bbbone& \\[-0.5ex]  
\end{array}
\right).
\end{equation}
Therefore we have explicitly
\begin{equation}
\mathbb{M}^{odd}_k= \left(
\begin{array}{c|ccccc}
  0 &   i\beta_1 & i \alpha_1 &\cdots & i\alpha_{k-2}  & i\bbbone \\ \hline
  i\beta_1^\dagger & &&\raisebox{-30pt}{{\huge\mbox{{$0$}}}} \\[-6ex]
  \beta_3^\dagger& \\[-0.5ex]
  \vdots & \\[-0.5ex]
\bbbone& \\
\alpha_{k-2}^\dagger& \\
[-0.5ex]  
\end{array}
\right), \quad
\mathbb{M}^{even}_k= \left(
\begin{array}{c|ccccc}
  0 &  i\sigma & i\beta_2  &  \cdots & i\alpha_{k-2} & i\bbbone  \\ \hline
  \beta_2^\dagger   & &&\raisebox{-30pt}{{\huge\mbox{{$0$}}}} \\[-6ex]
 \sigma^\dagger& \\[-0.5ex]
  \vdots & \\[-0.5ex]
\bbbone & \\
 \alpha_{k-2}^\dagger& \\[-0.5ex]  
\end{array}
\right).
\end{equation}
}

\noindent\textbf{Examples.}
In the simplest cases $k=3,4$, hence the $e^{-\frac{\lambda}{N^2}\Tr[(MM^\dagger)^{3}]}$ and  $e^{-\frac{\lambda}{N^3}\Tr[(MM^\dagger)^{4}]}$ models, we obtain the representations:
\begin{equation} 
Z_3( \lambda,N) =\int d\mu_X^c(\alpha_1,\beta_1)e^{-N \Tr\ln\bigl[  \bbbone^{ \otimes 4 }- g_3 \mathbb{M}_3(\alpha_1,\beta_1) \bigr] }, \quad g_3 = \frac{\lambda^{1/6}}{N^{1/3}}, \quad \mathbb{M}_3= \begin{pmatrix}
0  &  i\beta_1 & i\alpha_1 &i\bbbone  \\
i\beta_1^\dagger & 0 & 0 &0 \\
\un & 0 & 0 &0 \\
\alpha_1^\dagger & 0 & 0 & 0
\end{pmatrix}
\end{equation}
and
\begin{equation} Z_4( \lambda, N) = \int d\mu^c(\sigma) d\mu_X^c(\alpha_2,\beta_2)e^{-N \Tr\ln\bigl[  \bbbone^{ \otimes 5 }-g_4 \mathbb{M}_4(\alpha_2,\beta_2) \bigr] },
\quad g_4 = \frac{\lambda^{1/8}}{N^{3/8}}, \quad \mathbb{M}_4= \begin{pmatrix}
0  & i\sigma & i\beta_2 & i\alpha_2 & i\un \\
\beta_2^\dagger & 0 & 0 & 0 & 0 \\
\sigma\dagger & 0 & 0 & 0 & 0 \\
\un & 0 & 0 & 0 & 0 \\
\alpha_2\dagger & 0 & 0 & 0 & 0
\end{pmatrix}  .
\end{equation}

\subsection{Analyticity Domain and Borel Summability}
\label{bsmat}
\hspace{0.2cm}

Our main task in this subsection is to complete the proof of Theorem \ref{Borelmatrix}. We still have to prove
\begin{theorem} \label{shrinkmatrix} For any fixed $N$ the normalization $Z_k(\lambda, N)$ and
the free energy $F_k := N^{-2}\log Z_k(\lambda, N)$ are Borel-Le Roy summable
of order $m=k-1$, in the shrinking domain $D^m_{\rho_m (N)}$ with 
$ \rho_m (N) = N^{-1-2/m} r_m$, $r_m >0$ independent of $N$. \label{th1}
\end{theorem}

We shall in fact prove analyticity and uniform Taylor remainder estimates in a slightly larger (but similarly shrinking 
as $N \to \infty$) domain $E^m_{\rho_m (N)}$ 
consisting of all $\lambda$'s with $\vert \lambda \vert  < [\rho_m (N)]^m$,  and
$\vert \arg \lambda \vert < \frac{ m \pi }{2}$, hence for $\lambda^{1/m}$ in an open half-disk of radius $\rho_m (N)$, which obviously contains
the smaller tangent disk $D^m_{\rho_m (N)}$ of diameter $\rho_m (N)$
necessary for Theorem \ref{blrsok}.

Our strategy is to bound the determinant factor $e^{-N\Tr\ln\bigl[  \unk-g_k \mathbb{M}_k(\xi) \bigr] }$ in the Hermitian representation \eqref{PrtF2}
by computing the eigenvalues of the matrix $\unk-g_k \mathbb{M}_k$.
In order to have absolutely convergent formulas instead of formal expressions we need however to now
reinstall everywhere the correct contour integrals with the $\epsilon$ regulator. It means that 
in every integral over $\alpha$ and $\beta$ variables we have to return to the
$a + b$ or $a -b$ real and imaginary parts, which we call collectively the $\{a, b\}$ variables. Instead of being
formally integrated each over the real line with an oscillating factor $e^{- i a^2}$ or $e^{+ i b^2}$, we need
to integrate each such $a$ or $b$ variable with the appropriate $C_{\pm \epsilon}$
contour. This is equivalent to keep every $a$ or $b$ contour real but then first substitute $a \to a - i \epsilon \tanh a$
and $b \to b + i \epsilon \tanh b$ into all imaginary factors $e^{- i a^2}$ and $e^{+ i b^2}$, and second perform the same
substitution into all $a$ and $b$ linear-dependent coefficients of $\mathbb{M}_k (\{a,b\})$.
Remark that this destroys the Hermitian character of the matrix $H_k$. More precisely
since the $\mathbb{M}_k$ matrix depends linearly of all $\{a,b\}$ variables, this second substitution corresponds
to substitute  
\bee  \mathbb{M}_k( \{a,b\}) \to \mathbb{M}_k ( \{a,b\})  + \epsilon \mathbb{N}_k (\{a,b\})
\ee 
where $\mathbb{N}_k $ is $(k+1)N \times (k+1)N$ with matrix elements all zero except on the first $N $ rows and first
$N$ columns. In these rows and columns, remark that any non zero matrix element is of the form 
$\pm ( i ) \frac{1}{\sqrt 2}(\tanh a_{jk} \pm  \tanh b_{jk})$ for some $j,k$
where the factor $i$ may or may not be present. Hence\\ 

\begin{lemma} \label{lemnormnk}
The operator norm of $\mathbb{N}_k$ is uniformly bounded by $2\sqrt{k} N$, hence
\bee \Vert \epsilon g_k \mathbb{N}_k (\{a,b\})\Vert \le  2\epsilon \vert g_k \vert \sqrt{k} N  \quad \quad\forall \{a,b\}.
\ee
\end{lemma}

\prf Simply bound $\Vert \mathbb{N}_k \Vert $ by its Hilbert-Schmidt norm $\Vert \mathbb{N}_k \Vert_2 $ (we took into account that the first $N$ by $N$
block in $\mathbb{N}_k$ is zero). \qed\\

Returning to the $\mathbb{M}_k ( \{a,b\}) $ matrix, we now study its spectrum by
computing its characteristic polynomial in term of the Hermitian matrix $H_k$.\\
\begin{lemma} 
\label{block_mat}
Considering $\mathbb{M}_k$, $H_k$ and $\eta(k)$ as defined before, 
the characteristic polynomial of the $(k+1)N\times(k+1)N$ matrix $\unk - g_k \mathbb{M}_k$ is
\bee 
\det [ (1-x) \unk - g_k \mathbb{M}_k ]  \ =\ (1-x)^{(k-1)N}\det [  (1-x)^2\bbbone_{N}  -  g_k^2 (iH_k -\eta(k)\beta_1\beta_1^\dagger)]
\ee
\end{lemma}

\prf It follows from the $(k+1)N\times(k+1)N$ square matrix identity

\[ \left(
\begin{array}{c|c|c|c}
  (1-x)^2\un    &   -  g_k A_1                &             \cdots     &   -   g_k A_k \\ [+1ex]\hline
 - (1-x)  g_k B_1  &  (1-x)\un            &       0     &                 0       \\[+1ex]\hline
  \vdots                             &     0      &  \ddots        &                 \vdots            \\[+1ex]\hline
 - (1-x) g_k B_k &  0                       &  \cdots &       (1-x)\un      \\[+1ex]\hline
\end{array}
\right)= 
\left(\begin{array}{c|c|c|c}
 U    &  -  g_k A_1                &           \cdots     &     -    g_k A_k \\ [+1ex]\hline
 0  &  (1-x)\un  &          0          &                  0       \\[+1ex]\hline
   \vdots                       &           0                      &  \ddots        &             \vdots            \\[+1ex]\hline
  0 &  0                       &        \cdots      &    (1-x)\un \\[+1ex]
\end{array}
\right)
\left(\begin{array}{c|c|c|c}
  \un   &   \quad   0  \quad        &           \cdots     &       \quad   0 \quad\\ [+1ex]\hline
- g_k B_1  &  \un  &          0           &                  0       \\[+1ex]\hline
   \vdots                       &              0                     &  \ddots        &             \vdots            \\[+1ex]\hline
- g_k B_k &  0                       &        \cdots      &    \un \\[+1ex]
\end{array}
\right),
\]\\
where $U=(1-x)^2\un-g_k^2\sum_{j=1}^k A_{j}B_{j} $. Choosing $0, A_1, \cdots , A_k$, as the first row of $\mathbb{M}_k$
and  $0, B_1, \cdots , B_k$ as the first column of $\mathbb{M}_k$, and taking the determinant of this identity, one obtains that 
\bee(1-x)\det[(1-x)\unk -g_k \mathbb{M}_k]=\det(U)(1-x)^{kN},
\ee
so that 
\be
\det[(1-x)\unk-g_k \mathbb{M}_k]=\det(U)(1-x)^{(k-1)N}.
\ee
More precisely, for $k$ odd we choose $A_1 = i \beta_1 $, $A_2 = i \alpha_1 $, \dots  $A_{k-1} =  i \alpha_{k-2} $, $A_{k} = i \un$
and $B_1 = i \beta_1^\dagger $, $B_2 = \beta_3^\dagger $, \dots  $B_{k-1} =  \un  $, $B_{k} = \alpha_{k-2}^\dagger$, and for $k$ even
we choose $A_1 = i \sigma $, $A_2 = i \beta_2 $, \dots  $A_{k-1} =  i \alpha_{k-2} $, $A_{k} = i \un$
and $B_1 = \beta_2^\dagger $, $B_2 = \sigma^\dagger  $, \dots  $B_{k-1} =  \un  $, $B_{k} = \alpha_{k-2}^\dagger$. In all cases we find
\be
U=(1-x)^2\un-g_k^2\sum_{j=1}^k A_{j}B_{j} = (1-x)^2\un-g_k^2[iH_k-\eta(k)\beta_1\beta_1^\dagger].
\ee\qed\\

To prove Borel-Le Roy summability of order $m$ of $Z_k$ and $\log Z_k$ of order $m=k-1$ in the intermediate field representation \eqref{PrtF2}
the key step is an upper bound on the norm of the resolvent $[1-g_k \mathbb{M}_k(\{a,b\}) ]^{-1}$. This bound must be uniform both in $\lambda$
in that domain and uniform in the intermediate fields along the contours.

\begin{lemma} For $\lambda \in E^m_{\rho_m(N)}$ we have
\begin{equation}
\Vert (\unk-g_k \mathbb{M}_k )^{-1}\Vert \le [ \sin \frac{\pi}{4k} ]^{-1}. \label{resobound1}
\end{equation} \label{analytlemma1}
\end{lemma}
\prf
Let us compute the spectrum of the matrix $\unk-g_k \mathbb{M}_k (\{a,b\})$. By the previous Lemma
it has a trivial eigenvalue $1$ with multiplicity $N(k-1)$ and an additional non trivial spectrum, which is exactly made
of the elements $x$ of the form $x= 1\pm g_k \sqrt{ y}$ where $y$ belongs to the spectrum of the matrix $iH_k-\eta(k)\beta_1\beta_1^\dagger$. But  $y$ belongs to the spectrum of that matrix if and only if
\bee  \det (-y + iH_k-\eta(k)\beta_1\beta_1^\dagger)=0 .
\ee
This is equivalent to the existence of an eigenvector $u$ with $\Vert u \Vert =1$ such that $(-y + iH_k-\eta(k)\beta_1\beta_1^\dagger)u=0$, hence  for which
$y = -\eta(k) \Vert \beta_1^\dagger u \Vert^2 + i <u, H_k u> $. Hence, since $H_k$ is Hermitian, no matter whether $\eta(k) = 0$ or 1, $y$ 
must be of the form $-v^2 +i w$ with $v$ and $w$ real. 
It means that any square root of $y$ (written as $\pm \sqrt y$) must be either 0 or 
have a complex argument in $ I=  [\frac{ \pi} {4} , \frac{3 \pi} {4} ] \cup [-\frac{3 \pi} {4} ,  -\frac{ \pi} {4}  ] $.
But in the domain $E^k_\rho$
the argument of $g_k$  is bounded by $\frac{ (k-1) \pi }{4k}$ hence the argument of 
$ \pm g_k  \sqrt{ y} $ (when $y \ne 0$) must lie in 
\bea I_k & =&[\frac{ \pi} {4} -  \frac{ (k-1) \pi }{4k}, \frac{3 \pi} {4} +  \frac{ (k-1) \pi }{4k} ] \cup [-\frac{3 \pi} {4}-  \frac{ (k-1) \pi }{4k} ,  -\frac{ \pi} {4} +  \frac{ (k-1) \pi }{4k} ]  \nonumber  \\
&=&  [\frac{\pi}{4k},  \pi - \frac{\pi}{4k} ] \cup [- \pi +\frac{\pi}{4k}, -\frac{\pi}{4k}],
\eea
hence in that domain the spectrum of $\unk-g_k\mathbb{M}_k$ lies out of the open disk of center 0 and radius $\sin \frac{\pi}{4k}$.
\qed\\\\

We denote $R_k=r_m^{\frac{m}{2k}}=r_{k-1}^{\frac{k-1}{2k}}$.

\begin{lemma} \label{normresl} For $\lambda \in E^m_{\rho_m(N)}$, choosing $\epsilon= R_k^{-1}\frac{\sin (\pi /4k) }{4 \sqrt k}$  we have 
\footnote{We decided to put the $N$ dependence on the Borel radius and none on $\epsilon$.
Other choices could include an $N$ dependence on $\epsilon$ but would lead to contour 
integrals no longer defined in the large $N$ limit.}
\begin{equation}
\Vert [\unk-g_k (\mathbb{M}_k +\epsilon\mathbb{N}_k )]^{-1}\Vert \le 2 [ \sin \frac{\pi}{4k} ]^{-1}. \label{resobound2}
\end{equation} \label{analytlemma2}
\end{lemma}
\prf We recall that for $\lambda \in E^m_{\rho_m(N)}$, $\vert \lambda \vert \le \rho_m(N)^{m} $. Since $\rho_m (N) = N^{-1-\frac{2}{m}}r_m$,  it implies
\bee
\vert g_k \vert =  \vert \lambda \vert^{\frac{1}{2k}}  N^{-\frac{m}{2k}} \le \rho_m(N)^{\frac{m}{2k}} N^{-\frac{m}{2k}} = N^{-\frac{m}{2k} -\frac{1}{k}} N^{-\frac{m}{2k}} r_m^{\frac{m}{2k}} =N^{-\frac{m}{k} -\frac{1}{k}} R_k= N^{-1}R_k.
\ee
Hence by Lemma \ref{lemnormnk} we have
$\Vert \epsilon g_k \mathbb{N}_k \Vert\le \frac{1}{2}\sin \frac{\pi}{4k}$. Since 
\bee   [\unk-g_k (\mathbb{M}_k +\epsilon\mathbb{N}_k )]^{-1} = 
[\unk- (\unk-g_k \mathbb{M}_k)^{-1} g_k \epsilon\mathbb{N}_k ]^{-1} (\unk-g_k \mathbb{M}_k)^{-1},
\ee
it implies
\bee  
\Vert [\unk-g_k (\mathbb{M}_k +\epsilon\mathbb{N}_k )]^{-1} \Vert \le  \bigl( 1-  [ \sin \frac{\pi}{4k} ]^{-1} 
\frac{1}{2}\sin \frac{\pi}{4k}  \bigr)^{-1}  [ \sin \frac{\pi}{4k} ]^{-1} =  2 [ \sin \frac{\pi}{4k} ]^{-1}.
\ee
\qed\\

Remark that this bound also implies a bound on the factor
$e^{-N \Tr\ln\bigl[  \unk -g_k (\mathbb{M}_k +\epsilon\mathbb{N}_k ) (\xi) \bigr] }$ in 
\eqref{PrtF2}, namely
\bee
e^{-N \Tr\ln\bigl[  \unk -g_k  (\mathbb{M}_k +\epsilon\mathbb{N}_k )(\xi) \bigr] } = 
\vert {\rm det}^{-N} [\unk-g_k(\mathbb{M}_k+\epsilon\mathbb{N}_k)] \vert  \le e^{2N^2 \vert \log \frac{\sin \frac{\pi}{4k}}{2}  \vert} =  
2^{2N^2} [ \sin \frac{\pi}{4k} ]^{-2N^2},
\ee
since there are only at most $2N$ non zero eigenvalues of $\mathbb{M}_k+\epsilon \mathbb{N}_k$. It
means a uniform upper bound on the integrand in \eqref{PrtF2}. Since the integration in $d \nu$ is over $2(k-1)N^2$ integration contours,
taking absolute values for each of them with $\epsilon= R_k^{-1}\frac{\sin (\pi /4k) }{4 \sqrt k}$ leads to a loss of $O(1)k^{3/2}$ per contour. Hence since
$[\sin \frac{\pi}{4k} ]^{-1} \le O(1) k$ in the shrinking domain we have the uniform bound
\be
\label{PrtF3}
\vert Z_k(\lambda,N) \vert \le  O(1)^{k^{3/2}N^2}   \quad \implies \quad \vert F_k(\lambda,N) \vert \le  O(1)^{k^{3/2}}  .
\ee
Analyticity of $Z_k$ and $F_k$ (for any finite $N$, but non-uniformly in $N$) then follows by the standard theorem that a uniformly
convergent integral of an analytic integrand is analytic. 

For any finite $N$ the perturbation theory of $Z_k$ in the intermediate field 
representation is identical to the standard one in the direct representation (this just follows from the fact that they are
independent of the $\epsilon $ regulator, see \eqref{compper1}).
As usual, Borel-LeRoy Taylor remainder bounds just correspond to the factorial growth 
of perturbation theory and follow easily. Indeed they correspond to insert $q$ additional vertices (with a $1/q!$ symmetry factor)
in the functional integral for $Z$, hence $qk$ pairs of fields. It is by now well known that this adds simply $qk$ resolvents to the functional integral
\cite{Gurau:2013pca}, together with a factor $(qk)!$ for pairing the arguments into the resolvents. But since our bounds for the determinant in $Z$
precisely followed from a uniform bound on such resolvents (Lemma \ref{normresl}), we obtain a bound in $O(1)^{kq}$ for addition of such resolvents.
Combining the two factorials leads to a bound in $[O(1)\vert\lambda\vert]^q  \frac{qk!}{q!}$, hence since $m=k-1$, this bound is
exactly of the desired type \eqref{borelrem}.

By unicity of the Borel sum, we can claim that representation \eqref{PrtF2}, although derived by some formal
computations at $\epsilon =0$, is in fact convergent when $\epsilon$-dependent contours are used, and that it defines
non-perturbatively the \emph{same} partition function and free energy that the direct initial representation. These results
extend easily to free energy with sources, hence to cumulants of the theory.
\vspace{1cm}

\section{Positive tensor models}\label{tenrep}

\subsection{Random tensor models}
\vspace{0.5cm}

We include here for self-containedness a brief remainder about invariant (uncolored) tensor models,
essentially reproduced from \cite{Bonzom:2012hw}.

Let $\cH_1,\dotsc,\cH_D$ be complex Hilbert spaces of dimensions $N_1,\dotsc,N_D$. A rank $D$ 
covariant tensor $T_{n_1\dotsc n_D}$ is a collection of $\prod_{i=1}^{D} N_i$ complex numbers 
supplemented with the requirement of covariance under independent change of basis in each $\cH_i$. 
The complex conjugate tensor $\bar T_{  n_1 \dotsc n_D }$ is then a rank $D$ contravariant tensor. 
Under independent unitary base change $U^{(i)}$ in each $U(N_i)$, $T$ and $\bar T$ transform as
\bea
 T'_{a_1\dotsc a_D} = \sum_{n_1,\dotsc,n_D} U^{(1)}_{a_1n_1}\dotsm U^{(D)}_{a_Dn_D}\ T_{n_1\dotsc n_D}  \; ,\qquad
 \bar T'_{  a_1\dots  a_D} = \sum_{n_1,\dotsc,n_D} \bar U^{(D)}_{  a_D n_D}\dotsm \bar U^{(1)}_{a_1 n_1}\ \bar T_{n_1\dots n_D}  \; .
\eea

From now on we shall restrict to the case where all $N_i$, $i = 1, \cdots D$ are equal to $N$.
 A \emph{trace invariant} is a connected monomial in $T$ and $\bar T$ invariant under that action of the external tensor product of the $D$  independent unitary groups $U(N)$, namely $U(N)^{\otimes D}$. It is built by contracting all tensor indices two by two, a tensor entry always with a conjugate tensor entry,  respecting the positions of indices. Note that a trace invariant has necessarily the same number of
$T$ and $\bar T$.
Any trace invariant is then represented by a $D$-bubble, which is a D-regular edge-colored bipartite graph:
\begin{definition}
A {\bf closed $D$-colored graph}, or {\bf $D$-bubble}, is a \emph{connected} graph $\cB = (\cV,\cE)$ with vertex set $\cV$ and line set $\cE$
such that
\begin{itemize}
\item  $\cV$ is bipartite, i.e. there exists a partition of the vertex set $\cV  = A \cup \bar A$, such that for any
element $l\in\cE$, then $ l = \{v,\bar v\}$ with $v\in A$ and $\bar v\in\bar A$. Their cardinalities
satisfy $|\cV| = 2|A| = 2|\bar A|$.
\item  The line set is partitioned into $D$ subsets $\cE = \bigcup_{i  =1}^{D} \cE^i$, where $\cE^i$ is the subset
of lines with color $i$, with $|\cE^i|=|A|$.
\item  It is $D$-regular (all vertices are $D$-valent) with all lines
incident  to a given vertex having distinct colors.
\end{itemize}
\end{definition}
\begin{figure}[h]
\begin{center}
 \includegraphics[width=8cm]{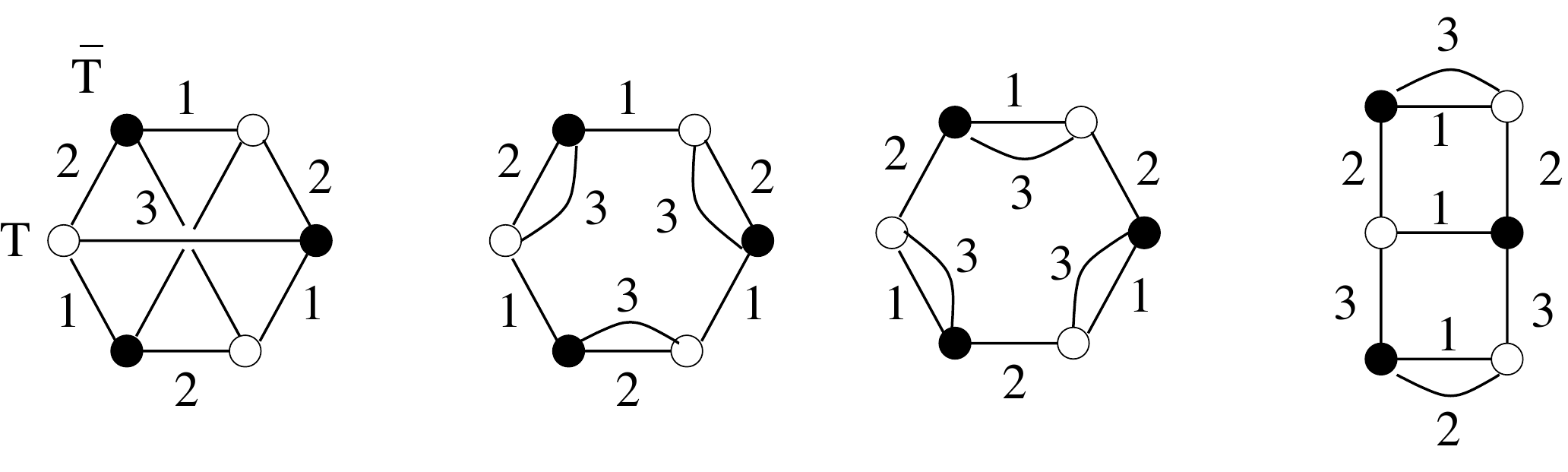}
\caption{Graphical representation of trace invariants.}
\label{fig:tensobs}
\end{center}
\end{figure}
To draw the graph associated to a trace invariant we
represent every $T$ by a white vertex $v$ and every $\bar T$ by a black vertex $\bar v$. Each position of an index
is represented as a color or number: $n_1$ has color $1$, $n_2$ has color $2$ and so on. 
The contraction of two indices $n_i$ and $\bar n_i$ of tensors
is represented by a line $l^i = (v,\bar v)$ connecting the corresponding two vertices. Lines inherit the color of the index, and
always connect a black and a white vertex. 
Examples of trace invariants for rank 3 tensors are represented in Figure \ref{fig:tensobs}. 

The trace invariant associated to the graph $\cB$ writes as
\bea
\Tr_{\cB}(T,\bar T ) = \sum_{\{\vec n^v,\bar{\vec{n}}^v\}_{v,\bar v \in \cV}}  \delta^{\cB}_{\{\vec{n}^v, \bar{\vec{n}}^{\bar v}\}}  \;
\prod_{v,\bar v \in \cB} T_{\vec n^v} \bar T_{\bar {\vec n }^{\bar v} }
\; ,\quad \text{with} \qquad \delta^{\cB}_{\{\vec{n}^v, \bar{\vec{n}}^{\bar v}\}} = \prod_{i=1}^D \prod_{l^i = (v,\bar v)\in \cB} \delta_{n_i^v \bar n_i^{\bar v}} \; ,
\eea
where $l^i$ runs over all the lines of color $i$ of $\cB$. $\delta^{\cB}_{\{\vec{n}^v, \bar{\vec{n}}^{\bar v}\}}$
is the product of delta functions encoding the index contractions of the trace invariant associated to the graph $\cB$. 
Notice that there exists a unique $D$-colored graph with
two vertices, namely the graph in which all the lines connect the two vertices. 
Its associated invariant is simply noted as a scalar product
\bea\label{eq:gaussian}
T \cdot \bar T  = \sum_{\vec n,\bar{\vec{n}}}\, T_{\vec n}\, \bar T_{\bar {\vec n} }\ \Bigl[\prod_{i=1}^D \delta_{n_i \bar n_i}\Bigr] \; .
\eea
For example the trace invariant associated with the $K_{3,3}$ example in the left of Figure \ref{fig:tensobs} is
\be
 \includegraphics[scale=1]{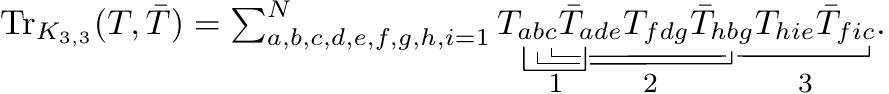}
\ee

The free action at rank $D$ is defined as the normalized Gaussian measure
\bee  d\mu_0 (T, \bar T)  =   \Bigl( \prod_n  \frac{ d \bar{T}_{{n}}  dT_n }{2 \imath \pi} \Bigr) \,
 e^{ -  T \cdot \bar T  } .
\ee

A generic tensor model with trace invariant interaction $\cB(T,\bar T)$ is given by the (invariant) normalized measure
\bee\label{eq:model}
d\mu_\cB(T, \bar T) = \frac{1}{Z_\cB (\lambda , N)} d\mu_0 (T, \bar T) 
e^{-\lambda N^{-s(\cB )} \cB(T,\bar T)} \;,
\ee
where $\lambda$ is the coupling constant and $s(\cB )$ an appropriate scaling power, which we keep undetermined at this stage. The 
normalization $Z_\cB (\lambda , N)$ and free energy $ F_\cB (\lambda , N)$ are defined by
\bee\label{eq:model1}
Z_\cB (\lambda , N)    = \int d\mu_0 (T, \bar T) 
e^{-\lambda N^{-s(\cB )} \cB(T,\bar T)}, \quad F_\cB (\lambda , N) =  N^{-D} \log Z_\cB (\lambda , N)  .
\ee
The cumulants of the model are then written in terms of the moment-generating function
\bee
F_\cB (\lambda , N, J, \bar{J})=\log \int  d\mu_0 (T, \bar T) \;\;  e^{-\lambda N^{-s(\cB )}  \cB(T,\bar T)   + J \dot \bar T  + T \dot \bar J }\;,
\ee
via the usual formulas 
\begin{align*}
& \kappa(T_{n_1}\bar{T}_{\bar{n}_1}...T_{n_k}\bar{T}_{\bar{n}_k})
=\frac{\partial^{(2k)} \Bigl( \ln Z(J,\bar J) \Bigr) }{\partial \bar{J}_{n_1}
\partial J_{\bar{n}_1}...\partial \bar{J}_{n_k}\partial J_{\bar{n}_k}} \Bigg{\vert}_{J =\bar J =0}.
\end{align*}

The nice properties of tensor models stem from their relationship to colored triangulations and crystallization theory \cite{crystal1,crystal2,crystal3}.
In particular they support a full $D$-homology and have a simple canonical definition of \emph{faces}\footnote{This is the crucial property from the point of view of gravity quantization since it allows to associate to the dual space a canonical discretized Einstein-Hilbert action.}.
Faces are simply subgraphs with two fixed colors. We denote them $\cF$.
For instance graphs with three colors have three types of faces, given by the subgraphs with lines of colors $12$, $13$ and $23$.
As every line belongs to exactly two faces (for instance a line of color $1$ belongs to a single face $12$ and to a single face $13$...), the graphs with
three colors can be represented as ribbon graphs, i.e. can be embedded into the sphere, giving a combinatorial map.

The analysis of tensor models at large $N$ and their relationship to quantum gravity relies on the existence of a non-negative integer,
the \emph{Gurau degree}, governing the $1/N$ tensor expansion \cite{Gurau:2010ba,Gurau:2011aq,Gurau:2011xq}. 
We recall briefly its definition and properties. 
First one needs the notion of \emph{jacket}.
\begin{definition}
Let $\cB$ be a $D$-bubble and $\tau$ be a cycle (up to orientation) on $\{1,\dotsc,D\}$. A {\bf jacket} $\cJ$ of $\cB$ is a ribbon graph
having all the vertices and all the lines of $\cB$, but only the faces with colors $(\tau^q(1),\tau^{q+1}(1))$, for $q=0,\dotsc,D-1$,
modulo the orientation of the cycle.
\end{definition}
Any jacket  $\cJ$ of $\cB$ is a ribbon graph containing all the vertices and all the lines of $\cB$. Each 
of the $(D-1)!/2$ jackets associated to a $D$-bubble defines therefore a compact oriented surface
which has therefore a well-defined genus $g_\cJ$, related to its Euler characteristic
by the usual relation $\chi_\cJ = 2 - 2 g_\cJ$.
The Gurau degree $\omega(\cB)$ of the $D$-bubble $\cB$ is then defined 
as the sum of the genera of its jackets, $\omega(\cB)=\sum_{\cJ} g_{\cJ}$.
Graphs with three colors are ribbon graphs, hence have  a single jacket. In that case the degree reduces
to the genus.  But for $D>3$ the degree provides a generalization of the genus which is {\it not} a topological invariant,
as it combines topological and combinatorial information about the graph. It is related to the total 
number of faces $\cF_\cB$ of a bubble $\cB$ with $|\cV|$
black or white vertices
through
\bea \label{eq:facesmese}
 \cF_\cB= \frac{(D-1)(D-2)}{2} |\cV| + (D-1) - \frac{2}{(D-2)!} \omega(\cB) \; ,
\eea
an equation simply obtained by combining Euler's formula for the genus of each jacket with the observation that any face
belongs always to the \emph{same number} of jackets, those for which the two colors of the face are \emph{adjacent} in any of the two cycles $\tau$
defining the jacket.
Observing that the  Gurau degree is a positive integer and reorganizing the perturbation expansion according to increasing values of that integer leads to the tensorial $1/N$ expansion. 

The main reason for physicists interest in the Gurau degree stems from quantum gravity. Since in any dimension \emph{faces} 
are dual to $D-2$ dimensional hinges, equation \eqref{eq:facesmese} means that the Gurau degree provides in any dimension a discretization of the Einstein-Hilbert action on equilateral triangulations. 

The graphs with zero Gurau degree are called \emph{melonic}. They can be exactly enumerated \cite{Gurau:2011xp}.
\subsection{Positivity}
\vspace{0.3cm}

\begin{definition}
A bubble $\cB(T,\bar T)$ is said to be {\it positive} if there exist an edge-cut $\cI$ which divides the graph into two connected components $F$ and $\bar F$ which are identical up to inversion of the black and white vertices in one of the components. 

To each vertex $v$ in $F$ is  therefore canonically associated a vertex $\bar v$ of the opposite color in $\bar F$. The connected components $F$ and $\bar F$ have boundaries. The positivity of $\cB$ requires one last constraint : the edge-cut must be without crossing, meaning that the permutation of $\cS_{\lvert \cI \rvert}$ induced by the edge-cut $\cI$ when identifying the vertices in the boundary of $F$ with their canonical companion in $\bar F$ is the identity.
\end{definition}

\begin{figure}[h!]
\includegraphics[scale=0.7]{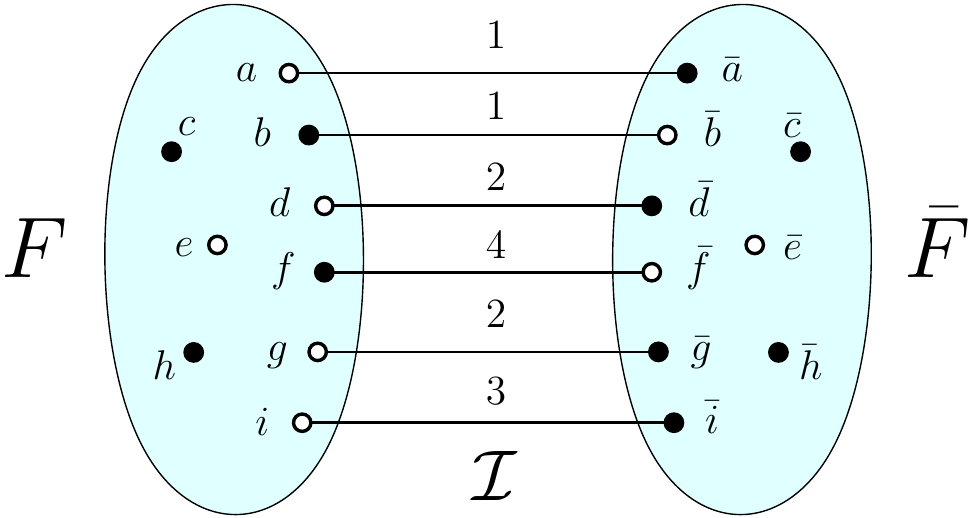}
\caption{\label{fig:FF}  Positive unitary tensorial invariant.}
\end{figure}

\begin{figure}[!htp]
  \centering
 \includegraphics[height=4cm]{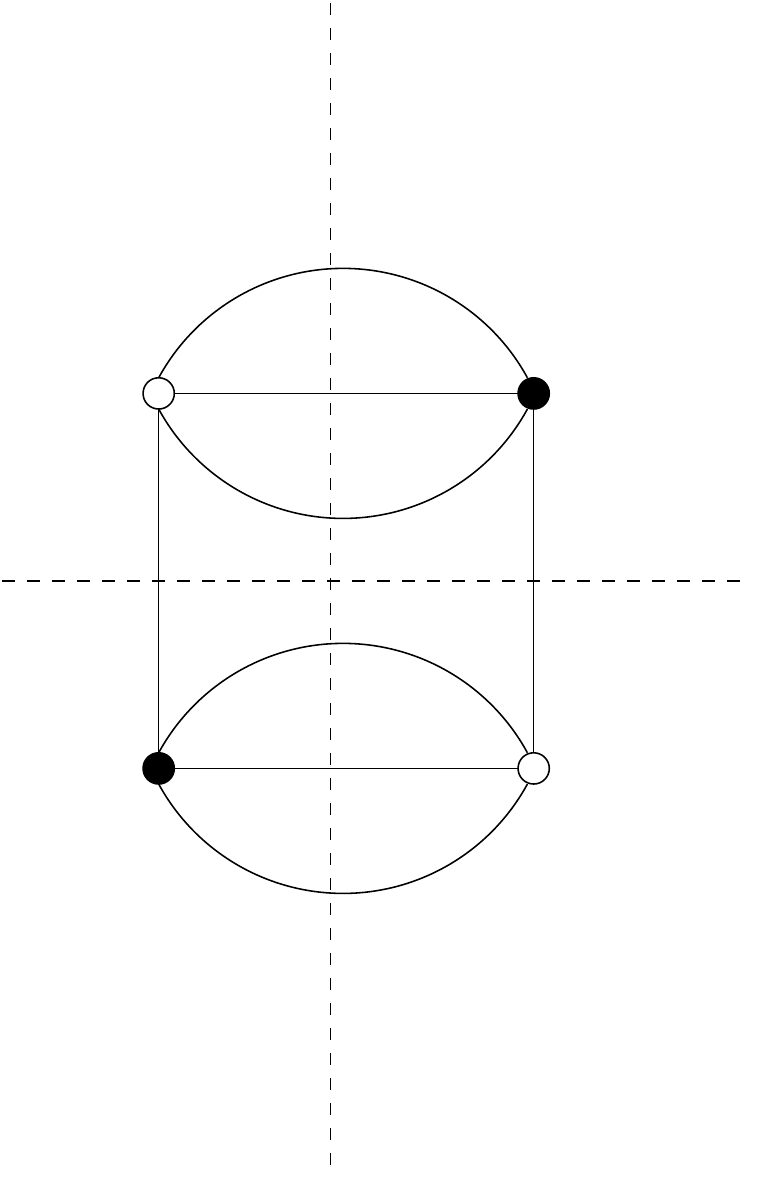} \hskip1cm \includegraphics[height=4cm]{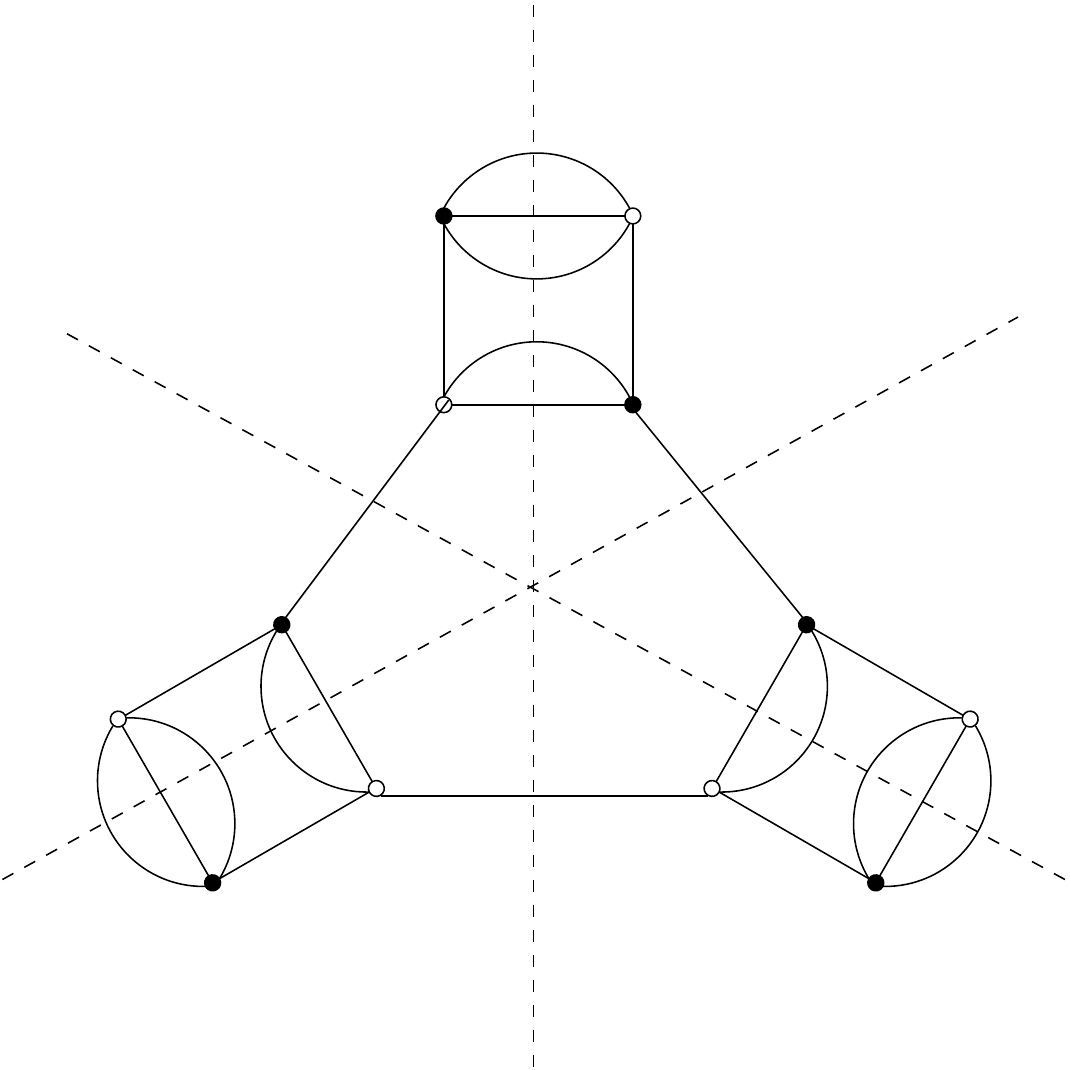}
  \caption{Some rank-four invariants with two or three Hermitian axis of symmetry, pictured as dotted lines.}
  \label{fig:symme}
\end{figure}

\begin{figure}[!htp]
  \centering
 \includegraphics[scale=0.7]{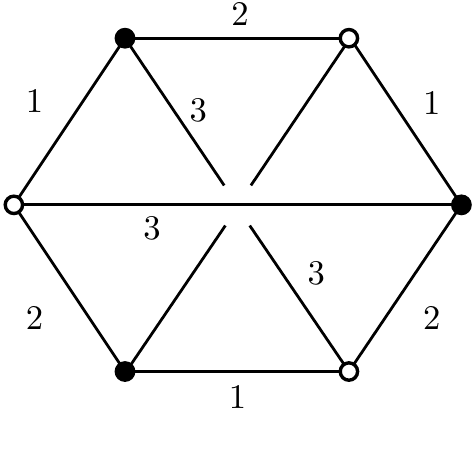} \hskip2cm \includegraphics[height=3.5cm]{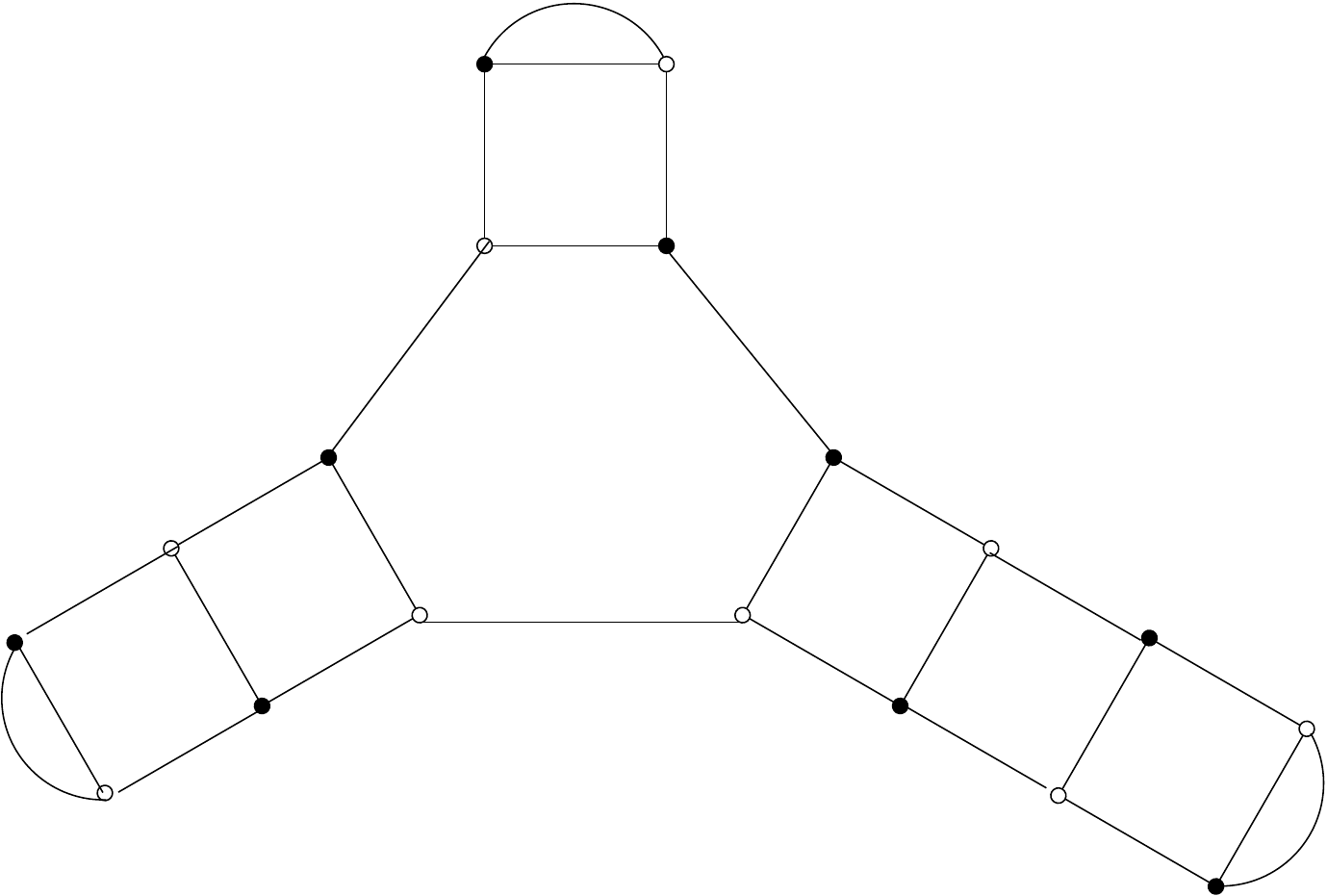}
  \caption{Some rank-six invariants with no Hermitian symmetry, respectively of order 6 (left) and 18 (right). The first one, the complete bipartite graph $K_{3,3}$, is not-melonic, but the second is melonic.}
\label{fig:nosymme}
\end{figure}

In other words positive graphs have therefore the edge-cut as symmetry axis. Some graphs of this type are pictured in Figure \ref{fig:symme}.
Remark that the edge-cut with the above properties may not be unique, see examples in Figure \ref{fig:symme}.
Some graphs without any such symmetry axis  are pictured in Figure \ref{fig:nosymme}.

\begin{figure}[h!]
\includegraphics[scale=0.7]{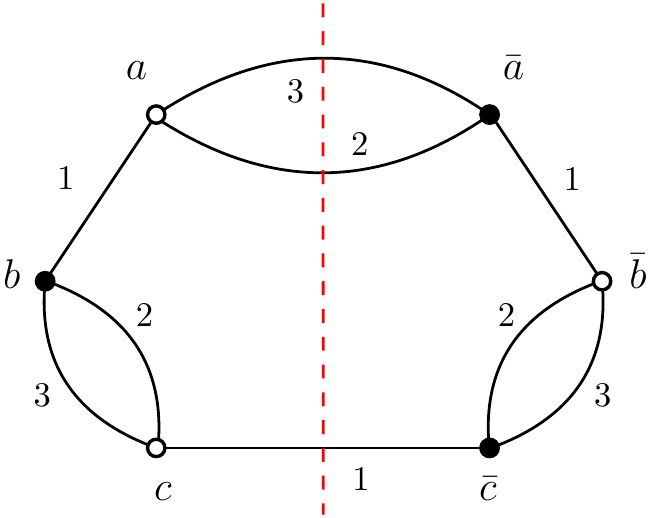}\hspace{2.5cm}\includegraphics[scale=0.7]{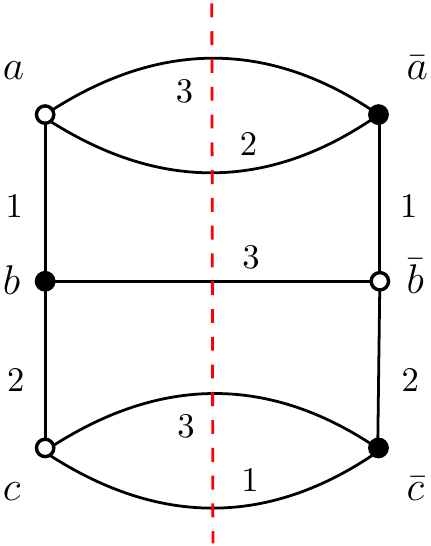}
\caption{\label{fig:3d6n}  Melonic $D=3$ bubbles with 6 vertices}
\end{figure}

Consider in more detail the $D=3$ bubbles with six vertices, pictured on the left of Figure \ref{fig:nosymme} and in Figure 
\ref{fig:3d6n}. The latest are positive and pictured along with a choice of edge-cut. They are also \emph{melonic} \cite{Gurau:2011xp}, which allows an easy identification of the $1/N$ expansion of the corresponding interacting models. The first one, the complete bipartite graph $K_{3,3}$, also called \emph{utility graph}, is non-positive, non-melonic and non-planar, and will not be studied further in this paper.\\
\\

\subsection{Intermediate Field Representation}

\begin{theorem}\label{inttensrep}
For any positive $\cB$ with $2 k( \cB)$ vertices
the partition function $Z_\cB(\lambda,N)$ has an HIF representation in the sense of Definition \ref{hifrepre}
\be
\label{PrtensF1}
Z_\cB(\lambda,N)=\int  d\nu(\xi)e^{-\Tr\ln\bigl[ \un^{\Gamma(\cB)}-g_\cB \frak{M}_\cB(\xi) \bigr] },
\ee
in which 
\begin{itemize}

\item
the Gaussian measure $d \nu ( \xi)$ is of the mixed Gaussian type in the sense of Definition \ref{defmixed},

\item
$g_\cB=\lambda{\frac{1}{2k(\cB)}} N^{\frac{-s(\cB)}{2k(\cB)}}$, 

\item
the matrix  $\frak{M}_\cB(\xi)=i{\bf C}_\cB.\frak{H}_\cB(\xi)$ is a $N^{\Gamma(\cB)}\times N^{\Gamma(\cB)}$ matrix, with $\Gamma(\cB)$ an integer depending on the following developments (see the end of subsection \ref{int_field_tens})

\item
${\bf C}_\cB$ is a mixed covariance, i.e. a direct sum of blocks of type $\un$ (size N identity matrix) 
and $\begin{pmatrix} 0  & - i\un_{i} \\ - i\un_i & 0  \end{pmatrix}$ factors ($\un_i$ being a size $i$ identity matrix), 

\item
the matrix $\frak{H}_\cB$ is  linear in the $\{\xi \}$ variables
and Hermitian when these variables are taken on undeformed contours, i.e. at $\epsilon = 0$.
\end{itemize}
\end{theorem}

\subsubsection{First step, along the cut}

We choose an arbitrary ordering of the vertices in $F$ and name them $\{a,b,c,\cdots\}$ accordingly. As the vertices in $\bar F$ each have a single associated vertex in $\{a,b,c,\cdots\}$, they inherit the order and we name them $\{\bar a,\bar b,\bar c,\cdots\}$, as shown in Figure \ref{fig:FF}. 
In this section, as in the section that deals with matrix invariants, each successive splitting will introduce new intermediate fields. The edge-cut $\cI$ will require the introduction of a tensor field $\sigma$ of rank $\lvert \cI \rvert$, that may have more than one index summed up e.g. with the first index of some tensor $T$ or $\bar T$. To distinguish between edges of the same color reaching such a vertex $\sigma$, we give new colors to the edges of the cut. An edge of color $i\in\llbracket 1,D\rrbracket$ now gets color $i_j$, where $j$ is the name of the tensor vertex they link. This goes back to distinguishing the corresponding copies of the Hilbert spaces $\cH_{i_j}$ for different values of $j$.
This is only necessary when more than one edge of the same color belongs to the cut. To an edge-cut $\cI$ as defined above is associated the set $I$ of colors of its edges.
For instance, the set $I$ corresponding to the edge-cut $\cI$ in Figure \ref{fig:FF} is $I=\{1_a,1_b,2_d,2_g,3,4\}$. \\ 

The tensor model associated with a positive invariant with chosen edge-cut $\cI$ can be rewritten as
\begin{equation}
Z_\cB(\lambda,N)=\int d\mu(T,\bar T) e^{-\lambda N^{-s(\cB} F(T,\bar T)._{I}\bar F(T,\bar T)},
\end{equation}
where $s$ is the appropriate scaling
associated to the invariant considered to ensure a non-trivial limit as $N\to \infty$ $d\mu(T,\bar T)=\prod_{i_1,...,i_D=1}^N dT_{i_1,...,i_D}d\bar T_{i_1,...,i_D}e^{-N^{D-1}T_{i_1,...,i_D}\bar T_{i_1,...,i_D}}$.\footnote{Beware however that the optimal scaling $s$ is not known for general tensor invariants \cite{BLR}.}\\

We generalize now the developments of the matrix section. Relations (\ref{eqref=HS1}) and (\ref{eqref=HS2}), which formally justify every upcoming intermediate field split, become
\be
\label{eqref=HS1_tens}
e^{-gA._I B}=\int d\mu^c(\Phi)e^{i\sqrt g\,( A._I \Phi + B._I \bar \Phi)},
\ee
and variations for intermediate fields of imaginary covariances $\pm i$
\be
\label{eqref=HS2_tens}
e^{-gA._I B}=\int d\mu^c_{\pm i}(\Phi)e^{i\sqrt g\,( A._I \Phi \mp B._I \bar \Phi)},
\ee
where now $A, B, \Phi$ are tensors.

Again, Gaussian imaginary integrals are considered in this section in the $\epsilon \to 0$ \emph{formal limit}, and the corresponding
computations will be justified by reinstating later the $\epsilon$ regulator, as in the matrix section.

\begin{figure}[h!]
\includegraphics[scale=0.7]{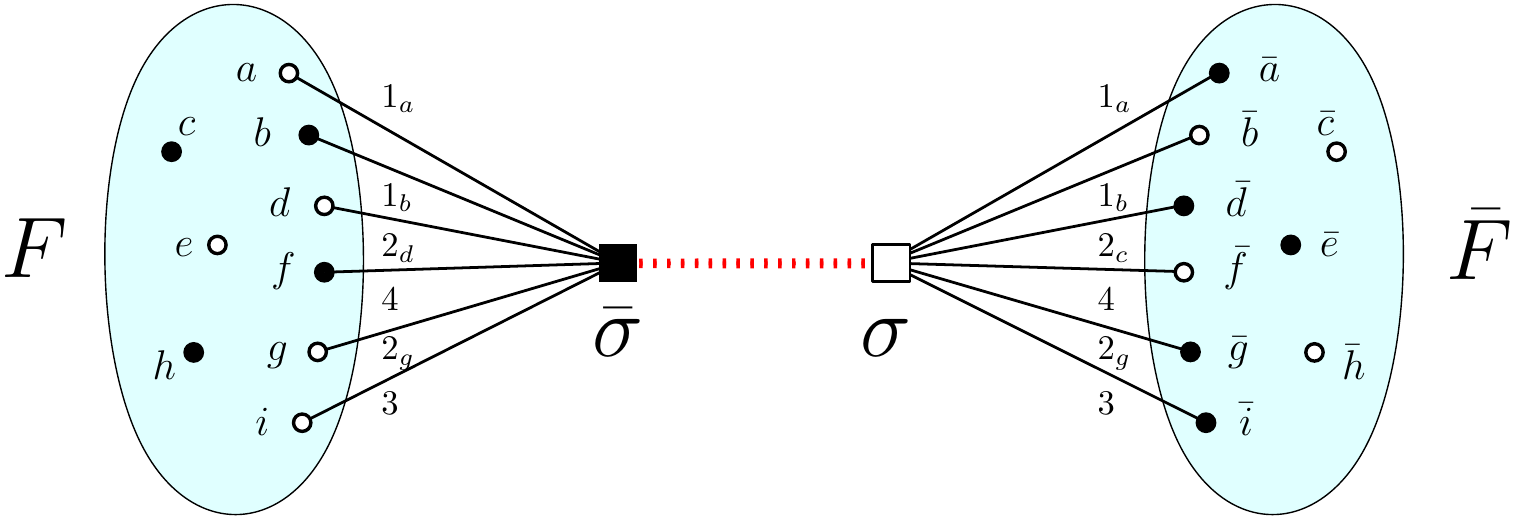}
\caption{\label{fig:FFcut} Intermediate field along the edge-cut $\cI$.}
\end{figure}

As sketched above, we decompose the bubble $\cB$ along the edge-cut $\cI$. This requires the introduction of an intermediate tensor field $\sigma$ of rank $\lvert I \rvert$. 


\begin{equation}
Z_\cB(\lambda,N)=\int d\mu(T,\bar T) e^{-\lambda N^{-s(\cB)} F._{I}\bar F}=\int d\mu(T,\bar T)d\mu(\sigma,\bar \sigma) e^{i\sqrt{\lambda N^{-s(\cB)}} \bigl[ F._{I}\bar \sigma+\sigma._{I}\bar F\bigr]},
\end{equation}
as pictured in Figure \ref{fig:FFcut}. The contraction of the $\kappa$'th index of $\sigma$ or $\bar \sigma$ with the $i$th index of some other tensor $T$ or $\bar T$ of the boundary of $\bar F$ denoted $j$ is graphically represented by an edge of color $\kappa=i_j\in I$. As before for matrices, the tensor $\sigma$ (resp. $\bar \sigma$) will be represented by a white (resp. black) square. For instance, the contraction of $\bar\sigma$ and the boundary of $F$ (if $T$ is of rank 4) in Figure \ref{fig:FFcut} is :
\begin{equation}
\sum_{i_{1_a},i_{1_b},i_{2_d},i_{2_g},i_{3},i_{4}} T_{i_{1_a}, h_2,h_3,h_4}\bar T_{i_{1_b}, j_2,j_3,j_4}T_{k_1,i_{2_d},k_3,k_4}\bar T_{l_1,i_{2_g},l_3,l_4}T_{m_1,m_2,i_3,m_4}T_{n_1,n_2,n_3,i_4} \bar \sigma_{i_{1_a},i_{1_b},i_{2_c},i_{2_e},i_{3},i_{4}}.
\end{equation}

\begin{figure}[h!]
\includegraphics[scale=0.8]{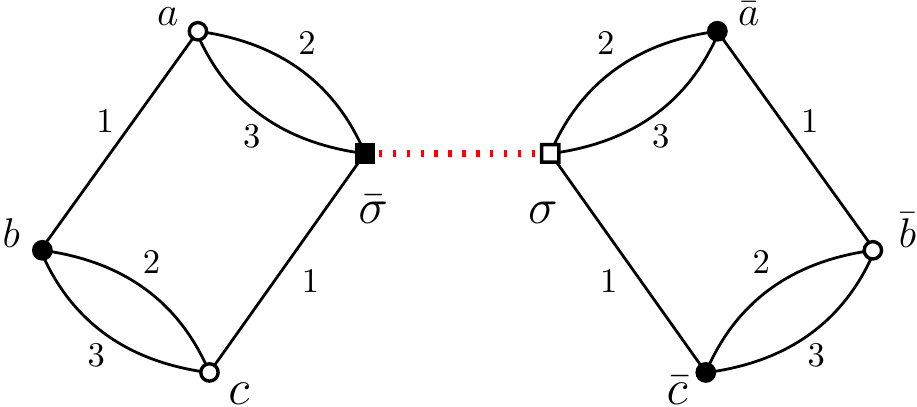}\hspace{2cm}\includegraphics[scale=0.85]{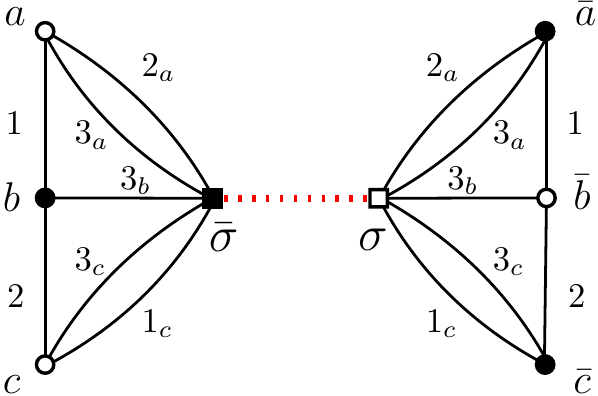}\hspace{0.2cm}
\caption{\label{fig:3d6ncut}  Positive $D=3$ bubbles with 6 vertices and intermediate field along the chosen cut.}
\end{figure}

The examples of the positive $D=3$ bubble with 6 vertices are shown in Figure \ref{fig:3d6ncut}.\\

\begin{figure}[h!]
\hspace{1cm}\includegraphics[scale=0.9]{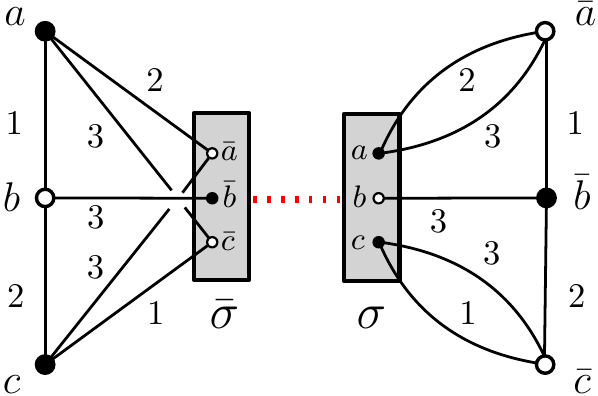}\hspace{3.5cm}\includegraphics[scale=0.9]{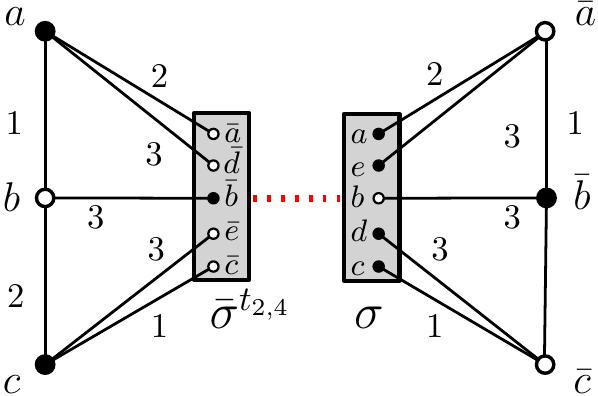}\caption{\label{fig:nonposcut}  Non-positive $K_{3,3}$ bubble. On the left is a unitary intermediate field decomposition, which produces a non-Hermitian term. On the right is a Hermitian term which is not unitary invariant.}
\end{figure}

\noindent
{\bf Remark} The non-crossing condition is important. When the product between $F$ and $\bar F$ crosses and we still implement the intermediate field decomposition, as 
illustrated in Figure \ref{fig:nonposcut} in the case of the $K_{3,3}$ graph, which is the only non-positive $D=3$ bubble with 6 vertices,
we get either unitary invariants or Hermiticity but not both.

\subsubsection{Full decomposition}
\label{int_field_tens}


We now pair $\bar \sigma$ with the first tensor $a$. The contraction $F._I\bar \sigma$, which is graphically represented by a connected graph, can be rewritten in order to make this pairing explicit. We denote $\cI_1$ the edge-cut which separates tensors $a$ and $\bar\sigma$ from the rest, $\bar F_1$, of its connected component, $\cJ_{1}$ the set of edges between tensor $a$ and $\bar F_1$, and the edges in $\tilde \cJ_1$ are the other edges reaching $a$, all contracted with $\sigma$. Note that $\cJ_1$ is not empty if the initial bubble has more than 2 vertices, which is the case considered here.\\

As before for $\cI$, we change the colors of the edges in $\cI_1$ by adding the vertex they reach in $\bar F_1$ as a subscript, and denote $I_1$ the corresponding set of colors. Note that if an edge in $\cI_1$ of color $i\in\llbracket 1,D\rrbracket$ reaches a vertex $j\in\{a,b,\cdots\}$ which previously belonged to the boundary of $F$, then that edge already carries color $i_j$ from the previous splitting. 

\begin{figure}[h!]
\includegraphics[scale=.72]{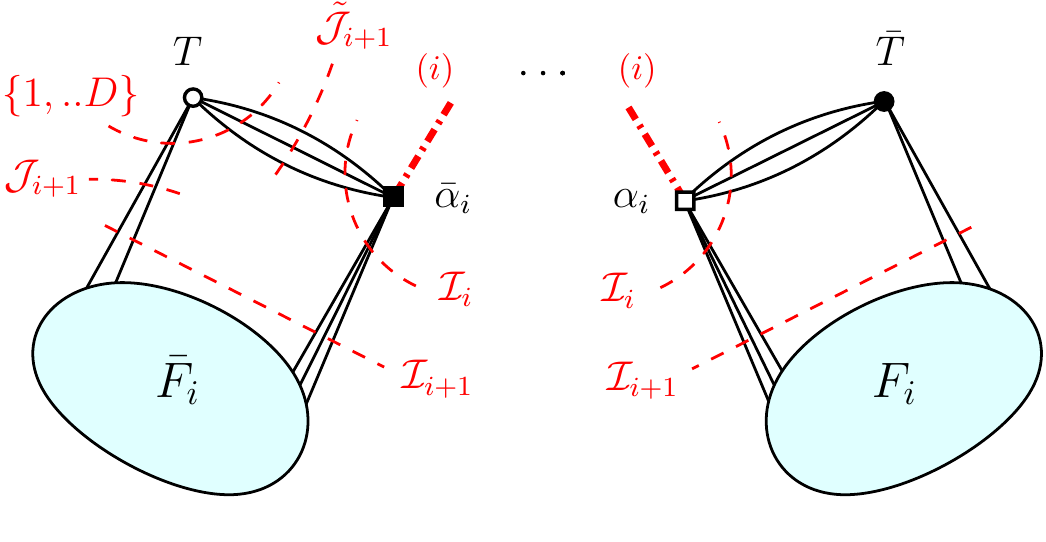}\hspace{0.7cm}\includegraphics[scale=.72]{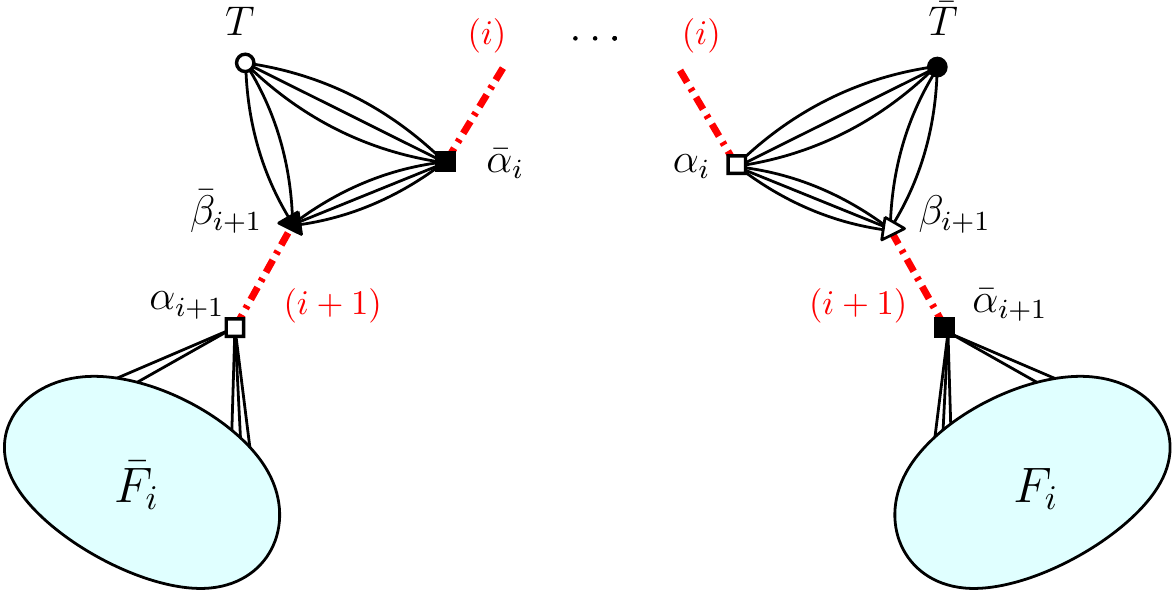}
\caption{\label{fig:odd_split_tens} 
Step $i$ of the intermediate field decomposition, which also applies for $i=0$, in which case $\alpha_0=\sigma$.
}
\end{figure}

To the subset $\cJ_1\subset\cI_1$ is associated the corresponding subset $J_1\subset I_1$. The edges in $\tilde \cJ_1$ are precisely those with color $i_a\in I$, where $i$ spans all the color of $\llbracket 1,D \rrbracket$ that are not in $I_1$. Those new sets of colors are such that 
$I=(I_1\setminus J_1)\sqcup \tilde J_1$, and $I_1=J_1\sqcup(I\setminus \tilde J_1)$, which is easily seen on Figure \ref{fig:odd_split_tens}, with the convention that $\alpha_0=\sigma$. This allows to re-express 
\be
F._I\bar \sigma=[\bar F_1._{J_1}T]._{I}\bar\sigma=\bar F_1._{I_1}[T._{\tilde J_1} \bar \sigma].
\ee
For an initial bubble $\cB$ with $2k(\cB)$ vertices, we define 
\be
g_\cB=\lambda^{\frac{1}{2k(\cB)}} N^{\frac{-s(\cB)}{2k(\cB)}},
\ee 
and generalize the trick we used in the matrix section, 
\bea
\label{eqref:squar_diff_tens}
\sqrt{\lambda N^{-s(\cB)}} \bigl[ F._{I}\bar \sigma+\sigma._{I}\bar F\bigr]&=&g_\cB^{k}\bigl[\bar F_1._{I_1}[T._{\tilde J_1} \bar \sigma]+[\bar T._{\tilde J_1} \sigma] ._{I_1}F_1\bigr]\\
&=&\frac{1}{2}\bigl[(g_\cB^{k-1}\bar F_1+g_\cB[\bar T._{\tilde J_1}\sigma])._{I_1}(g_\cB[T._{\tilde J_1}\bar \sigma]+g_\cB^{k-1}F_1)\nonumber\\
&&\hspace{3cm}+(g_\cB^{k-1}\bar F_1-g_\cB[\bar T._{\tilde J_1}\sigma])._{I_1}(g_\cB[T._{\tilde J_1}\bar \sigma]-g_\cB^{k-1}F_1)\bigr].
\eea

Now applying (\ref{eqref=HS2_tens}) with complex intermediate tensor fields $a_1$ and $b_1$ of covariances $-i$ and $+i$ respectively, 
\bea
&e^{-\lambda N^{-s(\cB)} F._{I}\bar F}&=\int d\mu^c(\sigma)d\mu^c_{\pm i}(a_1,b_1)e^{\frac{i}{\sqrt 2}\bigl[g_\cB^{k-1}\bar F_1._{I_1}(a_1+b_1) +g_\cB[T._{\tilde J_1}\bar \sigma]._{I_1}(\bar a_1-\bar b_1) + c.c. \bigr]}
\eea
where the measure is $d\mu^c_{\pm i}(a_1,b_1)=d\mu^c_{-i}(a_1)d\mu^c_{+i}(b_1)$ and $c.c.$ stands for {\it complex conjugate}.

As in the matrix subsection, we now change variables, 
\be
\alpha_1=\frac{a_1+b_1}{\sqrt 2}, \quad\beta_1=\frac{a_1-b_1}{\sqrt 2},
\ee
and complex conjugates. With those variables, 

\bea
&e^{-\lambda N^{-s(\cB)} F._{I}\bar F}&=\int d\mu^c(\sigma)d\mu^c_X(\alpha_1,\beta_1)e^{i\bigl[g_\cB^{k-1}\bar F_1._{I_1}\alpha_1 +g_\cB[T._{\tilde J_1}\bar \sigma]._{I_1}\bar\beta_1 + c.c. \bigr]},
\eea
the Gaussian measure $d\mu^c_X$ being defined by its moments, that all vanish apart from 
\be
\label{eqref:cov_tens}
\forall i_1,\cdots,i_{\lvert I_1\rvert} \in \{1,...,N\}, \quad<\alpha_{1\mid i_1,\cdots,i_{\lvert I_1\rvert}}\bar\beta_{1\mid i_1,\cdots,i_{\lvert I_1\rvert}}>_X=<\bar\alpha_{1\mid i_1,\cdots,i_{\lvert I_1\rvert}}\beta_{1\mid i_1,\cdots,i_{\lvert I_1\rvert}}>_X=-i.
\ee

The term $\bar F_1._{I_1}\alpha_1$ is of the exact same form as the term $\bar F ._I \sigma$ before. We apply therefore the same reasoning again. 
We couple $\alpha_1$ with the first black vertex $a'$ in the boundary of $\bar F_1$ such that there existed an edge $(a a')\in\cJ_1$ in the previous step. It is always possible as $\cJ_1$ is not empty. We denote $\cI_2$ the edge-cut which separates tensors $a'$ and $\alpha_1$ from the rest, $ F_2$, of its connected component, $\cJ_{2}$ the set of edges between tensor $a'$ and $F_2$, $\tilde\cJ_{1}$ those between $a'$ and $\alpha_1$, and as before we change the colors of the edges when necessary and name  $I_2$, $J_2$, $\tilde J_2$ the associated sets of colors. Our choice for vertex $a'$ ensures that $\tilde J_2\cap J_1 \neq \emptyset$.
As in the previous step, $\bar F_1._{I_1}\alpha_1=[ F_2._{J_2}\bar T]._{I_1}\alpha_1=F_2._{I_2}[\bar T._{\tilde J_2} \alpha_1]$, and after developments such as in (\ref{eqref:squar_diff_tens})-(\ref{eqref:cov_tens}),
\be
e^{ig_\cB^{k-1}\bar F_1._{I_1}\alpha_1}=\int d\mu^c_X(\alpha_2,\beta_2)e^{i\bigl[g_\cB^{k-2}F_2._{I_2}\bar\alpha_2 +g_\cB[\bar T._{\tilde J_2}\alpha_1]._{I_2}\beta_2 + c.c. \bigr]}.
\ee

Note that we could choose any black vertex in $\bar F_1$ but this choice ensures that $\tilde J_2$ is not empty.  We also stress that we
use pairings which are not necessarily optimal, in the sense that we do noty try to 
introduce intermediate fields of the smallest possible ranks.

We inductively apply the same reasoning until step $k-2$, which leaves us with $F_{k-1}$ having only two remaining tensor vertices. As we decomposed the initial interaction into a sum of connected interactions involving only three tensors each, we shall later be able to do a Gaussian integration over $T$ and other fields.  The partition function currently has the following expression
\bea
\label{eqref:full_int_dec}
Z_\cB(\lambda,N) &=&\int   d\mu^c(T)d\mu^c(\sigma)\prod_{j=1}^{k-2}d\mu^c_{X}(\alpha_j,\beta_j)
e^{  ig_\cB \bigl(  [\bar T._{\tilde J_1}\sigma]._{I_1}\beta_1\, +\, [\bar T._{\tilde J_2}\alpha_1]._{I_2}\beta_2 \,+ \,[\bar T._{\tilde J_{k-2}}\alpha_{k-3}]._{I_{k-2}}\beta_{k-2}\,+\cdots}\hspace{2cm}
\\  \nonumber && \hspace{10cm}
^{\cdots +\, g_\cB[\bar T._{J_{k-1}}T]._{I_{k-2}}\alpha_{k-2} \,+ c.c. \bigr)},
\eea
where we recall that $k$ is the number of vertices of $F$ and $c.c.$ stands for complex conjugate. The $i$'th splitting introduces the tensor intermediate fields $\alpha_{i+1}$ and $\beta_{i+1}$, with complex covariances as in \eqref{eqref:cov_tens} and is represented in Figure \ref{fig:odd_split_tens} (this may also apply for $\alpha_0=\sigma$). As in the two first steps, the sets of newly introduced colors are such that $J_i$ is non-empty, and
\be
\label{eqref:reltn_i}
\tilde J_{i+1} \cap J_i\neq\emptyset,\quad\textrm{and}\quad I_{i+1}\setminus J_{i+1} = I_i \setminus \tilde J_{i+1},
\ee
and in particular, 
\bee
\label{relatI}
I_{i}=(I_{i+1}\setminus J_{i+1})\sqcup \tilde J_{i+1} =J_{i}\sqcup(I_{i-1} \setminus \tilde J_{i}),
\ee 
the sets having empty intersections because two edges of the same color cannot reach the same tensor.\\

\begin{figure}[h!]
\includegraphics[scale=.8]{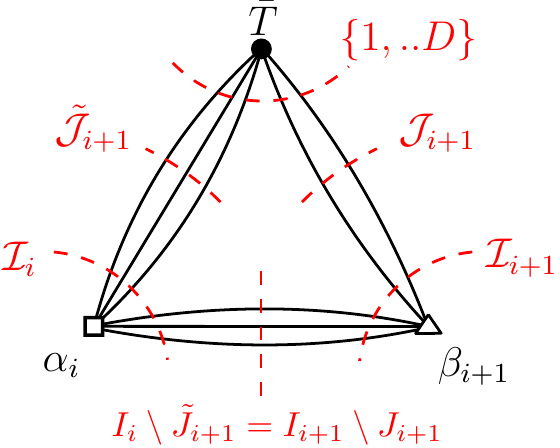}
\caption{\label{fig:triangle_tens} 
Triangular connected term in the full decomposition (\ref{eqref:full_int_dec}).}
\end{figure}

As each connected interaction is now a contraction of three tensors, $\bar T$, $\alpha_{i}$, and $\beta_{i}$ (or complex conjugates), we may now naturally organize each tensor $\alpha_i$ (resp. $\beta_i$) as a rectangular matrix of size $N^{\lvert \tilde J_{i+1}\rvert }\times N^{\lvert I_i\setminus \tilde J_{i+1}\rvert}$ (resp. $N^{\lvert J_i\rvert }\times N^{\lvert I_i\setminus J_i\rvert}$), in the sense that we specify its first and second sets of indices, respectively those contracted to $T$ and the remaining ones (see Figure \ref{fig:triangle_tens}).
 More precisely, we understand $\alpha_i$ and $\beta_i$ as matrices of linear maps :
\begin{equation}
\begin{array}{rll}
\otimes_{j\in {I_i\setminus \tilde J_{i+1}}}\cH_j&\longrightarrow &\otimes_{j\in \tilde J_{i+1}}\cH_j \\\nonumber\\\nonumber
X_{I_i\setminus \tilde J_{i+1}}&\longrightarrow &\sum_{I_i\setminus \tilde J_{i+1}}\alpha_{i\ \mid \ \tilde J_{i+1}\  ;\ I_i\setminus \tilde J_{i+1}} \ X_{I_i\setminus \tilde J_{i+1}} 
\end{array}
\  \textrm{and} \qquad
\begin{array}{rll}
\otimes_{j\in {I_{i}\setminus J_{i}}}\cH_j&\longrightarrow &\otimes_{j\in J_{i}}\cH_j \\\nonumber\\\nonumber
X_{I_{i}\setminus J_{i}}&\longrightarrow &\sum_{I_{i}\setminus J_{i}}\beta_{i\ \mid \ J_{i}\  ;\ I_i\setminus J_{i}} \ X_{I_i\setminus J_i} .
\end{array}
\end{equation}
For the last term, $[\bar T._{J_{k-1}}T]._{I_{k-2}}\alpha_{k-2}$, the convention is that $\alpha_{k-2}$ is a $N^{\lvert J_{k-1} \rvert}\times N^{\lvert \tilde J_{k-1} \rvert}$ square matrix, since $\lvert J_{k-1} \rvert=\lvert \tilde J_{k-1} \rvert=D-\lvert J_{k-1} \rvert$.
These conventions will be very useful at the end of the section. The sizes of the matrices are easily readable on the connected triangular graphs of the graphical representation of the full intermediate field decomposition (Figure \ref{fig:triangle_tens} and examples in subsection \ref{Examples}). The following relations are also verified:
\be
\label{eqref:trans_brackets}
\bar F_i._{I_i}\alpha_i\ =\ [\bar T._{J_{i+1}} F_{i+1}]._{I_i}\alpha_i\ =\ [\bar T._{\tilde J_{i+1}} \alpha_i]._{I_{i+1}}F_{i+1},
\ee
and similarly after the intermediate field split (Figure \ref{fig:triangle_tens}),

\be
\label{eqref:trans_brackets_2}
[\bar T._{J_{i+1}} \beta_{i+1}]._{I_i}\alpha_i\ =\ [\bar T._{\tilde J_{i+1}} \alpha_i]._{I_{i+1}}\beta_{i+1} \ = \ \bar T._{\llbracket 1,D\rrbracket}\ [ \beta_{i+1}\,\alpha_i^{\mathrm T}],
\ee
where with our new convention, $\beta_{i+1}\alpha_i^{\mathrm T}\ $ is a $N^{\lvert J_{i+1} \rvert }\times N^{\lvert \tilde J_{i+1} \rvert }$ matrix.\\

We shall now integrate over a subset of the intermediate fields using the following relation, 
\be
\label{eqref:tens_int}
\int d\mu^c_X(\alpha,\beta)e^{i(A._I\alpha+B._I\beta+C._I\bar \alpha+D._I\bar\beta)}=e^{i(A._ID+B._IC)}.
\ee
The integration is performed over $T$ and all $\alpha_{k-1-2j}$, $\beta_{k-1-2j}$, for 
$j\in\{1,...,\lfloor\frac{k}{2}\rfloor\}$, i.e. 
\begin{itemize}
\item for $k$ odd, over the $k$ intermediate tensor fields $T$, $\sigma$, all even $\alpha_{2j}$, $\beta_{2j}$, for $j\in\{1,..,\frac{k-3}{2}\}$ and complex conjugates. 

\item for $k$ even, over the $k$ intermediate tensor fields $T$, all odd $\alpha_{2j-1}$, $\beta_{2j-1}$, for $j\in\{1,..,\frac{k-2}{2}\}$ and complex conjugates. 
\end{itemize}

To use (\ref{eqref:tens_int}) we must rewrite, using relation (\ref{eqref:trans_brackets_2}), 
\be
 [\bar T._{\tilde J_{k-2j}}\alpha_{k-2j-1}]._{I_{k-2j}}\beta_{k-2j}=[\bar T._{ J_{k-2j}}\beta_{k-2j}]._{I_{k-2j-1}}\alpha_{k-2j-1}.
\ee
Each integration step is then done independently of the others: 
\bea
\int d\mu^c_X(\alpha_{k-1-2j},\beta_{k-1-2j})e^{ig_\cB\bigl([\bar T._{J_{k-2j}}\beta_{k-2j}]._{I_{k-2j-1}}\alpha_{k-2j-1} + [\bar T._{\tilde J_{k-2j-1}}\alpha_{k-2(j+1)}]._{I_{k-2j-1}}\beta_{k-2j-1} + c.c. \bigr)}
\nonumber\\
=\ e^{ig_\cB^2\bigl([\bar T._{J_{k-2j}}\beta_{k-2j}]._{I_{k-2j-1}} [T._{\tilde J_{k-2j-1}}\bar\alpha_{k-2(j+1)}]  +c.c. \bigr ) }, \ 
\eea
except for the $\sigma$ integration for $k$ odd, 
\bea
\int d\mu^c(\sigma)e^{ig_\cB\bigl( [\beta_1._{J_1}\bar T]._{I}\sigma +  c.c. \bigr)} =\ e^{-g_\cB^2\bigl([\bar T._{J_1}\beta_1]._{I}  [\bar\beta_1._{J_1}T] \bigr)}, \ 
\eea
which leaves us with the integration of the $T$ variable in 
\bea
Z_\cB(\lambda,N) =\int d\nu(\xi)  d\mu^c(T)
e^{  i g_\cB ^2 \biggl(  \sum_{j=1}^{\lfloor\frac{k}{2}\rfloor}[\bar T._{J_{k-2j}}\beta_{k-2j}]._{I_{k-2j-1}} [T._{\tilde J_{k-2j-1}}\bar\alpha_{k-2(j+1)}] }\nonumber \\
^{  +\, [\bar T._{J_{k-1}}T]._{I_{k-2}}\bar \alpha_{k-2} \,+\, c.c. \,+\, i\eta(k) [\bar T._{J_1}\beta_1]._{I}  [\bar\beta_1._{J_1}T] \biggr)},\nonumber\\
\eea
in which $\xi=(...\alpha_{k-2j}, \beta_{k-2j},...)$ is the vector containing the $k-1$ remaining variables, $\xi_{odd}=(\alpha_1,\beta_1,..., \alpha_{k-2}, \beta_{k-2})$ and $\xi_{even}=(\sigma, \alpha_2,\beta_2,..., \alpha_{k-2} , \beta_{k-2})$,
and  $\alpha_0=\sigma$, $\beta_k=1$, $\alpha_{<0}=0$ and $\beta_{\le0} = 0$, and $\eta(k)$ is 0 for $k$ even and 1 for $k$ odd.

In order to perform the integration over $T$, we must {\it factorize} 
\bea
[\bar T._{J_{k-1}}T]._{I_{k-2}}\bar \alpha_{k-2}&=&\sum_{J_{k-1},\tilde J_{k-1},  J'_{k-1}, \tilde J'_{k-1}} \bar T_{J_{k-1} \,; \,\tilde J_{k-1}} \un_{J_{k-1} \,;\, J'_{k-1}}\alpha_{k-2 \mid \tilde J_{k-1}\, ; \,  \tilde J'_{k-1}}T_{J'_{k-1}\, ;\, \tilde J'_{k-1}}\nonumber\\
&=&\ \ \bar T ._{\llbracket 1,D \rrbracket} \bigl(\ \alpha_{k-2}\otimes\un^{ \otimes\lvert J_{k-1}\rvert}\bigr)._{\llbracket 1,D \rrbracket} T,
\eea
since the indices that aren't contracted in $[\bar T._{J_{k-1}}T]$ are both the indices of $\bar T$ and of $T$ with colors in $\tilde J_{k-1}$. 
Also tensor $T$ does not see the subscript colors, and regardless of those, $J_{k-1}\sqcup\tilde J_{k-1}=\llbracket 1,D\rrbracket$.\footnote{We use a simplifying notation to have clearer expressions. The sum over a color set $J$ here means a sum for variables indexed with each color in $J$.} In these equations, $\alpha_{k-2}$ is to be understood as a square $N^{\lvert \tilde J_{k-1} \rvert}\times N^{\lvert \tilde J_{k-1} \rvert}$ matrix, as outlined before.\\

Similarly, as $I=(I_1\setminus J_1)\sqcup \tilde J_1$, the indices of $\beta_1$ in $[\bar T._{J_1}\beta_1]._{I}  [\bar\beta_1._{J_1}T]$ that are not contracted with $\bar T$ are precisely the indices in $I$ that are summed up with those of the same colors in $\bar \beta_1$, so that 
\bea
[\bar T._{J_1}\beta_1]._{I}  [\bar\beta_1._{J_1}T]&=&\sum_{J_1,\tilde J_1,J'_1,\tilde J'_1}\bar T _{J_1\,;\,\tilde J_1}[\beta_1._{I_1\setminus J_1}\bar \beta_1]_{J_1\,;\,J'_1} \un_{\tilde J_1\,;\,\tilde J'_1}T_{J'_1\,;\,\tilde J'_1}\nonumber\\
&=&\ \ \bar T ._{\llbracket 1,D \rrbracket}\bigl(\  \beta_1\beta_1^\dagger\otimes\un^{ \otimes\lvert\tilde J_1\rvert}\bigr)._{\llbracket 1,D \rrbracket} T,
\eea
in which $\beta_1\beta_1^\dagger$ is a square $N^{\lvert J_{1} \rvert}\times N^{\lvert J_{1} \rvert}$ matrix. 

\begin{figure}[h!]
\includegraphics[scale=.8]{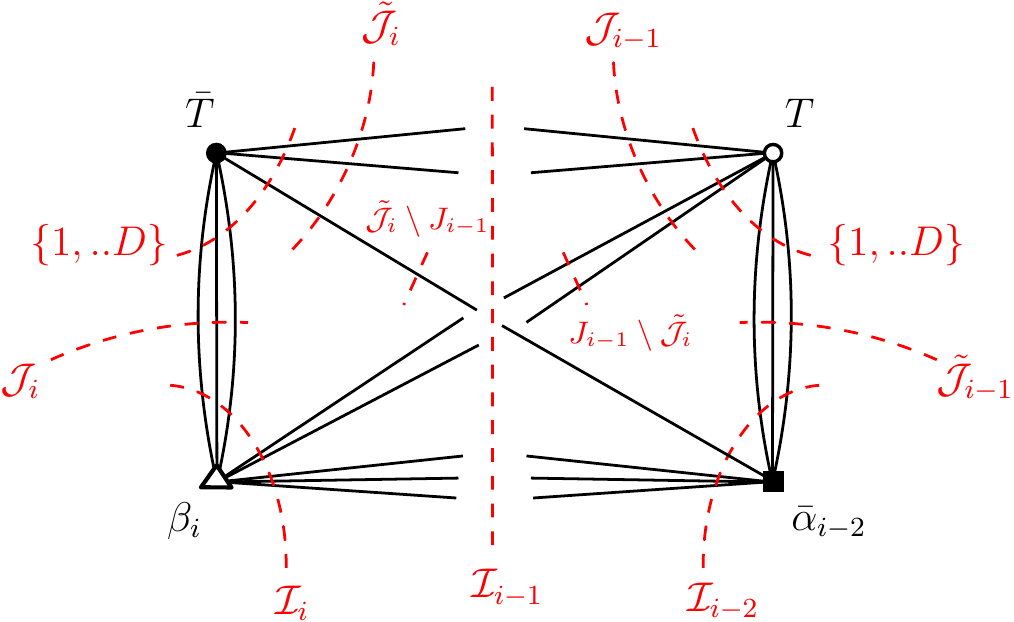}
\caption{\label{fig:int_tens} 
Terms in $\mathbf{H_k}$ after integration, taking $i=k-2j$.}
\end{figure}

The other terms of the sum require a slightly subtler treatment, as $J_{k-2j}$ and $\tilde J_{k-2j-1}$ may have a non-empty intersection (see Figure \ref{fig:int_tens}),
\bea
[\bar T._{J_{i}}\beta_{i}]._{I_{i-1}} [T._{\tilde J_{i-1}}\bar\alpha_{i-2}]&=&
\sum_{
\substack {
{  J_{i},\ \tilde J_{i}\cap J_{i-1},\ \tilde J_{i-1},\ \   }\\
\nonumber
{  \tilde J_{i}\setminus J_{i-1}, \ J_{i-1}\setminus \tilde J_{i}    }
}}
\nonumber\\
\nonumber
&&\bar T _{J_{i}  \,;\,     \tilde J_{i}\setminus J_{i-1}   \,;\,     \tilde J_{i}\cap J_{i-1}}
[\beta_i._{I_{i-1}\setminus (J_{i-1}\cup\tilde J_i) }\bar\alpha_{i-2}]_{J_i\,;\, \tilde J_{i}\setminus J_{i-1} \vert \tilde J_{i-1}\,;\,J_{i-1}\setminus \tilde J_{i}} 
T_{ \tilde J_{i-1}   \,;\,  J_{i-1}\setminus \tilde J_{i}    \,;\     \tilde J_{i}\cap J_{i-1}}\\
\nonumber\\
&=& \ \ \bar T ._{\llbracket 1,D \rrbracket}\biggl( [\beta_i._{I_{i-1}\setminus (J_{i-1}\cup\tilde J_i) }\bar\alpha_{i-2}]\otimes\un^{ \otimes\lvert\tilde J_i \cap J_{i-1}\rvert}\biggr)._{\llbracket 1,D \rrbracket} T,
\eea
where  $[\beta_i._{I_{i-1}\setminus (J_{i-1}\cup\tilde J_i) }\bar\alpha_{i-2}]$ might be understood as a $N^{D-\lvert  \tilde J_i \cap J_{i-1}\rvert} \times N^{D-\lvert  \tilde J_i \cap J_{i-1}\rvert}$ square matrix with its first half indices contracted to sub-indices of $\bar T$ and the other half to sub-indices of $T$, since  $J_i \sqcup \tilde J_i\setminus J_{i-1}$ corresponds to the edges reaching $\bar T$ that are not in $\tilde J_i \cap J_{i-1} $, and  $\tilde J_{i-1}\sqcup J_{i-1}\setminus \tilde J_{i}$ to the edges reaching $T$ that are not in $\tilde J_i \cap J_{i-1} $ (Figure \ref{fig:int_tens}). This might be rewritten using the matrix conventions introduced earlier,

\be \label{incompr}
[\bar T._{J_{i}}\beta_{i}]._{I_{i-1}} [T._{\tilde J_{i-1}}\bar\alpha_{i-2}]=\ \bar T ._{\llbracket 1,D \rrbracket}\biggl(\  [\beta_i\otimes\un^{\lvert \tilde J_i\rvert }].[\alpha_{i-2}^\dagger\otimes\un^{ \otimes\lvert  J_{i-1}\rvert}]\ \biggr)._{\llbracket 1,D \rrbracket} T,
\ee
where the tensorial products might be understood as a Kronecker product of matrices. The central dot in 
\eqref{incompr} stands for the usual matrix product. Recall indeed that  $\beta_i\otimes\un^{\lvert \tilde J_i\rvert }: \otimes_{j\in (I_{i-1}\setminus\tilde J_i)\sqcup\tilde J_i}\cH_j  \longrightarrow  \otimes_{j\in J_i\sqcup\tilde J_i}\cH_j$, and $\alpha_{i-2}^\dagger\otimes\un^{ \otimes\lvert  J_{i-1}\rvert} : \otimes_{j\in \tilde J_{i-1}\sqcup  J_{i-1}}\cH_j \longrightarrow   
\otimes_{j\in (I_{i-1}\setminus J_{i-1})\sqcup J_{i-1}}\cH_j$.

The complex conjugate term gives 
\bea
[\bar T._{\tilde J_{i-1}}\alpha_{i-2}]._{I_{i-1}} [T._{J_{i}}\bar\beta_{i}]&=&\bar T ._{\llbracket 1,D \rrbracket}\biggl( [\alpha_{i-2}._{I_{i-1}\setminus (J_{i-1}\cup\tilde J_i) }\bar\beta_i]\otimes\un^{ \otimes\lvert\tilde J_i \cap J_{i-1}\rvert}\biggr)._{\llbracket 1,D \rrbracket} T\nonumber\\
&=&\bar T ._{\llbracket 1,D \rrbracket}\biggl( [\beta_i._{I_{i-1}\setminus (J_{i-1}\cup\tilde J_i) }\bar\alpha_{i-2}]\otimes\un^{ \otimes\lvert\tilde J_i \cap J_{i-1}\rvert}\biggr)^\dagger._{\llbracket 1,D \rrbracket} T\nonumber\\
&=&\ \bar T ._{\llbracket 1,D \rrbracket}\biggl(\  [\alpha_{i-2}\otimes\un^{ \otimes\lvert  J_{i-1}\rvert}].[\beta_i^\dagger\otimes\un^{\lvert \tilde J_i\rvert }]\ \biggr)._{\llbracket 1,D \rrbracket} T,
\eea\\
and the $\beta_1$ term may also be written as
\be
[\bar T._{J_1}\beta_1]._{I}  [\bar\beta_1._{J_1}T]=\ \bar T ._{\llbracket 1,D \rrbracket}\biggl(\  [\beta_1\otimes\un^{\lvert \tilde J_1\rvert }].[\beta_1\otimes\un^{\lvert \tilde J_1\rvert }]^\dagger\ \biggr)._{\llbracket 1,D \rrbracket} T.
\ee
so that if we denote the $N^D\times N^D$ Hermitian matrix 
\bea
\mathbf{H}_\cB(\xi)&= &\sum_{j=0}^{\lfloor\frac{k}{2}\rfloor}\ \  [\beta_{k-2j}._{I_{k-2j-1}\setminus (J_{k-2j-1}\cup\tilde J_{k-2j}) }\bar\alpha_{k-2(j+1)}]\otimes\un^{\otimes \lvert\tilde J_{k-2j} \cap J_{k-2j-1}\rvert} + c.t.\nonumber \\
&= &\sum_{j=0}^{\lfloor\frac{k}{2}\rfloor}\ \  [\alpha_{i-2}\otimes\un^{ \otimes\lvert  J_{i-1}\rvert}].[\beta_i^\dagger\otimes\un^{\lvert \tilde J_i\rvert }] + c.t.,
\eea
the integration over $T$ leads to the following expression of the partition function, 

\be
\label{PrtF1_tens}
Z_\cB(\lambda,N)=\int  d\nu(\xi)e^{-\Tr\ln\bigl[\,  \un^{\otimes D} -\,g_\cB^2 \bigl(\, i \mathbf{H}_\cB(\xi) \ -\ \eta(k)\beta_1\beta_1^\dagger\otimes\un^{ \otimes\lvert\tilde J_1\rvert} \,\bigr)\ \bigr] },
\ee\\
where $d \nu$ factorizes over the measures $d\mu_o^c$ of each $\alpha,  \beta $ pair plus the measure $d\mu^c (\sigma)$ 
for $k$ even, and $\eta(k)$ is 0 for $k$ even and 1 for $k$ odd.\\

In the previous expression however, it may be possible to factorize some identity factors of the sum of tensorial products when they act on the same color $i$. Here, only the color as defined in the first place matters, i.e. identity factors acting on spaces $\cH_{i_a}$ and $\cH_{i_b}$ for the same $i$ are factorized. We denote $\Theta_{\cB, \{\cI_i\}}$ the number of tensorial products of the $N\times N$ identity one can factorize in the sum $\ i \mathbf{H}_\cB(\xi) \ -\ \eta(k) \beta_1\beta_1^\dagger\otimes\un^{ \otimes\lvert\tilde J_1\rvert}\,$. In the matrix case
this $\Theta_{\cB, \{\cI_i\}}$ was always exactly 1, leading to the factor $N$ in front of the logarithm in \eqref{PrtF1a}.
In the tensor case we always choose to pair tensor $\alpha_i$ with a vertex of the boundary of the edge-cut $\cI_i$ which belonged to $\cJ_{i}$, and we know that $\forall i, \lvert\tilde J_i \cap J_{i-1}\rvert \ge 1$ and $\lvert \tilde J_1 \rvert \ge 1$, so that $\Theta_{\cB, \{\cI_i\}}\ge 1$.
As $\det(\un^{\otimes\Theta}\otimes M)=\det(M)^{N^{\Theta }}$, this factorisation lays out a global factor $N^{\Theta_{\cB, \{\cI_i\}}}$,
\be
\label{PrtF12_tens}
Z_\cB(\lambda,N)=\int  d\nu(\xi)e^{-N^{\Theta_{\cB, \{\cI_i\}}} \Tr\ln\biggl[\,  \un^{\otimes \bigl(D-\Theta_{\cB, \{\cI_i\}}\bigr)} -\,g_\cB^2 \biggl(\, i H_\cB(\xi) \ -\ \eta(k) \beta_1\beta_1^\dagger\otimes\un^{ \otimes\bigl(\lvert\tilde J_1\rvert-\Theta_{\cB, \{\cI_i\}}\bigr)}\,\biggr)\ \biggr]} ,
\ee\\
where we denoted $H_\cB$ the Hermitian $N^{D-\Theta_{\cB, \{\cI_i\}}}\times N^{D-\Theta_{\cB, \{\cI_i\}}}$ matrix that verifies $H_\cB\otimes \un^{\otimes(\Theta_{\cB, \{\cI_i\}})}=\mathbf{H}_\cB$.
 It is straightforward to determine $\Theta$ graphically. The fields that are not integrated and thus remain in the final representation have the parity of $k$. In the graphical decomposition of the interaction into connected graphs with three vertices, among which a single $T$ or $\bar T$, the identity tensor products are the edges between $T$ (or $\bar T$) and the field which is to be integrated. $\Theta$ is the number of colors for which such an edge exists in every connected graph of order 3. See the example in subsection $\ref{Examples}$.

Obviously, $\Theta_{\cB, \{\cI_i\}}$ depends heavily on the successive choices of edge-cuts $\cI_i$. Some splittings will give a maximal value $\Theta_{\cB}=\max_{\{\cI_i\}}\Theta_{\cB, \{\cI_i\}}$. We shall treat in detail the melonic examples of order 6 in $D=3$ for which 
we give this maximal value. From now on we write simply $\Theta$ for $\Theta_{\cB, \{\cI_i\}}$.

\vspace{0.3cm}
\noindent{\bf Linear Hermitian Intermediate Field Representation}\\
\hspace{0.3cm}

{\it With a similar proof, Lemma \ref{block_mat} generalizes to rectangular matrices, so that expression \eqref{PrtF1_tens} can be reformulated using a $N^{\Gamma(\cB)}\times N^{\Gamma(\cB)}$ determinant. $\Gamma$ is an integer that depends on the choices of edge cuts $\{I_j\}$, $\Gamma=3D+\lvert I\rvert + 2(\lvert I_2\rvert+\lvert I_4\rvert+\cdots \lvert I_{k-3} \rvert)$ for $k$ odd, and $\Gamma=3D+2(\lvert I_1\rvert+\lvert I_3\rvert+\cdots \lvert I_{k-3} \rvert)$ for $k$ even
\be
\label{PrtF2T}
Z_\cB(\lambda,N)=\int  d\nu(\xi)e^{-\Tr\ln\bigl[  \un^{\otimes \Gamma(\cB)} -g_\cB \frak{M}_\cB(\xi) \bigr] },
\ee
where $\frak{M}_\cB(\xi)=i{\bf C}_\cB.\frak{H}_\cB(\xi)$, ${\bf C}_\cB$ is the complex square symmetric covariance of size $N^{\Gamma(\cB)}\times N^{\Gamma(\cB)}$of the integrated fields
\begin{eqnarray}
\label{eqref=Codda}
\newcommand*{\temp}{\multicolumn{1}{c|}{}}
\newcommand*{\tempideux}{\multicolumn{1}{c|}{-i\bbbone^{\otimes \lvert I_2\rvert}}}
\newcommand*{\tempitrois}{\multicolumn{1}{c|}{-i\bbbone^{\otimes \lvert I_4\rvert}}}
\newcommand*{\tempo}{\multicolumn{1}{c|}{0}}
{\bf C}_{odd}=\left(\begin{array}{ccccccc}
\bbbone^{\otimes D}&\multicolumn{1}{c|}{0} &&&&& \\ 
0&\multicolumn{1}{c|}{\bbbone^{\otimes \lvert I \rvert}} &&&&& \\ 
\cline{1-4}
& \temp & 0 &  \tempideux &  &\Large\mbox{{$0$}} & \\ 
& \temp & - i\bbbone^{\otimes \lvert I_2\rvert} & \tempo &  & & \\ 
\cline{3-6} 
&&& \temp & 0 & \tempitrois & \\
& &\Large\mbox{{$0$}}& \temp & -i \bbbone^{\otimes  \lvert I_4\rvert}& \tempo & \\
\cline{5-6} 
&&&&&&\ddots
\end{array}\right),
\end{eqnarray}
\begin{eqnarray}
\label{eqref=Coddab}
\newcommand*{\temp}{\multicolumn{1}{c|}{}}
\newcommand*{\tempiun}{\multicolumn{1}{c|}{-i\bbbone^{\otimes \lvert I_1\rvert}}}
\newcommand*{\tempitrois}{\multicolumn{1}{c|}{-i\bbbone^{\otimes \lvert I_3\rvert}}}
\newcommand*{\tempo}{\multicolumn{1}{c|}{0}}
{\bf C}_{even}=\left(\begin{array}{cccccc}
\multicolumn{1}{c|}{\bbbone^{\otimes D}} &&&&& \\ 
\cline{1-3}
 \temp & 0 &  \tempiun &  &\Large\mbox{{$0$}} & \\ 
 \temp & - i \bbbone^{\otimes \lvert I_1\rvert}& \tempo &  & & \\ 
\cline{2-5} 
&& \temp & 0 & \tempitrois & \\
 &\Large\mbox{{$0$}}& \temp & -i \bbbone^{\otimes \lvert I_3\rvert}& \tempo & \\
\cline{4-5} 
&&&&&\ddots
\end{array}\right).\nonumber \\
\end{eqnarray}
The matrix $\frak{H}_\cB$ is Hermitian, and has two different forms for $k$ odd or even 
\begin{equation}
\frak{H}^{odd}_\cB= \left(
\begin{array}{c|ccccc}
  0 &   \beta_1 \otimes\un^{\otimes\lvert\tilde J_1\rvert} &  \alpha_1\otimes\un^{ \otimes\lvert  J_{2}\rvert} &\cdots & \alpha_{k-2} \otimes\un^{ \otimes\lvert  J_{k-1}\rvert} & \bbbone^{\otimes D} \\ \hline
 \raisebox{-4pt}{ $\beta_1^\dagger\otimes\un^{\otimes\lvert\tilde J_1\rvert}$ }& &&\raisebox{-30pt}{{\huge\mbox{{$0$}}}} \\[-3ex]
  \alpha_1^\dagger\otimes\un^{ \otimes\lvert  J_{2}\rvert} & \\
  \vdots & \\[0.5ex]
  \alpha_{k-2}^\dagger\otimes\un^{ \otimes\lvert  J_{k-1}\rvert} & \\[1ex]
\bbbone^{\otimes D}& \\
[1ex]
\end{array}
\right), 
\end{equation}
\begin{equation}
\frak{H}^{even}_\cB= \left(
\begin{array}{c|ccccc}
  0 &  \sigma\otimes\un^{ \otimes\lvert  J_{1}\rvert}  & \beta_2 \otimes\un^{\otimes\lvert\tilde J_2\rvert}  &  \cdots & \alpha_{k-2}\otimes\un^{ \otimes\lvert  J_{k-1}\rvert} & \bbbone  ^{\otimes D}  \\ \hline
  \raisebox{-4pt}{ $  \sigma^\dagger \otimes\un^{ \otimes\lvert  J_{1}\rvert}$} & &&\raisebox{-30pt}{{\huge\mbox{{$0$}}}} \\[-3ex]
  \beta_2^\dagger\otimes\un^{\otimes\lvert\tilde J_2\rvert} & \\
  \vdots & \\[0.5ex]
  \alpha_{k-2}^\dagger\otimes\un^{ \otimes\lvert  J_{k-1}\rvert} & \\[1ex]
\bbbone^{\otimes D}  & \\[1ex]
\end{array}
\right).
\end{equation}

In this block matrix, the various blocks may not have the same size. The $(k+1)$ blocks of the first raw are $N^D\times N^{\lvert I_j\rvert}$ rectangular matrices, where $j=i-1$ when it regards $\beta_i$ and $j=i+1$ when it comes to $\alpha_i$, which are Kronecker products
\begin{align}
&[\alpha_i \otimes\un^{\otimes\lvert J_{i+1}\rvert}]_{m_1,\cdots, m_D\ ;\ n_1\cdots, n_{\lvert I_{i+1}\rvert}}=[\alpha_i]_{m_{i_1},\cdots\ m_{i_{\lvert \tilde J_{i+1} \rvert}}\ ;\ n_{j_1}\cdots n_{j_{\lvert I_{i+1}\setminus  J_{i+1}\rvert}}} \delta_{m_{i'_1},n_{j'_1}}\cdots \delta_{m_{i'_{\lvert J_{i+1} \rvert}}, n_{j'_{\lvert J_{i+1}\rvert}}},\\
&[\beta_i \otimes\un^{\otimes\lvert \tilde J_i\rvert}]_{m_1,\cdots, m_D\ ;\ n_1\cdots, n_{\lvert I_{i-1}\rvert}}=[\beta_i]_{m_{i_1},\cdots m_{i_{\lvert J_i \rvert}}\ ;\ n_{j_1}\cdots n_{j_{\lvert I_{i-1}\setminus \tilde J_i\rvert}}} \delta_{m_{i'_1},n_{j'_1}}\cdots \delta_{m_{i'_{\lvert \tilde J_i \rvert}}, n_{j'_{\lvert \tilde J_i\rvert}}}.
\end{align}
or using the simplified notations introduced previously, 
\begin{align}
&[\alpha_i \otimes\un^{\otimes\lvert J_{i+1}\rvert}]_{J_{i+1}\sqcup\tilde J_{i+1}\ ;\ I_{i+1}}=\alpha_{i\ \mid\ \tilde J_{i+1} \ ;\  I_{i+1}\setminus J_{i+1}}\ \delta_{J_{i+1}\ ;\  J_{i+1}},\\
&[\beta_i \otimes\un^{\otimes\lvert \tilde J_i\rvert}]_{J_i\sqcup\tilde J_i\ ;\  I_{i-1}}=[\beta_i]_{ J_i \ ;\  I_{i-1}\setminus \tilde J_i} \ \delta_{\tilde J_i\ ;\  \tilde J_i}.
\end{align}
Note that the first block is of size $N^D\times N^D$, as the two last ones. This is because $I_{k-1}$ always has $D$ colors, as it contains all the colors that reach a tensor vertex.
Therefore we have explicitly
\begin{equation}\label{equamodd}
\frak{M}^{odd}_\cB= \left(
\begin{array}{c|ccccc}
  0 &   i\beta_1\otimes\un^{\otimes\lvert\tilde J_1\rvert} & i \alpha_1\otimes\un^{ \otimes\lvert  J_{2}\rvert} &\cdots & i\alpha_{k-2}\otimes\un^{ \otimes\lvert  J_{k-1}\rvert}  & i\bbbone^{\otimes D} \\ \hline
   \raisebox{-4pt}{ $i\beta_1^\dagger \otimes\un^{\otimes\lvert\tilde J_1\rvert} $}& &&\raisebox{-30pt}{{\huge\mbox{{$0$}}}} \\[-3ex]
  \beta_3^\dagger\otimes\un^{\otimes\lvert\tilde J_3\rvert} & \\
  \vdots & \\[0.5ex]
\bbbone^{\otimes D}& \\[1ex]
\alpha_{k-2}^\dagger\otimes\un^{\otimes\lvert J_{k-1}\rvert} 
\end{array}
\right),
\end{equation} 
\begin{equation}\label{equameven}
\frak{M}^{even}_k= \left(
\begin{array}{c|ccccc}
  0 &  i\sigma\otimes\un^{ \otimes\lvert  J_{1}\rvert} & i\beta_2\otimes\un^{\otimes\lvert\tilde J_2\rvert}   &  \cdots & i\alpha_{k-2} \otimes\un^{ \otimes\lvert  J_{k-1}\rvert}& i\bbbone^{\otimes D}  \\ \hline
  \raisebox{-4pt}{ $ \beta_2^\dagger \otimes\un^{\otimes\lvert\tilde J_2\rvert} $}  & &&\raisebox{-30pt}{{\huge\mbox{{$0$}}}} \\[-3ex]
 \sigma^\dagger\otimes\un^{ \otimes\lvert  J_{1}\rvert}& \\
  \vdots & \\[0.5ex]
\bbbone^{\otimes D} & \\[1ex]
 \alpha_{k-2}^\dagger\otimes\un^{ \otimes\lvert  J_{k-1}\rvert}& \\[1ex]  
\end{array}
\right).
\end{equation}\\
}

Again, some tensorial products of the $N\times N$ identity are redundant and might be factorized, as explained for $\mathbf{H}_\cB$.
Please notice that in the matrix case the notation ${\bf \mathbb H} $ and   ${\mathbb M} $ was used for matrices after factorizing 
one $N$ by $N$ identity factor. In the tensor case the different notation ${\frak H} $ and   ${\frak M} $ 
is used since we have not yet performed any similar factorization.

Returning to the factorization \eqref{PrtF12_tens} we have a representation in terms of slightly smaller matrices
\be
\label{PrtF2Tfact}
Z_\cB(\lambda,N)=\int  d\nu(\xi)e^{-N^{\Theta} \Tr\ln\bigl[  \un^{\otimes \Gamma(\cB) - (k+1) \Theta } -g_\cB \mathbb{M}_\cB(\xi) \bigr] },
\ee
where $\mathbb{M}_\cB = iC \mathbb{H}_\cB$ are  now matrices similar to $\frak{M}_\cB = i{\bf C} \frak{H}_\cB$
but of smaller size.
\vspace{0.3cm}

\subsection{Analyticity Domain and Borel Summability}
\vspace{0.3cm}

This section is devoted to reintroduce the $\epsilon $ regulators and prove the following theorem
confirming non-perturbatively the previous representation. The arguments mirror exactly those of Section \ref{bsmat}
but with different powers of $N$.

\begin{theorem}\label{Boreltensor}
The partition function $Z_\cB(\lambda,N)$ 
is Borel-LeRoy summable of order $m=k(\cB)-1$, in the sense of Theorem \ref{blrsok}.
More precisely it is analytic in $\lambda$ in the \emph{shrinking} domain 
$D^m_{\rho_m(N)} =\{\lambda\in \mathbb C:  \Re \lambda^{-1/m}> [\rho_m(N)]^{-1}\}$ with
$ \rho_m (N) = N^{-u(\cB)} r_m$, $r_m >0$
 independent of $N$ and 
\bee
\label{defu}
u(\cB) := \frac{2 t(\cB) k(\cB )- s(\cB)}{k(\cB )-1}, \quad 
t(\cB) :=\frac{1}{2}\max\bigl[ \sup_{i=0, \cdots k-2} \vert I_i\vert + \vert  J_{i+1} \vert\ - \Theta ;\ D- \Theta\bigr] .
\ee
In that domain 
$Z_\cB(\lambda,N)$ admits the convergent HIF representation \eqref{PrtF2T} with all integration contours regularized
in the manner of Section \ref{imag}.
\end{theorem}
\hspace{0.cm}

Remark that we could also study the free energy $F_\cB(\lambda,N) =N^{-D}  \log Z_\cB(\lambda,N)$, but 
since we have not yet a sufficiently strong estimate on the 
scaling behavior in $N$ to prove a constant bound (independent of $N$) on $F_\cB$ in the most general case for
$\cB$, we postpone this to a future study.

Again we shall in fact prove analyticity and uniform Taylor remainder estimates in a slightly larger (but similarly shrinking 
as $N \to \infty$) domain $E^m_{\rho_m (N)}$ 
consisting of all $\lambda$'s with $\vert \lambda \vert  < [\rho_m (N)]^m$,  and
$\vert \arg \lambda \vert < \frac{ m \pi }{2}$ containing
the smaller tangent disk $D^m_{\rho_m (N)}$ of diameter $\rho_m (N)$.

We reintroduce again the $\epsilon $ regulators, substituting $a \to a - i \epsilon \tanh a$
and $b \to b + i \epsilon \tanh b$ into all imaginary factors $e^{- i a^2}$ and $e^{+ i b^2}$, and 
into all $a$ and $b$ linear-dependent coefficients of $\mathbb{M}_\cB (\{a,b\})$.
Hence the matrix $ \mathbb{M}_\cB$ becomes
\bee  \mathbb{M}_\cB( \{a,b\}) \to \mathbb{M}_\cB( \{a,b\})  + \epsilon \mathbb{N}_\cB (\{a,b\})
\ee 
where $\mathbb{N}_\cB $ is a 
$N^{\Gamma(\cB) - (k+1) \Theta }\times N^{\Gamma(\cB) - (k+1) \Theta }$
 matrix with any non zero matrix element of the form 
$\pm ( i ) \frac{1}{\sqrt 2}(\tanh a_{jk} \pm  \tanh b_{jk})$ for some $j,k$
where the factor $i$ may or may not be present. Hence we have the following generalization of Lemma \ref{lemnormnk}\\

\begin{lemma} 
\label{lemnormnka}
The norm of $\mathbb{N}_\cB$ is uniformly bounded by $2\sqrt{k(\cB)} N^{t(\cB)} $, hence
\bee \Vert \epsilon g_\cB \mathbb{N}_\cB (\{a,b\})\Vert \le  2\epsilon \vert g_\cB \vert \sqrt{k(\cB)}  N^{t(\cB)}  \quad \quad\forall \{a,b\}.
\ee
\end{lemma}
\prf Simply bound $\Vert \mathbb{N}_\cB \Vert $ by its Hilbert-Schmidt norm. 
Each rectangular matrix $\alpha_i\otimes \un^{\otimes \lvert J_{i+1}\rvert - \Theta}$ has at most $N^{\vert \tilde J_{i+1}\vert + \vert I_{i+1} \setminus J_{i+1}\vert + \lvert J_{i+1}\rvert - \Theta }=N^{\vert I_i\vert + \vert J_{i+1}\vert - \Theta}$ non-zero coefficients (where we made use of (\ref{eqref:reltn_i})), and each $\beta_i\otimes \un^{\otimes \lvert \tilde J_i\rvert- \Theta}$ has at most $N^{\vert  J_i\vert + \vert I_{i-1} \setminus \tilde J_i\vert + \lvert \tilde J_i\rvert- \Theta}=N^{\vert I_{i-1}\vert + \vert J_i\vert- \Theta}$ non-zero coefficients. The last block has $N^{D-\Theta}$ non zero coefficients, as it is a tensor product of identities. This implies that $\bN_\cB$ has at most $2[\sum_{i=0}^{k-2} N^{\vert I_i\vert + \vert J_{i+1}\vert- \Theta} +N^{D- \Theta}]\le 2k(\cB)N^{2t(\cB)}$ non-zero coefficients, and each one of them has a squared module smaller than 2.
\qed

\medskip

We compute again the characteristic polynomial of $\mathbb{M}_\cB ( \{a,b\})$ 
by generalizing 
the identity
\[ \left(
\begin{array}{c|c|c|c}
  (1-x)^2\un^{\otimes n_0}    &   -  g_\cB A_1                &             \cdots     &   -   g_\cB A_k \\ [+1ex]\hline
 - (1-x)  g_\cB B_1  &  (1-x)\un^{\otimes n_1}            &       0     &                 0       \\[+1ex]\hline
  \vdots                             &     0      &  \ddots        &                 \vdots            \\[+1ex]\hline
 - (1-x) g_\cB B_k &  0                       &  \cdots &       (1-x)\un^{\otimes n_k}      \\[+1ex]\hline
\end{array}
\right)= 
\left(\begin{array}{c|c|c|c}
 U    &  -  g_\cB A_1                &           \cdots     &     -    g_\cB A_k \\ [+1ex]\hline
 0  &  (1-x)\un^{\otimes n_1}  &          0          &                  0       \\[+1ex]\hline
   \vdots                       &           0                      &  \ddots        &             \vdots            \\[+1ex]\hline
  0 &  0                       &        \cdots      &    (1-x)\un^{\otimes n_k} \\[+1ex]
\end{array}
\right)
\left(\begin{array}{c|c|c|c}
  \un^{\otimes n_0}    &   \quad   0  \quad        &           \cdots     &       \quad   0 \quad\\ [+1ex]\hline
- g_\cB B_1  &  \un^{\otimes n_1}   &          0           &                  0       \\[+1ex]\hline
   \vdots                       &              0                     &  \ddots        &             \vdots            \\[+1ex]\hline
- g_\cB B_k &  0                       &        \cdots      &    \un^{\otimes n_k} \\[+1ex]
\end{array}
\right),
\]\\
to rectangular matrices $A_j$ of sizes $N^{n_0}\times N^{n_j}$ and  $B_j$ of sizes $N^{n_j}\times N^{n_0}$,
where $U=(1-x)^2\un^{\otimes n_0}-g_\cB^2\sum_{j=1}^k A_{j}B_{j}  =  (1-x)^2 \un^{\otimes n_0} -\,g_\cB^2 \bigl(i H_\cB(\xi) - \eta(k(\cB)) \beta_1\beta_1^\dagger\otimes\un^{ \otimes(\lvert\tilde J_1\rvert-\Theta)} \bigr) $, since the $A_j$ and $B_j$ are taken in the first generalized
row and column of \eqref{equamodd}-\eqref{equameven}. \\

It follows that the characteristic polynomial of $\un^{\otimes (\Gamma-\Theta)} - g_\cB \mathbb{M}_\cB$ is
\bee 
\det [ (1-x) \un^{\otimes (\Gamma-\Theta)} - g_\cB \mathbb{M}_\cB ]   = (1-x)^{
N^{\Gamma-\Theta -2D}}
\det 
\biggl[(1-x)^2 \un^{\otimes (D-\Theta)} -\,g_\cB^2 \bigl( i H_\cB(\xi)  -\eta(k(\cB)) \beta_1\beta_1^\dagger\otimes\un^{ \otimes(\lvert\tilde J_1\rvert-\Theta)} \,\bigr)\biggr]
\ee

We deduce in exactly the same way an upper bound on the resolvent:\\

\begin{lemma} For $\lambda \in E^m_{\rho_m(N)}$ we have
\begin{equation}
\Vert (\un^{\otimes (\Gamma-\Theta)}-g_\cB \mathbb{M}_\cB )^{-1}\Vert \le [ \sin \frac{\pi}{4k(\cB)} ]^{-1}. \label{resobound1a}
\end{equation} \label{analytlemma1a}
\end{lemma}
\prf
By the previous Lemma
the non-trivial eigenvalues of $\un^{\otimes (\Gamma-\Theta)}-g_\cB \mathbb{M}_\cB$ 
must be of the form $x= 1\pm g_\cB \sqrt{ y}$ 
where $y$ belongs to the spectrum of the matrix $i H_\cB(\xi)  - \eta(k(\cB)) \beta_1\beta_1^\dagger\otimes\un^{ \otimes(\lvert\tilde J_1\rvert-\Theta)} $. But  $y$ belongs to the spectrum of that matrix if and only if
\bee  \det (-y + i H_\cB(\xi) - \eta(k(\cB)) \beta_1\beta_1^\dagger\otimes\un^{ \otimes(\lvert\tilde J_1\rvert-\Theta)} )=0 .
\ee
Therefore, since in the domain $E^m_{\rho_m (N)}$
the argument of $g_\cB$  is bounded by $\frac{ (k(\cB)-1) \pi }{4k(\cB)}$,  the argument of 
$ \pm g_\cB  \sqrt{ y} $ (when $y \ne 0$) must lie in 
\bea I_{k(\cB)} & =&[\frac{ \pi} {4} -  \frac{ (k(\cB)-1) \pi }{4k(\cB)}, \frac{3 \pi} {4} +  \frac{ (k(\cB)-1) \pi }{4k(\cB)} ] \cup [-\frac{3 \pi} {4}-  \frac{ (k(\cB)-1) \pi }{4k(\cB)} ,  -\frac{ \pi} {4} +  \frac{ (k(\cB)-1) \pi }{4k(\cB)} ]  \nonumber  \\
&=&  [\frac{\pi}{4k(\cB)},  \pi - \frac{\pi}{4k(\cB)} ] \cup [- \pi +\frac{\pi}{4k(\cB)}, -\frac{\pi}{4k(\cB)}].
\eea
We conclude then in exactly the same way as for Lemma \ref{analytlemma2}. \qed\\\\


As before we introduce $R_k=r_{k-1}^{\frac{k-1}{2k}}$.

\begin{lemma} For $\lambda \in E^m_{\rho_m(N)}$, choosing again $\epsilon= R_k^{-1}\frac{\sin (\pi /4k(\cB)) }{4 \sqrt {k_\cB}}$   we have
\begin{equation}
\Vert [\un^{\otimes (\Gamma-\Theta)}-g_\cB (\mathbb{M}_k +\epsilon\mathbb{N}_k )]^{-1}\Vert \le 2 [ \sin \frac{\pi}{4k(\cB)} ]^{-1}. \label{resobound2a}
\end{equation} \label{analytlemma2a}
\end{lemma}
\prf We recall that for $\lambda \in E^m_{\rho_m(N)}$, $\vert \lambda \vert \le \rho_m(N)^{m} $. Since $\rho_m (N) = N^{-u(\cB)}r_m$ and 
$m=k-1$,  it implies
\bee
\vert g_\cB \vert =\vert  \lambda\vert^{\frac{1}{2k}} N^{\frac{-s}{2k}} \le \rho_m(N)^{\frac{m}{2k}} 
N^{-\frac{s}{2k}} = N^{-\frac{u[k-1]  + s}{2k}}r_{k-1}^{\frac{k-1}{2k}} = N^{-t}R_k,
\ee
where $k,\, m,\, s,\, t,\, u$ all depend on $\cB$ (see \eqref{defu}).
Hence by Lemma \ref{lemnormnka} we have
$\Vert \epsilon g_\cB \mathbb{N}_\cB \Vert\le \frac{1}{2}\sin \frac{\pi}{4k}$. Since 
\bee   [\un^{\otimes (\Gamma-\Theta)}-g_\cB (\mathbb{M}_\cB +\epsilon\mathbb{N}_\cB )]^{-1} = 
[\un^{\otimes (\Gamma-\Theta)}- (\un^{\otimes (\Gamma-\Theta)}-g_\cB \mathbb{M}_\cB)^{-1} g_\cB \epsilon\mathbb{N}_\cB ]^{-1} (\un^{\otimes (\Gamma-\Theta)}-g_\cB \mathbb{M}_\cB)^{-1}
\ee
it implies
\bee  
\Vert [\un^{\otimes (\Gamma-\Theta)}-g_\cB (\mathbb{M}_\cB +\epsilon\mathbb{N}_\cB )]^{-1} \Vert \le  \bigl( 1-  [ \sin \frac{\pi}{4k} ]^{-1} 
\frac{1}{2}\sin \frac{\pi}{4k}  \bigr)^{-1}  [ \sin \frac{\pi}{4k} ]^{-1} =  2 [ \sin \frac{\pi}{4k} ]^{-1}.
\ee
\qed

The rest of the proof of Theorem \ref{Boreltensor} then parallels the end of Section \ref{bsmat}. 

\hspace{0.3cm}

\subsection{Explicit Example: Melonic Sixth Order Interactions and a Non-Planar Tenth Order Interaction}
\label{Examples}

\hspace{0.3cm}

At any rank $D$ there are two types of melonic invariants $\cB_6$ of order 6, pictured in Figure \ref{fig:3d6n} for D=3.
The first type contains $D$ invariants $\cB^1_{c}$, one for each color $c$.
$\cB^1_{c}$ is obtained by picking a color $c$, and performing a partial trace 
\bee
A_c(T)= [\bar T._{\hat c} T]
\ee
where the notation $\hat c$ stands for  all colors except $c$. The matrix $A_c$
is therefore a matrix acting on $\cH_c$ and 
$\cB^1_{c}$ is obtained by tracing the cube of this matrix
\bee
\cB^1_{c}=: \Tr_{c} [A_c (T)]^3.
\ee

The corresponding partition function is 
\bee  
Z_{\cB^1} (\lambda, N) = \int d\mu (T) e^{- \lambda N^{-4} \cB^{1}} .
\ee

\begin{figure}[h!]
\includegraphics[scale=.5]{cycle-eps-converted-to.pdf}\hspace{0.5cm}\includegraphics[scale=.6]{cycle_cut2-eps-converted-to.pdf}\hspace{0.3cm}\includegraphics[scale=.8]{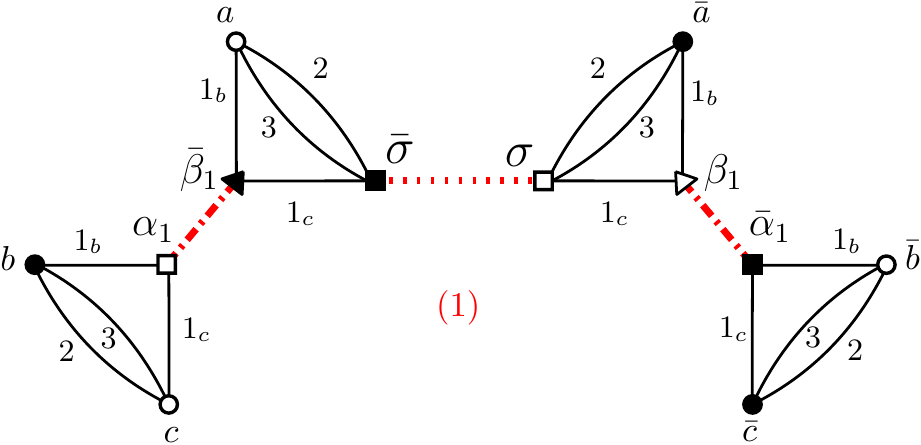}
\caption{\label{fig:cycle_dec} 
Graphical decomposition for the simpler melonic graph of order six in D=3.}
\end{figure}

The edge-cut $\cI$ is shown in Figure \ref{fig:cycle_dec}, as well as the full graphical decomposition. The successive chosen color sets are $I=\{1,2,3\}$, and $I_1=\{1_b,1_c\}$. Also, $J_1=\{1_b\}$, $\tilde J_1 = \{2_a,3_a\}$, and $J_2=\{2_c,3_c\}$. The expression of the partition function corresponding to these choices is 
\bee
Z_{\cB^1}(\lambda,N) =\int   d\mu^c(T)d\mu^c(\sigma)d\mu^c_{X}(\alpha_1,\beta_1)
e^{  ig_{\cB^1} \bigl(  [\bar T._{\tilde J_1}\sigma]._{I_1}\beta_1  + g_{\cB^1}[\bar T._{J_2}T]._{I_1}\alpha_1 + c.c. \bigr)}.
\ee
After the $\sigma$ integration, one obtains 
\bea
Z_{\cB^1}(\lambda,N) &=&\int   d\mu^c(T)d\mu^c_{X}(\alpha_1,\beta_1)
e^{ g_{\cB^1}^2\bigl( i [\bar T._{J_2}T]._{I_1}\alpha_1 + c.c. -  [\bar T._{1_b}\beta_1]._{I}  [\bar\beta_1._{1_b}T] \bigr)}\\
&=&\int   d\mu^c(T)d\mu^c_{X}(\alpha_1,\beta_1)
e^{ g_{\cB^1}^2 
\bar T ._{\llbracket 1,D \rrbracket} \bigl(
 i (\alpha_1+\alpha_1^\dagger)\otimes\un^{\otimes(2)} -  [\bar \beta_1._{1_c} \beta_1]\otimes\un^{ \otimes(2)} \bigr)._{\llbracket 1,D \rrbracket} T },
\eea
so that, taking into account that $\alpha_1$ and $\beta_1$ are actually matrices with first index $1_b$ and second index $1_c$, the integration over tensor $\sigma$ gives 
\be
Z_{\cB^1}(\lambda,N)=\int  d\mu^c_{X}(\alpha_1,\beta_1)e^{-N^2\Tr\ln\bigl[\,  \un -\,g_{\cB^1}^2 \bigl(\ i(\alpha_1+\alpha_1^\dagger) \ -\  \beta_1\beta_1^\dagger \bigr)\ \bigr] }.
\ee
With the notation of the previous sections, this model has $k=3,\,m=2,\,t=1,\,s=4$, and $u=1$. The only differences with the matrix invariant of order six are the squared factor $N^2$ in front of the trace and the power of $N$ in $g_{\cB^1}$, leading to a different value of $u$ :
\bee
Z_{\cB^1}( \lambda,N) =\int d\mu_X^c(\alpha_1,\beta_1)e^{-N^2 \Tr\ln\bigl[  \bbbone^{ \otimes 4 }- g_{\cB^1}{\mathbb{M}}_{\cB^1}(\alpha_1,\beta_1) \bigr] }, \quad g_{\cB^1} = \frac{\lambda^{1/6}}{N^{2/3}}, \quad {\mathbb{M}}_{\cB^1}= \begin{pmatrix}
0  &  i\beta_1 & i\alpha_1 &i\bbbone  \\
i\beta_1^\dagger & 0 & 0 &0 \\
\un & 0 & 0 &0 \\
\alpha_1^\dagger & 0 & 0 & 0
\end{pmatrix}.
\end{equation}
Here $\Theta(\cB_1)=2$ is optimal and the $\Theta(\cB_1)$ identity tensorial factors have been factorized in ${\mathbb{M}}_{\cB^1}$, giving the $N^2$ factor.\\

\begin{figure}[h!]
\hspace{0.5cm}\includegraphics[scale=.6]{mel2-eps-converted-to.pdf}\hspace{1.5cm}\includegraphics[scale=.7]{mel2_cut-eps-converted-to.pdf}\hspace{0.5cm}\includegraphics[scale=.8]{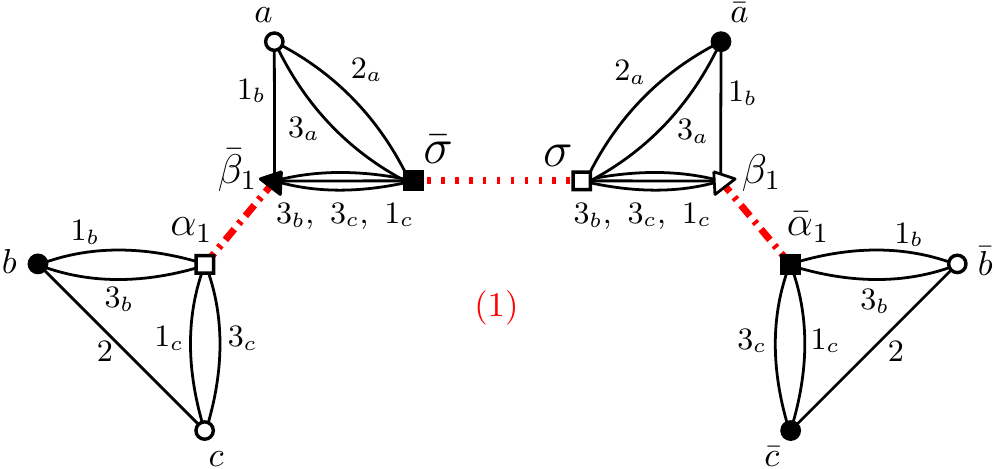}
\caption{\label{fig:mel6_dec} 
Graphical decomposition for the other melonic graph of order six in D=3.}
\end{figure}

The second type of invariant is obtained by picking two colors $c, c'$, hence there are $d(d-1)/2$
such invariants $\cB^2_{c,c'}$. We define the $N^2\times N^2$ matrix (acting on $\cH_{cc'}$), 
\bee  A _{c,c'}(T)=  [\bar T  ._{\widehat{ \{c,c'\}}} T ], 
\ee
with first indices those of $T$ left free in the above summation, and second indices the free indices of $\bar T$.
The previously defined $A_c $ and $A_{c'}$ act on $\cH_{c}$ and
$\cH_{c'}$, so their matrix tensor product $[A _{c} \otimes A _{c'} ]$
is a matrix also acting on $\cH_{cc'}$. Then $\cB^2_{c,c'}$ is defined by
\bee
\cB^2_{c,c'}=: \Tr_{cc'} \bigl( [A _{c}(T) \otimes A _{c'}(T) ] . A _{c,c'}(T)\bigr) ,
\ee
and corresponds to the graph in the left of Figure \ref{fig:mel6_dec}  (in the case $D=3$). The associated partition function is 
\bee  
Z_{\cB^2} (\lambda, N) = \int d\mu (T) e^{- \lambda  N^{-4}\cB^{2}}.
\ee

We choose the edge-cut $\cI$ ias on the left of Figure \ref{fig:mel6_dec}. The successive chosen color sets are $I=\{1_a,2_c,3_a,3_b,3_c\}$, and $I_1=\{1_b,1_c,3_b,3_c\}$. Here, $J_1=\{1_b\}$ and $\tilde J_1 = \{2_a,3_a\}$, but $J_2=\{2_c\}$. The expression of the partition function corresponding to these choices is 
\bee
Z_{\cB^2}(\lambda,N) =\int   d\mu^c(T)d\mu^c(\sigma)d\mu^c_{X}(\alpha_1,\beta_1)
e^{  ig_{\cB^2} \bigl(  [\bar T._{\tilde J_1}\sigma]._{I_1}\beta_1  + g_{\cB^2}[\bar T._{2_c}T]._{I_1}\alpha_1 + c.c. \bigr)},
\ee
where now $\alpha_1$ and $\beta_1$ are rank 4 tensors. We will however understand $\alpha_1$ as a $N^2\times N^2$ square matrix $\alpha_{1\mid 1_b,3_b\ ;\ 1_c,3_c}$ and $\beta_1$ as a rectangular $N\times N^3$ matrix, $\beta_{1\mid 1_b\ ;\ 3_b , 1_c , 3_c }$.
After the $\sigma$ integration, one obtains 
\bea
Z_{\cB^2}(\lambda,N) &=&\int   d\mu^c(T)d\mu^c_{X}(\alpha_1,\beta_1)
e^{ g_\cB^2\bigl( i [\bar T._{2_c}T]._{I_1}\alpha_1 + c.c. -  [\bar T._{1_b}\beta_1]._{I}  [\bar\beta_1._{1_b}T] \bigr)}\\
&=&\int   d\mu^c(T)d\mu^c_{X}(\alpha_1,\beta_1)
e^{ g_\cB^2 
\bar T ._{\llbracket 1,D \rrbracket} \bigl(
 i (\alpha_1+\bar \alpha_1)\otimes\un^{\otimes(1)} -  [\bar \beta_1._{\{1_c,3_b,3_c\}} \beta_1]\otimes\un^{ \otimes(2)} \bigr)._{\llbracket 1,D \rrbracket} T },
\eea
\be
Z_{\cB^2}(\lambda,N)=\int  d\mu^c_{X}(\alpha_1,\beta_1)e^{-N\Tr\ln\bigl[\,  \un -\,g_{\cB^2}^2 \bigl(\ i(\alpha_1+\alpha_1^\dagger) \ -\   \beta_1\beta_1^\dagger\otimes\un^{ \otimes(1)}]\bigr)\ \bigr] }.
\ee
This example has $k=3,\,m=2,\,t=3,\,s=4$, and $u=7$. It
exhibits rectangular matrices and identity factors that are not factorable, which is the case for general symmetric tensor invariants. One can express this result in terms of a linear Hermitian matrix, 

\setlength{\arrayrulewidth}{.6pt}
$$
\newcommand*{\temp}{\multicolumn{1}{c|}{}}
\newcommand*{\tempi}{\multicolumn{1}{c|}{i\bbbone}}
\newcommand*{\tempo}{\multicolumn{1}{c|}{0}}
\newcommand*{\tempun}{\multicolumn{1}{c|}{\bbbone}}
Z_{\cB^2}( \lambda, N) = \int d\mu_X^c(\alpha_1,\beta_1)e^{-N \Tr\ln\bigl[  \bbbone^{ \otimes (10) }-g_{\cB^2}{\mathbb M}_{\cB^2}(\alpha_1,\beta_1) \bigr] }
\quad \mathrm{and} \quad
{\bM}_{\cB^2}=\left(\begin{array}{cclccccccc}
&\temp & \multicolumn{1}{c|}{\raisebox{-2ex}[0.2cm][0.2cm]{\makebox[2cm][c]{{$i\beta_1\otimes \un$}}} } & \raisebox{-1.5ex}[0.2cm][0.2cm]{\makebox[0.7cm][r]{{$i\alpha_1$}}} & \temp & \tempi & 0 \\ 
\cline{6-7}
            &\temp &    \temp &      \qquad \      & \temp & \tempo & i\un \\ 
\hline
 &\temp & \\[0ex]
\raisebox{-2ex}[0.2cm][0.2cm]{\makebox[0.3cm][c]{{$\ \ \ \beta_1^\dagger\otimes \un$}}} &\temp&&  \raisebox{-12ex}[0.2cm][0.2cm]{\huge\mbox{{$0$}}} \\ 
 &\temp & \\[0ex]
  &\temp & \\[0ex]
\cline{1-2}
 \tempun & \tempo \\
  \cline{1-2}
 \tempo & \tempun\\
 \cline{1-2}
 \raisebox{-1.7ex}[0.2cm][0.2cm]{\makebox[0.4cm][r]{{$\alpha_1^\dagger$}}}& \temp \\
 & \temp 
\end{array}\right), \quad
$$
where $g_{\cB^2} = \frac{\lambda^{1/6}}{N^{2/3}}$,
$\Theta(\cB_2)=1$ is optimal and the $\Theta(\cB_2)$ identity tensorial factors have been factorized, giving rise to the $N$ factor before the trace. We recall that here $\beta_1\otimes\un$ is a $N^2\times N^4$ matrix, and $\alpha_1$ a $N^2\times N^2$ matrix, $\un$ being the $N\times N$ identity, as usual.\\

The previous $k=3$ examples are a bit special, in particular because all
positive tensor invariants at $k=3$ are planar, and in fact melonic. We could worry what happens when the initial invariant, hence also the initial decomposition step $F.\bar\sigma$ is a non-planar graph. Hence in our last explicit example 
we treat an example of this type with $k=5$. 

\begin{figure}[h!]
\hspace{1.5cm}\includegraphics[scale=0.7]{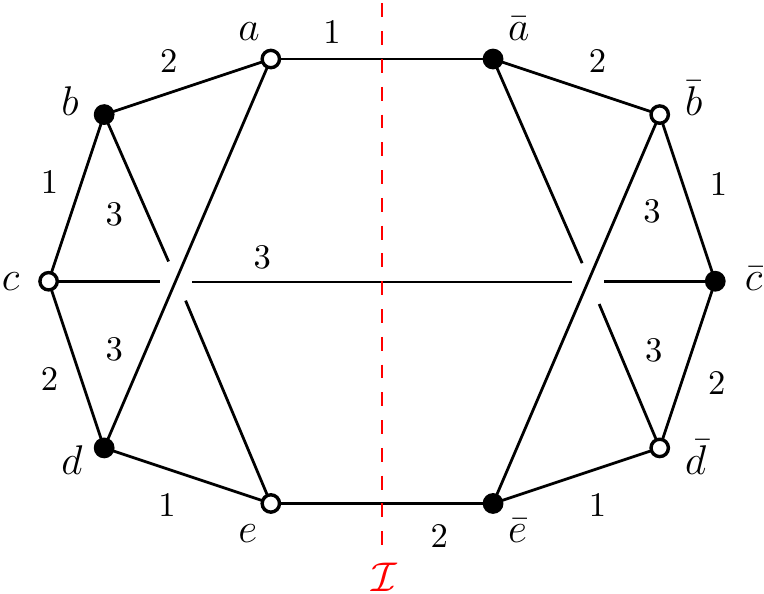}\hspace{1.5cm}\includegraphics[scale=0.7]{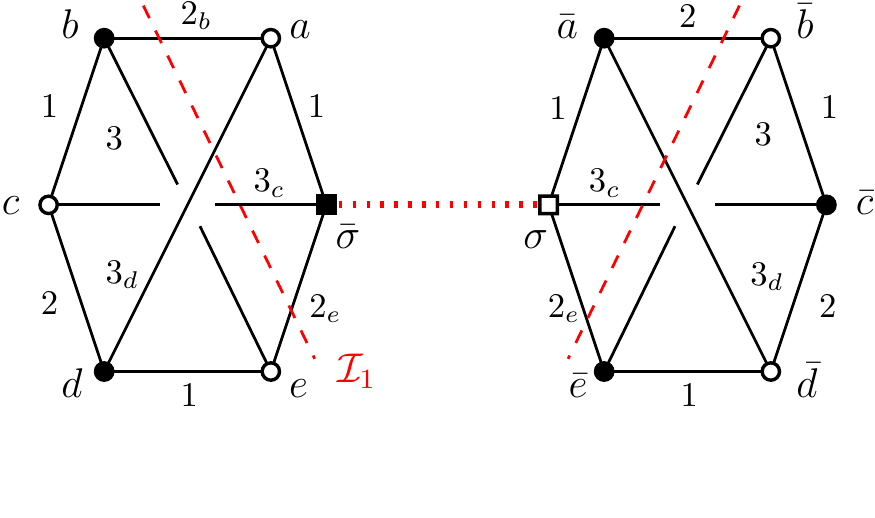}
\caption{\label{fig:k5}  A non-planar positive $k=5$ tensor invariant and its initial intermediate field step.}
\end{figure}

We consider the partition function $Z_{\cB} (\lambda, N) = \int d\mu (T) e^{- \lambda  N^{-s}\cB},$ 
where $\cB$ is represented on the left of Figure \ref{fig:k5} together with its axis of symmetry. Note that the correct scaling $s$ for a non-trivial perturbative $1/N$ expansion for this invariant is not known.  However, we know that for $s=D-1$ the $1/N$ expansion is at least defined, although possibly trivial.

\begin{figure}
\includegraphics[scale=0.8]{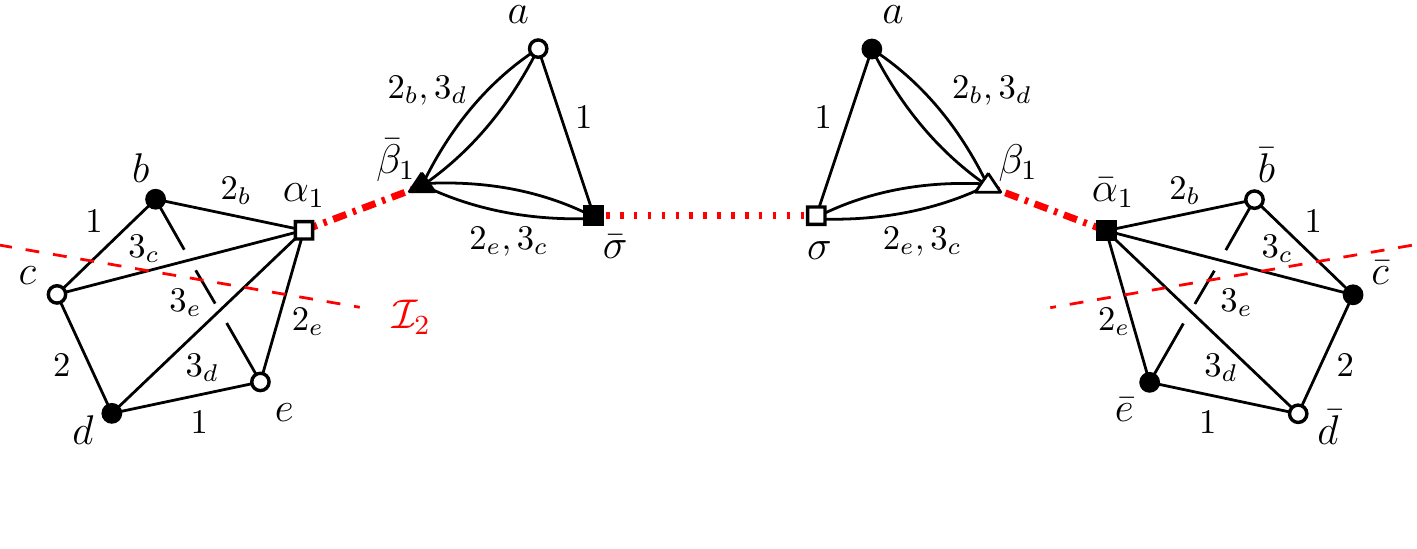}
\\
\includegraphics[scale=0.8]{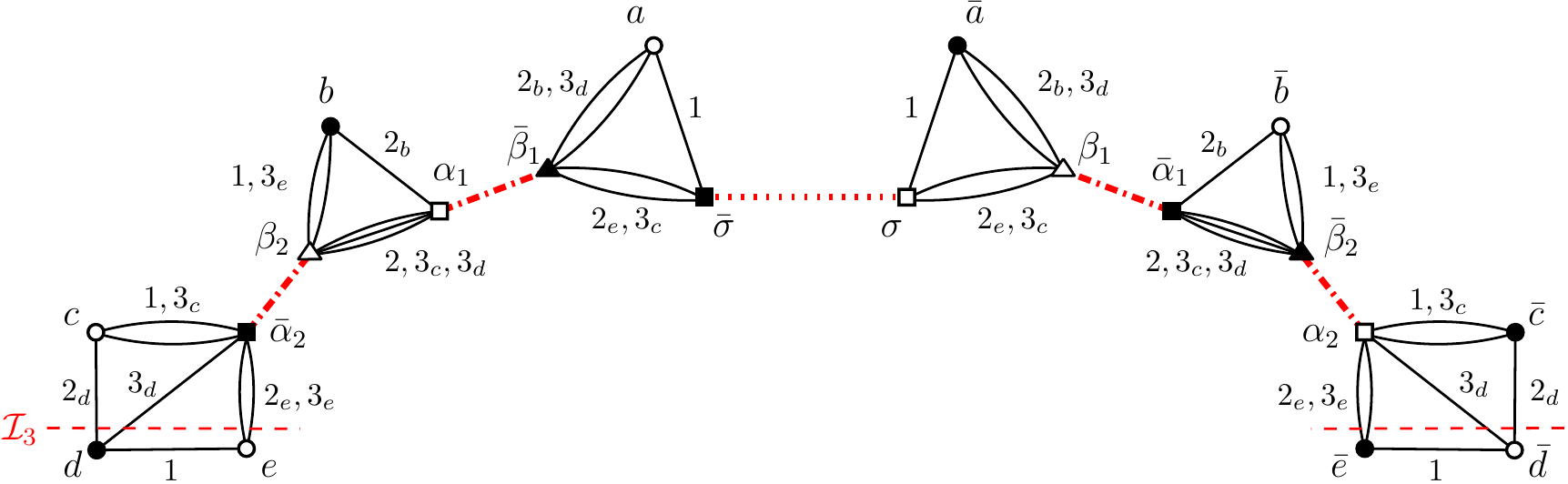}
\label{fig:k5_cut1}  
\end{figure}
\begin{figure}[h!]
\includegraphics[scale=0.8]{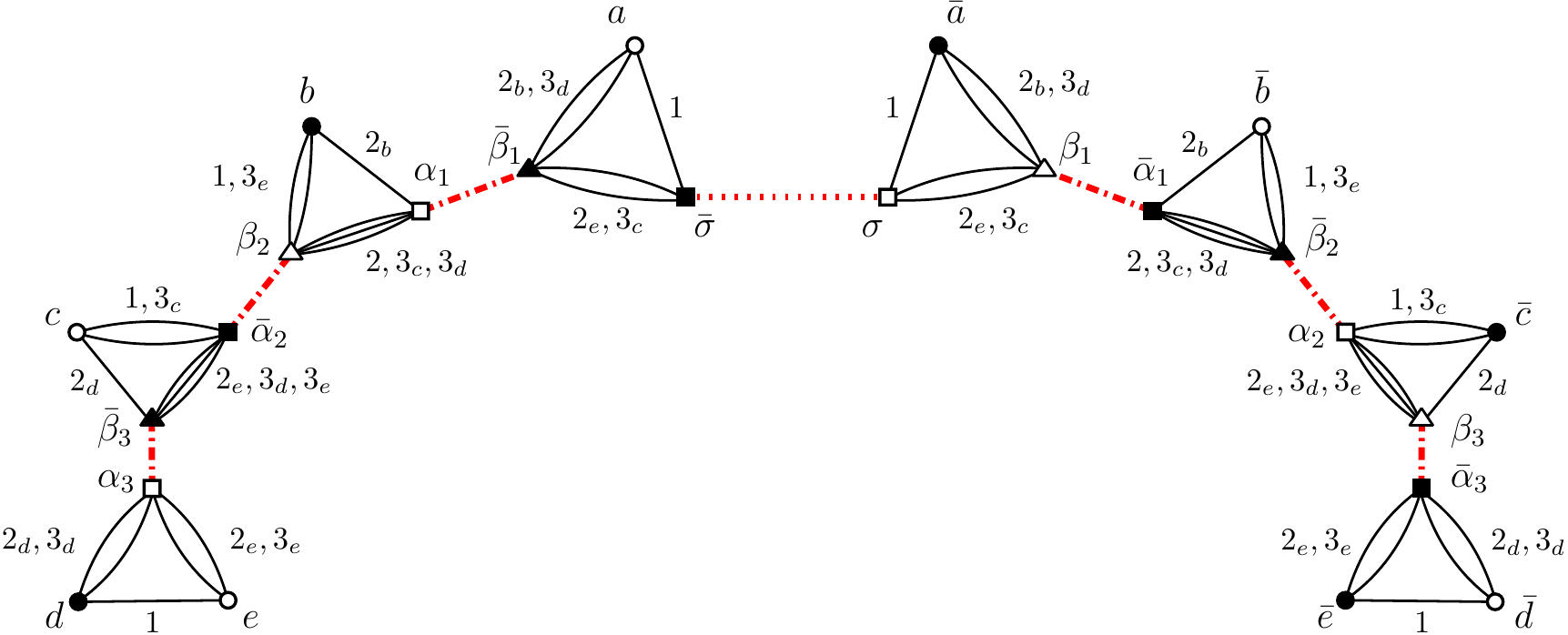}
\caption{\label{fig:k5_cut2} Full intermediate field decomposition, before integration of the field with even indices. }
\end{figure}
For this example we only provide a graphical decomposition and the resulting expression of the partition function. By looking at the triangular graphs of the intermediate field decomposition, one can read the sizes of the involved rectangular matrices, together with the colors of the spaces in which they act. As $k$ is odd in this case, the remaining fields after integration are those labeled with odd indices. One can see that  $\alpha_1$ is $N\times N^3$, $\beta_1$ and $\alpha_3$ are $N^2\times N^2$,  $\beta_3$ is $N\times N^3$, all acting on spaces of color 2,3. Color 1 is therefore factorable in the sum of tensor products $\ i \mathbf{H}_\cB(\xi) \ -\ \eta(k)\beta_1\beta_1^\dagger\otimes\un^{ \otimes\lvert\tilde J_1\rvert} \,\bigr)$. $\Gamma(\cB)$ is $3\times3+ 5+5+5=24$, and the factorization of the identity acting on color 1 leaves a size $\Gamma-(k+1)\times \Theta=24-6\times 1=18$ linear matrix,
\setlength{\arrayrulewidth}{.6pt}
\begin{eqnarray}
\newcommand*{\temp}{\multicolumn{1}{c|}{}}
\newcommand*{\tempi}{\multicolumn{1}{c|}{i\bbbone}}
\newcommand*{\tempo}{\multicolumn{1}{c|}{0}}
\newcommand*{\tempun}{\multicolumn{1}{c|}{\bbbone}}
Z_{\cB}( \lambda, N) = \int d\mu_X^c(\alpha_1,\beta_1)d\mu_X^c(\alpha_3,\beta_3)e^{-N \Tr\ln\bigl[  \bbbone^{ \otimes (18) }-g_{\cB}{\mathbb M}_{\cB}\bigr] },  \quad
g_{\cB} = (\lambda N^{-s})^{1/10},
\\\nonumber\\\nonumber\\
\newcommand*{\temp}{\multicolumn{1}{c|}{}}
\newcommand*{\tempi}{\multicolumn{1}{c|}{i\bbbone}}
\newcommand*{\tempo}{\multicolumn{1}{c|}{0}}
\newcommand*{\tempun}{\multicolumn{1}{c|}{\bbbone}}
\bM_{\cB}=\left(\begin{array}{cclccccccccccccc}
&\temp &  \multicolumn{1}{c|}{\raisebox{-2ex}[0.2cm][0.2cm]{\makebox[2cm][c]{{$i\beta_1\otimes \un$}}} }  &  \multicolumn{1}{c|}{\raisebox{-2ex}[0.2cm][0.2cm]{\makebox[2cm][c]{{$i\alpha_1\otimes \un$}}} }  &  \multicolumn{1}{c|}{\raisebox{-2ex}[0.2cm][0.2cm]{\makebox[2cm][c]{{$i\beta_3\otimes \un$}}} } & \multicolumn{1}{c|}{\raisebox{-1.5ex}[0.2cm][0.2cm]{\makebox[1.2cm][c]{{$i\alpha_3$}}}} & \tempi & 0 \\ 
\cline{7-8}
            &\temp &  \temp  & \temp & \temp & \temp & \tempo & i\un \\ 
\hline
 &\temp & \\[0ex]
\raisebox{-2ex}[0.2cm][0.2cm]{\makebox[0.3cm][c]{{$\ \ \ \beta_1^\dagger\otimes \un$}}} &\temp\\ 
 &\temp & \\[0ex]
  &\temp & \\[0ex]
\cline{1-2}       
 &\temp & \\[0ex]
\raisebox{-2ex}[0.2cm][0.2cm]{\makebox[0.3cm][c]{{$\ \ \ \beta_3^\dagger\otimes \un$}}} &\temp&& \raisebox{-12ex}[0.2cm][0.2cm]{\huge\mbox{{$0$}}} \\ 
 &\temp & \\[0ex]
  &\temp & \\[0ex]
\cline{1-2}
 &\temp & \\[0ex]
\raisebox{-2ex}[0.2cm][0.2cm]{\makebox[0.3cm][c]{{$\ \ \ \alpha_1^\dagger\otimes \un$}}} &\temp \\ 
 &\temp & \\[0ex]
  &\temp & \\[0ex]
\cline{1-2}
 \tempun & \tempo \\
  \cline{1-2}
 \tempo & \tempun\\
 \cline{1-2}
 \raisebox{-1.7ex}[0.2cm][0.2cm]{\makebox[0.4cm][r]{{$\alpha_3^\dagger$}}}& \temp \\
 & \temp 
\end{array}\right). \quad
\end{eqnarray}\\

For $s=D-1=2$, we obtain $k=5,\,m=4,\,t=3$, and $u=7$.


\end{document}